\documentclass[twocolumn]{aa}
\usepackage{epsfig}
\usepackage{graphicx}
\usepackage[latin1]{inputenc}
\def\simlt{\ \raise -2.truept\hbox{\rlap{\hbox{$\sim$}}\raise5.truept   %
\hbox{$<$}\ }}
\def\simgt{\ \raise -2.truept\hbox{\rlap{\hbox{$\sim$}}\raise5.truept   %
\hbox{$>$}\ }}                                                          %

\def\be{\begin{equation}}
\def\ee{\end{equation}}
\def\newline{\hfil\break}

\def\la{\mathrel{\hbox{\rlap{\hbox{\lower4pt\hbox{$\sim$}}}\hbox{$<$}}}}
\def\ga{\mathrel{\hbox{\rlap{\hbox{\lower4pt\hbox{$\sim$}}}\hbox{$>$}}}}

\begin{document}
\title{A multi-frequency study of the SZE in giant radio galaxies}
   \author{S. Colafrancesco\inst{1,2}, P. Marchegiani\inst{1}, P. de Bernardis\inst{3} and
   S. Masi\inst{3}}
   \offprints{S. Colafrancesco}
   \institute{INAF - Osservatorio Astronomico di Roma
              via Frascati 33, I-00040 Monteporzio, Italy.
              Email: sergio.colafrancesco@oa-roma.inaf.it
\and
              University of the Witwatersrand, Private Bag 3, 2054 South
              Africa.
              Email: sergio.colafrancesco@wits.ac.za
\and
              Dipartimento di Fisica, Universit\`a di Roma La Sapienza, P.le A. Moro 2, Roma, Italy
%
             }
\date{Received / Accepted }
\authorrunning {S. Colafrancesco et al.}
\titlerunning {SZ effect in giant radio galaxies}
\abstract
   {Radio-galaxy (RG) lobes contain relativistic electrons embedded in a tangled magnetic
   field that produce, in addition to low-frequency synchrotron
   radio emission, inverse-Compton scattering (ICS) of the cosmic microwave
   background (CMB) photons. This produces a relativistic, non-thermal Sunyaev--Zel'dovich
   effect (SZE).}
   {
   We study the spectral and spatial properties of the non-thermal SZE in
   a sample of radio galaxies and make predictions
   for their detectability in both the negative and the positive part of the SZE,
   with space experiments like Planck, OLIMPO, and Herschel-SPIRE. These cover a wide range of
   frequencies, from radio to sub-mm.
   }
  {
  We model the SZE in a general formalism that is equivalent to the relativistic covariant one
  and describe the electron population contained in the lobes of the radio galaxies
  with parameters derived from their radio observations, namely, flux, spectral index,
  and spatial extension.
  We further constrain the electron spectrum and the magnetic field of the RG lobes using
  X-ray, gamma-ray, and microwave archival observations.
  }
   {
    We determine the main spectral features
    of the SZE in RG lobes, namely, the minimum, the crossover, and the maximum  of the SZE.
    We show that these typical spectral features fall in the frequency ranges probed by
    the available space experiments. We provide the most reliable predictions for the amplitude and spectral shape of the
    SZE in a sample of selected RGs with extended lobes. In three of these objects, we
    also derive an estimate of the magnetic field in the lobe at the $\sim \mu$G level
    by combining radio (synchrotron) observations and X-ray (ICS) observations.
    These data, together with the WMAP upper limits, set constraints on the minimum momentum of the electrons
    residing in the RG lobes and allow realistic predictions for the visibility
    of their SZE to be derived with Planck, OLIMPO, and Herschel-SPIRE.
   }
  {
  We show that the SZE from several RG lobes can be observed with mm and sub-mm experiments like
  Planck, OLIMPO, and Herschel-SPIRE, as well as with ground-based telescopes that have $\simlt$ mJy
  sensitivity and sub-arcmin spatial resolution. These measurements will be crucial to disentangle
  the relativistic electron distribution from that of the magnetic field in RG lobes
  and to constrain the properties of their ICS emission, which is also visible at very high
  X-ray and gamma-ray energies.}

 \keywords{Cosmology; theory, CMB; Galaxies: active, radio galaxies, theory}
 \maketitle

\section{Introduction}
 \label{sec.intro}

The lobes of (giant) radio galaxies are unique laboratories for
constraining the plasma processes that accelerate and diffuse
relativistic electrons within large intergalactic volumes. Studies
of radio-galaxy (RG) lobes (e.g., Harris \& Krawczynski 2002,
Kataoka et al. 2003, Kronberg et al. 2004, Croston et al. 2005,
Blundell et al. 2006, and references therein) have shown, in fact,
that these extended structures contain relativistic electrons
embedded in a tangled magnetic field that produce both low-frequency synchrotron radio emission and
inverse-Compton scattering (ICS) of cosmic microwave
background (CMB) photons, a mechanism hereafter referred to as ICCMB, as
well as other radiation background.

Synchrotron radio emission and polarization are ubiquitously
observed in all the extended lobes of radio galaxies and
testify to the presence of a plasma of relativistic particles
embedded in a tangled magnetic field. Radio emission polarization
and Faraday Rotation measures also provide detailed information on
the structure of the magnetic field in and around the RG lobes
(see, e.g., Guidetti et al. 2011 and references therein). These
observations show, for instance, a typical amplitude $B_0 \approx
1.8 $ $\mu$G of the central B-field in the lobes of the RG
0208+35, assuming a typical variation of the field strength with
radius in the intracluster atmosphere around the RG.

The ICCMB mechanism up-scatters the CMB photons over a
wide range of frequencies, allowing the detection of such
emission even at high energies, where it can be observable in
X-rays and gamma-rays.
Extended X-ray emission has been, in fact, observed by
Chandra in the lobes of several powerful radio galaxies (see,
e.g., the cases of 3C 432, 3C 294 and 3C 191 discussed by Erlund
et al. 2006) and is due to ICCMB photons by the relativistic
electrons confined in the RG lobes. This is consistent
with previous findings of X-ray emission from the lobes of FR-II
radio galaxies (like 3C 223 and 3C 284, see Croston et al. 2004)
attributed to the same ICCMB mechanism.
The Fermi-LAT gamma-ray detection of a spatially resolved
gamma-ray glow emanating from the giant radio lobes of Centaurus A
(Abdo et al. 2010) also indicates that the radio-lobe gamma-ray
emission is mainly due to ICCMB, with the additional contribution,
at higher energies, from the infrared-to-optical extragalactic
background light.

As a consequence of the pervading presence of the ICCMB
mechanism in extended RG lobes, a Sunyaev-Zel'dovich
effect (SZE) from the lobes of radio galaxies is inevitably
expected (Colafrancesco 2008).
Such a non-thermal, relativistic SZE  has a specific spectral shape
that depends on the shape and on energy extent of the spectrum
of the electrons residing in the RG lobes.\\
The SZE from RG lobes has a wide astrophysical and cosmological
relevance, which allows the study of, e.g., the energetics and
pressure structure of the RG lobes, the detailed structure of
their B-field in combination with radio synchrotron observations,
and the evolution of the CMB temperature $T_{CMB}(z)$ at the RG
redshift that provides important cosmological constraints (see
Colafrancesco 2008 for an extended discussion).\\
The SZE in RG lobes has not yet been detected and so far only loose upper
limits have been derived to the SZE from these sources (see
McKinnon et al. 1991; see also the recent attempt to detect this
effect at radio wavelengths in the giant radio galaxy B1358+305 by
Yamada et al. 2010).

On the theoretical side, it has been shown that the SZE emission
is expected to be co-spatial with the relative ICS X-ray emission
(see Fig.1 and discussion in Colafrancesco 2008).
The spectral properties of the SZE are also strongly related to
those of the relative ICS X-ray (and gamma-ray) emission. In fact,
the spectral slope of the ICS X-ray emission $\alpha_X = (\alpha -
1)/2$ (where $F_{ICS} \sim E^{-\alpha_X}$) can be used to set the
electron energy spectral slope $\alpha$ (where $N_{e} \sim
E^{-\alpha}$) necessary to compute the SZE spectrum and to check
its consistency with the synchrotron radio spectral index
$\alpha_r = (\alpha - 1)/2$ (where $F_{Synch.} \sim
E^{-\alpha_r}$), which is expected to have the same value of $\alpha_X$ (see,
e.g., Longair 2003).
Observations of the ICCMB spectrum at high energies
(X-rays or gamma-rays) can also set the absolute normalization
(assuming a unique spectral slope $\alpha$) of the electron
spectrum used for the SZE calculations (see Colafrancesco \&
Marchegiani 2011).
Given the shape of the electron spectrum,
the total intensity and the spectral shape of the SZE in RG lobes
depend on the
minimum energy of the electron distribution (see, e.g.,
Colafrancesco 2008, Colafrancesco \& Marchegiani 2011).\\
In this context, it has been shown that the ICCMB X-ray emission
provides only marginal constraints to the minimum energy of the
electron population in the lobes (Colafrancesco 2008).
Since we can approximate that the ICCMB provides a photon
with energy $E_X\sim 8 (E_e/GeV)^2 keV$ (see, e.g., Longair 1993),
we can calculate 
that an X-ray instrument operating in the energy band $[E_{X,min} \div E_{X,max}]$
can only study the electrons with minimum energy
$E_{min} \approx 0.35 \, GeV (E_{X,min}/keV)^{1/2}$. This corresponds to
$\gamma_{min} \approx 484$ (here $E= \gamma m_e c^2$) for
an X-ray instrument, such as Chandra,
operating in the energy range $E_X=0.5 \div 3.0$ keV. Similar
considerations apply to X-ray instruments with sensitivity in a
different energy band.
The X-ray estimate of $\gamma_{min}$ hence provides only upper
limits to the realistic values of $\gamma_{min}$, which are
already observed to be in the range $ \sim 1 - 10^2$ (see,
e.g., Comastri et al. 2003; Croston et al. 2004; Kataoka \& Stawarz
2005; Hardcastle \& Croston 2005).
Therefore, unless the true value of $\gamma_{min}$ is
higher than $\simgt 2.2 \times 10^3$ (corresponding to a maximum
X-ray energy of 10 keV), as derived in several cases (see, e.g.,
Blundell et al. 2006, Blundell \& Fabian 2011), the value of the
RG lobe energy derived from X-ray observations (typically
sensitive to $\gamma_{min} \sim 10^3$) turns out to be a
lower limit to the true value. 
Therefore, because
the total energy value of the electron population depends on the
shape and overall extension of the electron spectrum, it
might increase by even a large factor for low values of
$\gamma_{min}$ and steep electron spectra (see Colafrancesco 2008
and Colafrancesco \& Marchegiani 2011 for a discussion).

The detection of the SZE from RG lobes can provide a determination
of the total energy density and pressure of the electron
population in the lobes. It thus allows the value of $\gamma_{min}$ to be determined more
precisely, once the slope of the
electron spectrum is determined from the radio and/or X-ray
observations (see Harris \& Grindlay 1979, Govoni \&
Feretti 2004, Colafrancesco 2008, Colafrancesco \& Marchegiani
2011).
To obtain this information, we need to determine the spectral
details of the SZE from RG lobes over a wide frequency range, from
radio to sub-mm frequencies, and the spatial extent of the SZE
signal throughout the RG lobes. The theoretical description and
the observational strategy aimed at studying the SZE in RG lobes
are the main subjects of this work.\\
The paper is organized as follows: we describe in Sect.2 the
theoretical approach to the calculation of the SZE and its
spectral and spatial properties. In this section, we also discuss
the relation between the SZE and the relative synchrotron and ICS
emission from the same relativistic particles. The general
properties of non-thermal SZE in a sample of giant radio galaxies
are derived in Sect.3, and we perform a more refined analysis of
the SZE for a subsample of the most interesting cases in Sect.4.
We discuss the multi-frequency prescriptions of the ICS emission
from RG lobes in Sect.5 and use the available X-ray data of the
RG lobes to better normalize the SZE signal from the RG lobes we
want to study. We discuss in Sect.6 the constraints imposed on the
RG lobe SZE by the WMAP data in the microwave range, and
present detailed predictions for the expected SZE signals for a
subset of the best RG lobe candidates that are observable in the
microwave, sub-mm, and mm ranges with Planck, OLIMPO, and
Herschel-SPIRE. We finally discuss our results and draw our
conclusions in Sect.7

Throughout the paper, we use a flat, vacuum--dominated
cosmological model with $\Omega_m = 0.3$, $\Omega_{\Lambda} = 0.7$,
and $H_0 = 70$ km s$^{-1}$ Mpc$^{-1}$.

\section{Theoretical approach}

In this section, we derive the spectral and spatial features of the
non-thermal, relativistic SZE produced by ICCMB photons off the
relativistic electrons (hereafter electrons) populating the RG
lobes. To provide a complete multi-frequency description of the
ICS emission in RG lobes, we also compute the high-energy ICS
spectral energy distribution (SED) as well as the synchrotron
emission due to the relativistic electrons.

The crucial quantities that are used to describe the
multi-frequency SED for RG lobes are the electron spectrum
$N_e(p,r)$, the CMB spectrum $I_0(\nu)$, and the magnetic field
distribution $B(r)$ in the RG lobes.
We consider an electron population with separable spatial and
spectral distributions as given by
\begin{equation}
N_e(p,r)=n_e(r) f_e(p) \; ,
 \label{spettro1}
\end{equation}
where $p=\beta\gamma$ is the normalized momentum and $n_e(r)$ is
the electron density at distance $r$ (in modulus) from the center of the
radio lobe, since $f_e(p)$ is normalized so as to give
$\int_0^\infty f_e(p) dp=1$.\\
We describe here the relativistic electron plasma within the RG
lobe with a single power-law momentum spectrum
\begin{equation}
 \label{leggep1}
f_{\rm e}(p;p_1,p_2,\alpha)=A(p_1,p_2,\alpha) p^{-\alpha} ~;
\qquad p_1 \leq p \leq p_2,
\end{equation}
with normalization given by
\begin{equation}
A(p_1,p_2,\alpha) = \frac{(\alpha-1)}
{p_1^{1-\alpha}-p_2^{1-\alpha}} \; .
 \label{eq.norm}
\end{equation}
An alternative way to write the power-law electron spectrum, which
we will use in the following, is to define
\begin{equation}
N_e(p,r)=k_0 g_e(r) p^{-\alpha},
  \label{spettro1bis}
\end{equation}
where $g_e(r)$ describes the spatial distribution of the electrons
and $k_0$ is a normalization factor. Comparing the two expressions
given in Eqs. (\ref{spettro1}) and (\ref{spettro1bis}), one finds
that
\begin{equation}
 n_e(r)=k_0 g_e(r) A(p_1,p_2,\alpha)^{-1}.
 \label{eq.k0}
\end{equation}
We describe the magnetic field in the RG lobes as
\begin{equation}
 B(r) = B_0 g_B(r),
\end{equation}
where $B_0$ is central value of the magnetic field and $g_B(r)$ is
the (three-dimensional) spatial distribution function that
expresses the spatial dependence of the magnetic field.\\
We finally consider the unscattered CMB spectrum $I_0(x)$ as
given by
\begin{equation}
 I_0(x) = \frac{2 (k_{\rm B} T_{CMB})^3 }{ (h c)^2} \frac{ x^3}{e^x -1},
\end{equation}
where $x \equiv h \nu /k_{\rm B} T_{CMB}$, $k_B$ is the Boltzmann
constant, $h$ is the Planck constant, and $T_{CMB}= 2.726$ K is the
CMB temperature today.

\subsection{The Sunyaev-Zel'dovich effect in RG lobes}

Following Colafrancesco (2008), we describe the spectral distortion
of the CMB spectrum observable in the direction of a RG lobe as
 \begin{equation}
\Delta I_{\rm lobe}(x)=2\frac{(k_{\rm B} T_{CMB})^3}{(hc)^2}y_{\rm
lobe} ~\tilde{g}(x) ~,
 \label{eq.deltai}
\end{equation}
where $\Delta I_{\rm lobe}(x)= I_{\rm lobe}(x) - I_0(x)$, $I_{\rm
lobe}(x)$ is the CMB spectrum in the direction of the radio lobe,
and $I_0(x)$ is the unscattered CMB spectrum in the direction of
a sky area contiguous to the RG lobe.

The Comptonization parameter $y_{\rm lobe}$ is written as
\begin{equation}
y_{\rm lobe}=\frac{\sigma_T}{m_{\rm e} c^2}\int P_{\rm e} d\ell ~,
 \label{eq.y}
\end{equation}
in terms of the pressure $P_{\rm e}$ contributed by the electronic
population.
Here $\sigma_T$ is the Thomson cross section, $m_e$ is the
electron mass, and $c$ is the speed of light.
The spectral function $\tilde{g}(x)$ of the SZE is written as
\begin{equation}
 \label{gnontermesatta}
 \tilde{g}(x)=\frac{m_{\rm e} c^2}{\langle \varepsilon_{\rm e} \rangle} \left\{ \frac{1}{\tau_e} \left[\int_{-\infty}^{+\infty} i_0(xe^{-s}) P(s) ds-
i_0(x)\right] \right\}
\end{equation}
in terms of the photon redistribution function $P(s)$ and of the
quantity
\begin{equation}
i_0(x) = I_0(x)/[2 (k_{\rm B} T_{CMB})^3 / (h c)^2] = x^3/(e^x -1)
\; .
\end{equation}
The quantity
\begin{equation}
 \langle \varepsilon_{\rm e} \rangle  \equiv  \frac{\sigma_{\rm T}}{\tau_e}\int P_e d\ell
= \int_0^\infty dp f_{\rm e}(p) \frac{1}{3} p \; v(p)\; m_{\rm e}
c,
 \label{temp.media}
\end{equation}
where $v(p)=\beta c$ is the velocity of the electron (we
recall that $p\equiv \beta \gamma)$, is the average energy of the
electron plasma (see Colafrancesco et al. 2003).
The optical depth of the electron population within the lobe is
\begin{equation}
\tau_{\rm e}= \sigma_T \int d \ell n_{\rm e}
 \label{tau}
\end{equation}
and depends parametrically (see Eq.\ref{eq.k0}) on the value of the lowest
momentum/energy of the electrons through $n_e(r)$ (Colafrancesco 2008, Colafrancesco et al.
2003).
The photon redistribution function $P(s)$, with $s =
\ln(\nu'/\nu)$ in terms of the CMB photon frequency increase
factor $\nu' / \nu$, can be calculated by repeated convolution of
the single-scattering redistribution function $P_1(s)= \int dp
f_{\rm e}(p) P_{\rm s}(s;p)$,
where $P_s(s;p)$ depends on the physics of inverse Compton
scattering process (see, e.g., Colafrancesco et al. 2003 for
details).

In calculating the SZE from the RG lobes, the relevant
momentum is the minimum momentum $p_1$ of the electron
distribution, while the value of $p_2 \gg p_1$ is irrelevant for
power-law indices $\alpha > 2$. These are indicated by the
electron spectra observed in the RG lobes that we consider here.
In fact, for $\alpha > 2$ and $p_2 \gg p_1$, the normalization of
the electron spectrum is $A \to {(\alpha -1) \over p_1^{1-\alpha}}$
(see Eq.\ref{eq.norm}).\\
Once the normalization $k_0$ in Eq.(\ref{spettro1bis}) is set and
the spatial distribution function $g_e(r)$ is given, the value of
$p_1$ sets the value of the electron density $n_{\rm e}$
and the value of the other relevant quantities that depend
on it, namely, the optical depth $\tau_{\rm e}$, the pressure
$P_{\rm e}$, and the energy density ${\cal E}_{\rm e}$ of the
non-thermal population.

The pressure $P_{\rm e}$ in the case of an electron distribution
as in Eq.(\ref{leggep1}), is written as
\begin{eqnarray}
 \label{press_rel}
 P_{\rm e}&=&n_{\rm e} \int_0^\infty dp f_e(p) \frac{1}{3} p \; v(p)\; m_e c \\
  & =& \frac{n_{\rm e} m_e c^2 (\alpha
  -1)}{6[p^{1-\alpha}]_{\rm p_2}^{p_1}}
  \left[{\cal B}_{\frac{1}{1+p^2}}\left(\frac{\alpha-2}{2},
   \frac{3-\alpha}{2}\right)\right]_{\rm p_2}^{p_1} \nonumber,
\end{eqnarray}
where ${\cal B}_x(a,b)=\int_0^x t^{a-1} (1-t)^{b-1} dt$
(see, e.g., Ensslin \& Kaiser 2000, Colafrancesco et al. 2003).

The energy density for the same electron population is written as
\begin{eqnarray}
 \label{eneden_rel}
 {\cal E}_{\rm e} &=&n_{\rm e} \int_0^\infty dp f_e(p) \bigg[ \bigg(1+p^2 \bigg)^{1/2} -1\bigg] m_e c^2 \\
  &=& \frac{n_{\rm e} m_e c^2}{[p^{1-\alpha}]_{\rm p_2}^{p_1}}
  \bigg[{1 \over 2} {\cal B}_{\frac{1}{1+p^2}} \left(\frac{\alpha-2}{2}, \frac{3-\alpha}{2}\right) \nonumber \\
   & & + p^{1 -\alpha} \left( (1+p^2)^{1/2} -1 \right) \bigg]_{\rm p_2}^{p_1} \nonumber
\end{eqnarray}
and for a relativistic population of electrons ${\cal E}_{e} = 3
\cdot P_{e}$ (see also Longair 1993 and Miley 1980 for a
more general discussion).
For an electron population with a double power-law (or more
complex) spectrum, analogous results can be obtained (see
Colafrancesco et al. 2003 for details).

\subsection{Multi-frequency ICS emission from RG lobes}

Inverse Compton scattering of relativistic electrons on target CMB
photons gives rise to a spectrum of photons stretching from the
microwave up to gamma-ray frequencies, depending on the values
of $p_1$ and $\alpha$.
The ICS power is written as
\begin{equation}
P_{\rm ICS}\left(E_\gamma,E\right) = c \, E_\gamma \int d\epsilon
\, n(\epsilon) \, \sigma(E_\gamma, \epsilon, E) \, ,
 \label{eq:ICpower}
\end{equation}
where $E_\gamma$ and $E=\gamma m_e c^2$ are the energies of the
scattered photon and the electron, respectively. Equation
\ref{eq:ICpower} is obtained by folding the differential number
density of target photons $n(\epsilon)$ with the ICS cross section
$\sigma(E_\gamma, \epsilon, E)$ given by the Klein-Nishina
formula:
\begin{equation}
\sigma(E_\gamma, \epsilon, E) = \frac{3 \sigma_T}{4 \epsilon
\gamma^2}\, G\left(q,\Gamma_e\right) \, ,
\end{equation}
where
\begin{equation}
G\left(q,\Gamma_e\right) \equiv \left[  2 q \ln q + (1+2 q)(1-q) +
\frac{\left(\Gamma_e q \right)^2 (1-q)}{2 \left( 1+ \Gamma_e
q\right)}\right]
\end{equation}
with
\begin{equation}
 \Gamma_e= 4 \epsilon \gamma / (m_e c^2) \;\; ; \;\;\;\;\;   q = E_\gamma /
\left[ \Gamma_e \left( \gamma m_e c^2 - E_\gamma \right)\right]\;.
\end{equation}
Here $\epsilon$ is the energy of the target photons.
%
Folding the ICS power in Eq.(\ref{eq:ICpower}) with the spectrum
of the electrons, $N_e(E,r)$ (derived from by Eq.\ref{spettro1}
with a variable change), the local emissivity of ICS photons of
energy $E_\gamma$ results in
\begin{equation}
 j_{\rm ICS}\left(E_\gamma, r\right) = \int dE\, N_e(E,r) P_{\rm ICS}
 \left( E_\gamma,E \right)\;,
 \label{eq:ICemiss}
\end{equation}
from which the integrated flux density spectrum
\begin{equation}
F_{\rm ICS}(E_\gamma)= \int d V_{lobe} \, \frac{ j_{\rm
ICS}\left(E_\gamma, r\right)}{4 \pi\, D^2}\;
 \label{eq:ICflux}
\end{equation}
and the ICS brightness along the line of sight (los)
\begin{equation}
S_{\rm ICS}(E_\gamma)= \int d \ell \, j_{\rm ICS}\left(E_\gamma, r
\right)\; .
 \label{eq:ICbrigtness}
\end{equation}
Here $D$ is the luminosity distance to the RG lobe.
In Eqs.~(\ref{eq:ICpower}) and (\ref{eq:ICemiss}), the limits of
integration over $\epsilon$ and $E$ are set from the kinematics of
the ICS scattering, which restricts the values of the parameter $q$
in the range $1/(4 \gamma^2) \le q \le 1$.

\subsection{Synchrotron emission from RG lobes}

The synchrotron local emissivity produced by the same
population of relativistic electrons in RG lobes with a
spectrum $N_e(\gamma,r)$ interacting with the magnetic field
$B(r)$ is given by
\begin{equation}
j_{syn}(\nu,r)=\int d \gamma
N_e(\gamma,r) P_{syn}(\nu,\gamma,r) \label{emiss.radio},
\end{equation}
where $\nu$ is the frequency of the photon
emitted (Schlickeiser 2002). The synchrotron power is given by
\begin{eqnarray}
P_{syn} &=&\int_0^\pi d\eta p(\eta) 2\pi \sqrt{3} r_0 m_e c \nu_0
\sin \eta F_S({\bar x}/\sin \eta)
 \label{eq.psyn}\\
 \nu_0&=& \frac{e B}{2\pi m_e}=2.8\left(\frac{B}{\mu\mbox{G}}\right)\,\mbox{Hz}\\
 {\bar x}&=&\frac{2\nu}{3\nu_0 \gamma^2} \left[
1+ \left( \frac{\gamma \nu_p}{\nu} \right)^2 \right]^{3/2}\\
F_S(t)&=&t \int_t^\infty K_{5/3}(\zeta)d\zeta\\
p(\eta)&=&\frac{1}{2}\sin\eta \; ,
\end{eqnarray}
where $\nu_p=8980 (n_{ep}/\mbox{ cm}^{-3})^{1/2}$ Hz is the plasma
frequency, $n_{ep}$ is the electron density of the plasma,
$r_0=2.82\times 10^{-13}$ cm is the classical radius of the
electron, and $K_{5/3}(\zeta)$ is the Bessel modified function of the
second kind (see, e.g., Abramowitz \& Stegun 1965).\\
The observed synchrotron flux is given by
\begin{equation}
F_{syn}(\nu)= \int dV_{lobe} \frac{j_{syn}(\nu,r)}{4\pi
D^2} \; ,
\end{equation}
where $D$ is the luminosity distance of the source.

\subsection{Spatial behaviour of the SZE}

The geometry of the RG lobes is generally not spherical, and
therefore the optical depth effects produced by the electron
spatial distribution must be calculated in the more general case.

If the electron population has a spatial distribution of the
general form
\begin{equation}
n_e(\vec{r})=n_e(r_x,r_y,r_z),
\end{equation}
where we define the axis $r_z$ along the los
$\ell$, the electron optical depth along the los that passes
through the projected distance $(r_x,r_y)$ from the RG lobe center
is given by the expression
\begin{eqnarray}
\tau_e(r_x,r_y)&=&\sigma_T \int_\ell n_e(\vec{r}) d\ell= \nonumber\\
&=&2\sigma_T \int_0^{\bar{R}_z(r_x,r_y)}n_e(r_x,r_y,r_z) dr_z,
\end{eqnarray}
where $\bar{R}_z(r_x,r_y)$ is determined by the geometry of the
system. For the ellipsoidal geometry we assume for the RG lobes,
we find
\begin{equation}
\bar{R}_z(r_x,r_y)=r_{z,max} \sqrt{1-\left(\frac{r_x}{r_{x,max}}\right)^2-
\left(\frac{r_y}{r_{y,max}}\right)^2},
\end{equation}
where $r_{x,max}$, $r_{y,max}$, and $r_{z,max}$ are the
intersections of the ellipsoid with the axes $r_x$, $r_y$, and
$r_z$, respectively.

\section{SZE properties of a sample of giant radio galaxy lobes}

In this section, we first select a sample of RGs with extended
lobes that are suited for our study and then evaluate the
properties of the expected SZE signals. We perform in
Sect.4 a more detailed analysis of the SZE for a sub-set of the
most interesting cases.

\subsection{The sample of giant radio galaxies}

To discuss quantitatively the SZE signals that are expected in RG
lobes and select the most viable candidates for a detection with
the available and planned space experiments, we
considered an initial sample of 21 RGs for which we have
information on the lobe extent and on their spectra obtained from
radio observations. These observations are available from
the giant RG catalog of Ishwara \& Saikia (1999), the
catalog of giant RGs of Schoenmakers et al. (2000), and the NASA
Extragalactic Database (NED) archive.
We selected the 21 RGs that fall into the visibility region of an
OLIMPO stratospheric balloon flight launched from the Svalbard
Islands (see Masi et al. 2005, Nati et al. 2007), assuming that
its ground path will circumnavigate the north pole at constant
latitude ($\sim$ 78$^o$).
In this selection procedure, the Planck and Herschel-SPIRE instruments do
not set any stringent requirements because Planck provides an
all-sky survey and Herschel-SPIRE can only provide pointed observations
of the selected sources.

We did not consider in our analysis the CenA radio galaxy, which
has very extended lobes and is clearly resolved both in the WMAP
microwave frequency range and in the GeV range probed by Fermi.
This specific case merits a dedicated and more
extensive work, which will be presented in a future paper.

This sample of 21 RGs with extended lobes is not complete
and therefore cannot be used for statistical studies. However, it
represents a collection of sources that are viable candidates for
our study of the extended non-thermal SZE.

For each one of the selected RG lobes, we collect a set of physical
information that is used to predict both the spectrum and the
spatial extension of the associated SZE. These comprise the
spatial extension of the RG lobes, as obtained from
radio observations, the synchrotron radio flux of the lobes, the
synchrotron radio spectral index of the lobes, $\alpha_r$, and the
presence of one or two extended lobes.

Using this information, we calculate the non-thermal SZE produced
in the RG lobes, following the procedure described in Sect.2
above. This procedure works under the following initial assumptions:\\
i) we assume an electron population in the RG lobes with a single
power-law energy spectrum given in Eqs. (\ref{spettro1}) and
(\ref{leggep1}) with index $\alpha=2\alpha_r+1$ (related to the
radio synchrotron spectral index $\alpha_r$) and with a minimum
momentum $p_{1}=1$;\\
ii) since most the available information on the radio galaxy lobes
comes from the synchrotron radio emission (that is, degenerate
between the electron density and the magnetic field
distributions), we assume, for simplicity, a constant electron
density and a constant magnetic field with intensity of 1 $\mu$G
within the emission region of the RG lobe. We will discuss later
on cases with different spatial distributions of both electrons
and magnetic field;\\
iii) we approximate the emission region of the lobe as an
ellipsoid with major and minor axes derived from the observed
dimensions of the available radio lobe images. We also assume that
the third axis of the ellipsoid (i.e., the one along the los)
is equal to the minor axis of the RG lobe along the other
directions;\\
iv) we calculate the SZE from the RG lobe at first order
approximation in $\tau_e$ because it ensures the required
precision accuracy of the calculations for these systems,
where the low electron density and the relatively
(with respect to the galaxy cluster case) small size of the
lobes imply a value of the optical depth $\tau_e <<1$ (see
Colafrancesco et al. 2003 for a discussion on the consequences
of this fact).

These initial assumptions yield an evaluation of the SZE expected
in the central parts of the RG lobes that is (given the available
information on the RG lobe structure) a realistic order of
magnitude. We further discuss the effects of more refined RG lobe
structure in the following.

\begin{figure}[ht!]
\begin{center}
\vbox{
 \epsfig{file=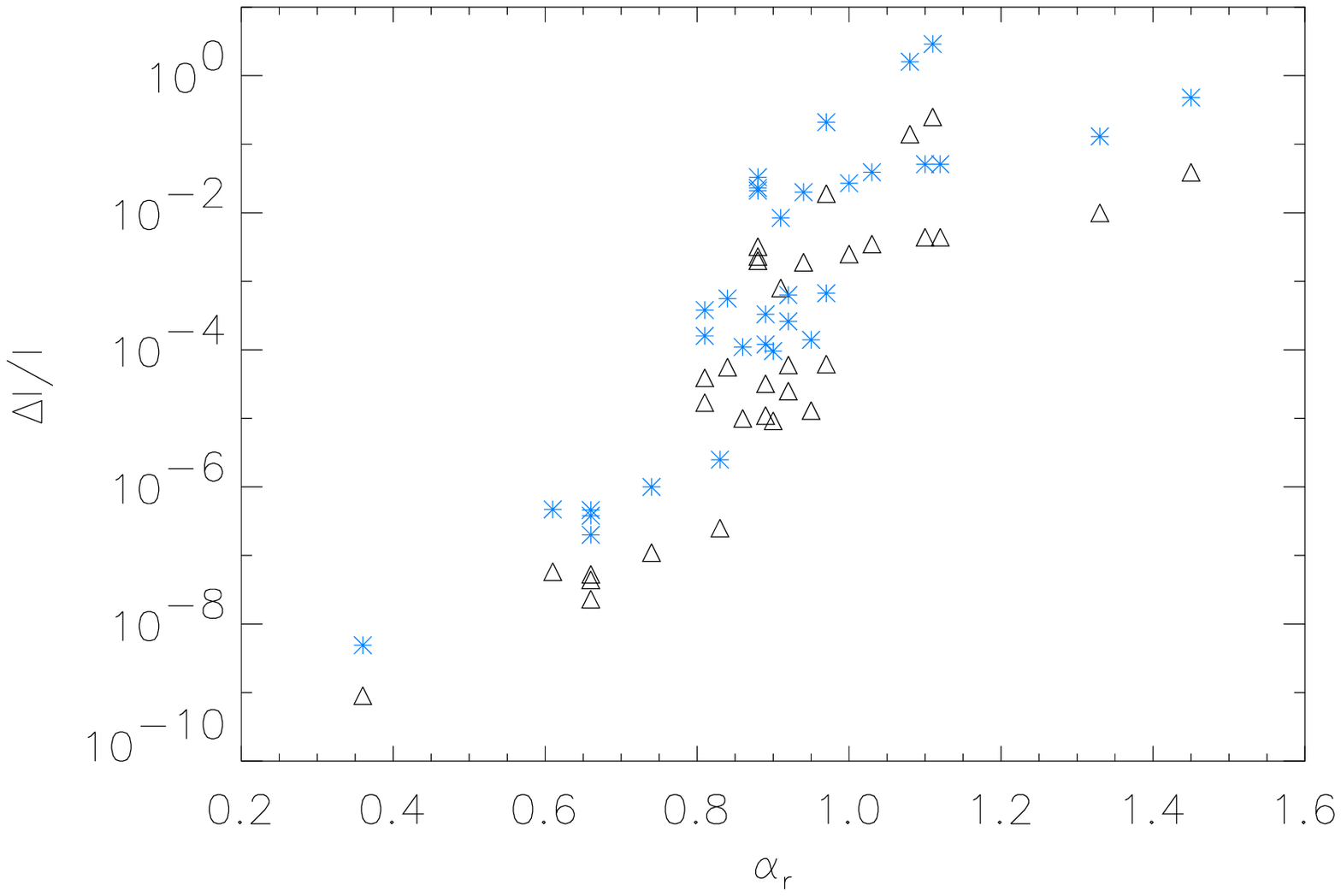,height=5.cm,width=9.cm,angle=0.0}
 \epsfig{file=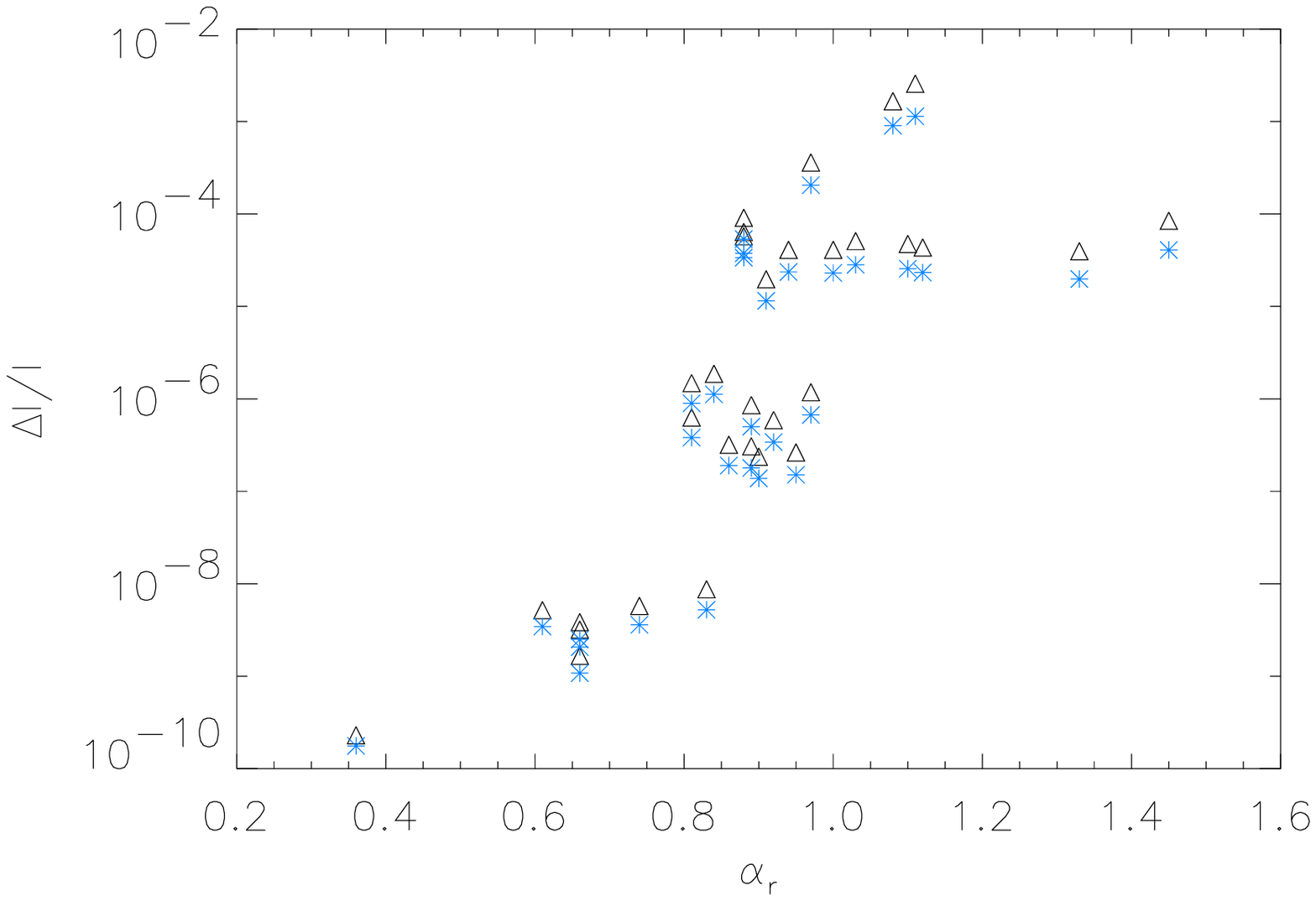,height=5.cm,width=9.cm,angle=0.0}
 \epsfig{file=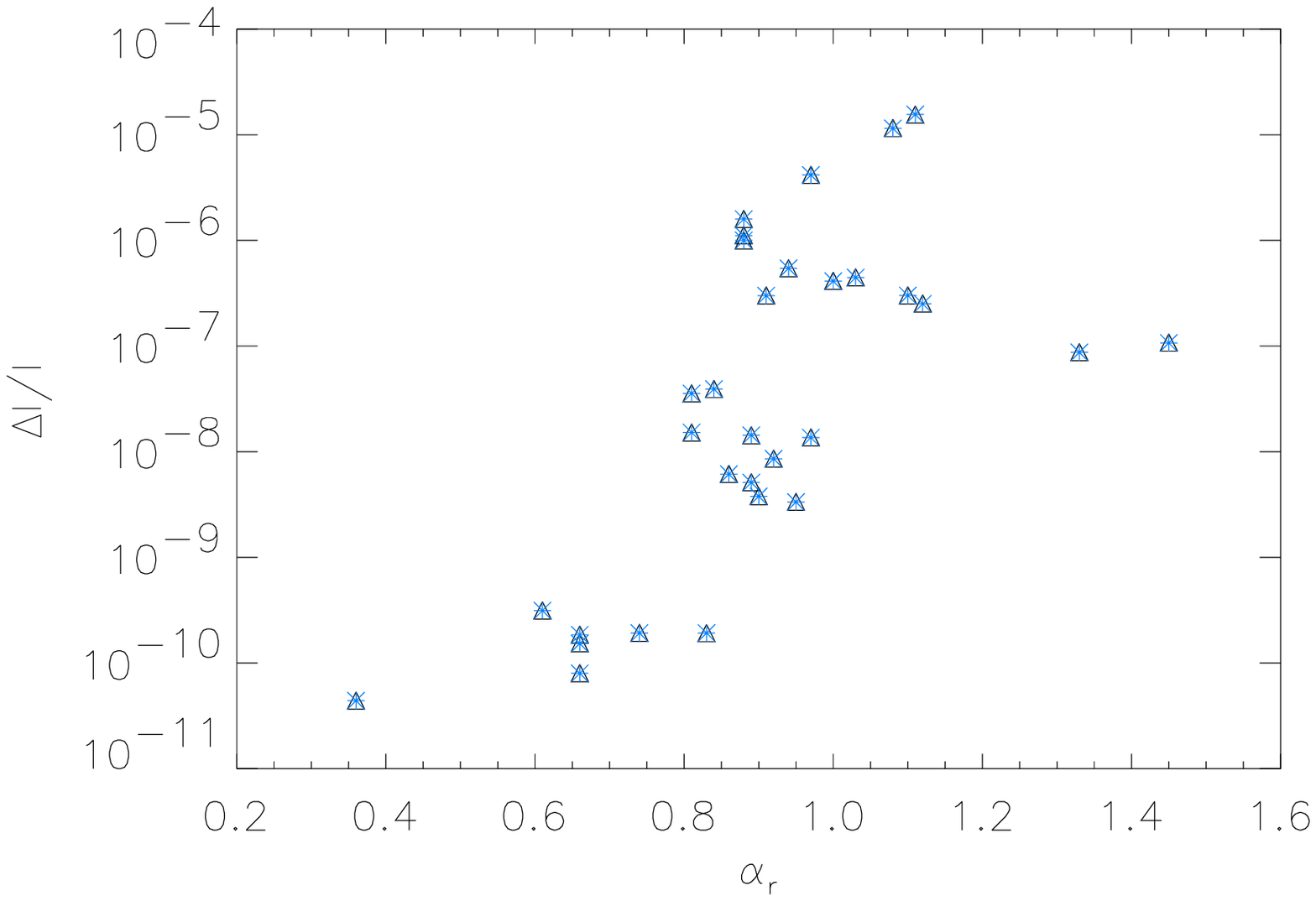,height=5.cm,width=9.cm,angle=0.0}
 \epsfig{file=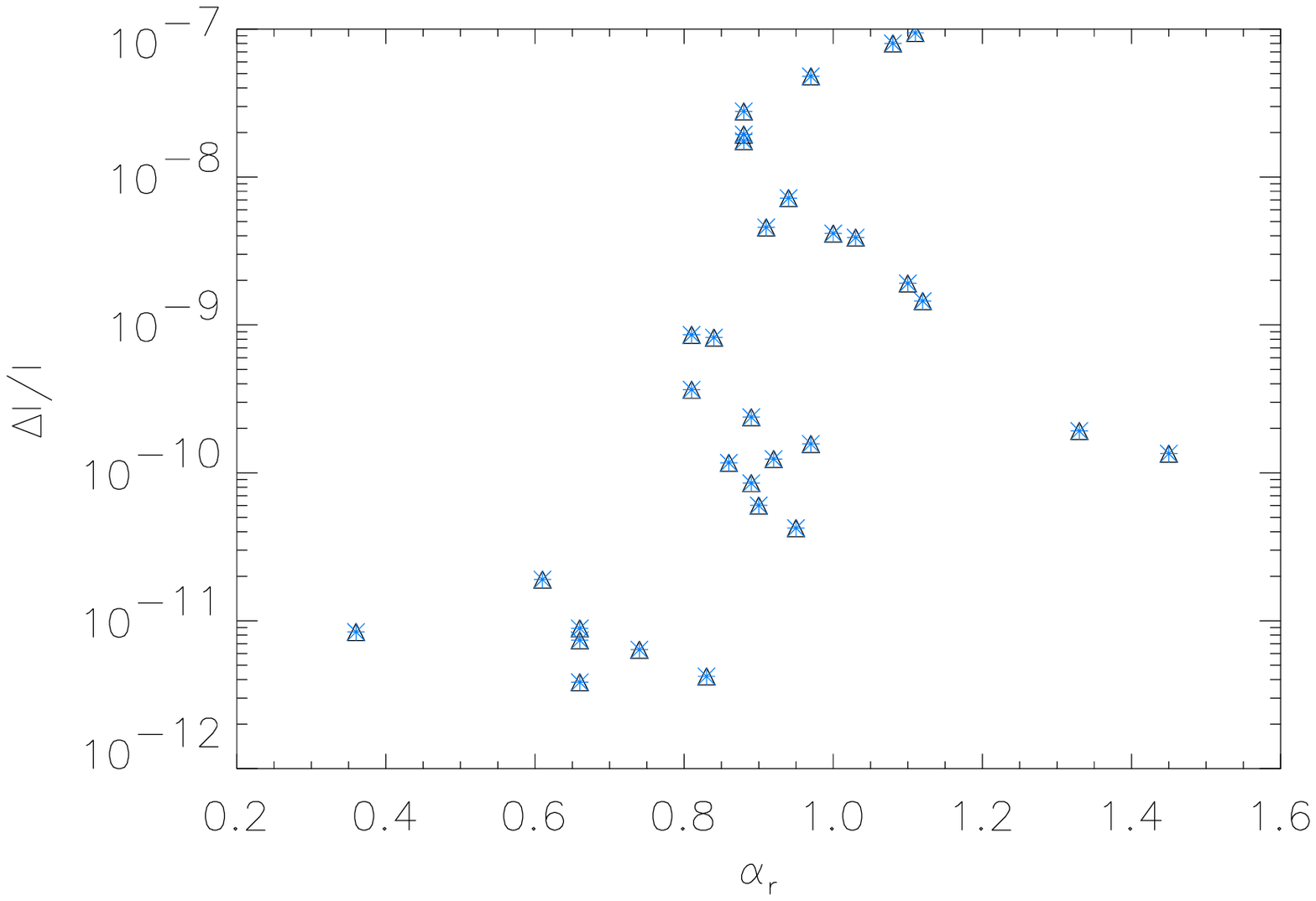,height=5.cm,width=9.cm,angle=0.0}
}
\end{center}
 \caption{\footnotesize{SZE brightness change $\Delta I / I$
 is shown as a function of the observed radio
 spectral index $\alpha_r$ for the RG lobes in Table \ref{tab.1}.
 The various panels refer to the cases $p=1,10,100,1000$ from top down.
 Black triangles refer to SZE decrement at 150 GHz (in modulus) while the blue
 asterisks refer to the SZE increment at 500 GHz for $p_1=1$ and
 to the SZE decrement at 500 GHz (in modulus) for $p_1=10$, 100, and 1000.
 The SZE signals in Table \ref{tab.1} are all evaluated for a fixed value of the
 RG lobe magnetic field ($B=1$ $\mu$G uniform in the emission region for
 all the objects considered).
 }}
 \label{fig.alphar_sz}
\end{figure}

In this framework, we calculated the SZE brightness change $\Delta
I/I$ from Eq.(\ref{eq.deltai}) at $\nu=$ 150 and 500 GHz for all
the 21 objects we selected: the results are reported in Table
\ref{tab.1}.
 This table reports the object name (column 1),
 equatorial coordinates RA (col.2) and DEC (col.3) (J2000), redshift (col.4),
 angular size (col.5), and SZE brightness change at 150 GHz (col.6) and 500 GHz
 (col.7) for the RGs we consider in this paper.
 The asterisk indicates the RGs for which we performed a more detailed
 study. For radio galaxies with more than one lobe, we indicate with an asterisk
 in col.5 the lobe that we consider in our analysis.
\begin{table*}[htb]{}
\vspace{2cm}
\begin{center}
\begin{tabular}{|*{7}{c|}}
\hline Object name         &  RA (J2000)  &    Dec (J2000)     & z
& Size (arcsec$^2$) &  -$\Delta I/I$ (150 GHz) & $\Delta I/I$ (500 GHz)\\ \hline CGCG
186-048 & 11h47m22.1s & +35d01m08s & 0.0629 & 275x169 &
$1.1\times10^{-7}$ & $1.0\times10^{-6}$\\
             &             &            &        & 388x141 & $2.5\times10^{-7}$ & $2.5\times10^{-6}$\\
B2 1158+35   & 12h00m48.7s & +34d50m10s &  0.55  & 70x42   & $2.5\times10^{-3}$ & $2.7\times10^{-2}$\\
             &             &            &        & 61x42   & $3.5\times10^{-3}$ & $3.9\times10^{-2}$\\
3C 270       & 12h19m18.6s & +05d49m26s & 0.007465 & 577x269 & $3.2\times10^{-5}$ & $3.3\times10^{-4}$\\
87GB 121815.5+635745 *& 12h20m37.7s & +63d41m07s & 0.2 & 924x528 & $3.9\times10^{-2}$ & $4.8\times10^{-1}$\\
M 87         & 12h30m49.4s & +12d23m28s & 0.00436 & 360x153 & $2.3\times10^{-3}$ & $2.3\times10^{-2}$\\
3C 274.1 *    & 12h35m26.6s & +21d20m35s & 0.422 &  89x20 & $1.9\times10^{-2}$ & $2.1\times10^{-1}$\\
4C +69.15    & 13h13m58.8s & +69d37m18s & 0.106 & 822x414 & $2.3\times10^{-8}$ & $2.0\times10^{-7}$\\
3C 292 *      & 13h50m41.8s & +64d29m31s & 0.71  & 63x21 * & $3.2\times10^{-3}$ & $3.3\times10^{-2}$\\
             &             &            &       & 64x36 & $2.0\times10^{-3}$ & $2.1\times10^{-2}$\\
B2 1358+30C *  & 14h00m43.4s & +30d19m19s & 0.206 & 384x264 & $1.3\times10^{-5}$ & $1.4\times10^{-4}$\\
             &             &            &       & 408x180 * & $1.0\times10^{-2}$ & $1.3\times10^{-1}$\\
3C 294 *      & 14h06m44.0s & +34d11m25s & 1.779 & 29x12.5 & $1.4\times10^{-1}$ & $1.6$\\
PKS 1514+00  & 15h16m40.2s & +00d15m02s & 0.052489 & 519x260 & $9.0\times10^{-10}$ & $4.9\times10^{-9}$\\
GB1 1519+512 & 15h21m14.5s & +51d05m01s & 0.37 & 312x60 & $5.6\times10^{-5}$ & $5.6\times10^{-4}$\\
3C 326       & 15h52m09.1s & +20d05m24s & 0.0895 & 684x267 & $9.2\times10^{-6}$ & $9.6\times10^{-5}$\\
             &             &            &        & 338x329 & $1.0\times10^{-5}$ & $1.1\times10^{-4}$\\
7C 1602+3739 * & 16h04m23.4s & +37d31m49s & 0.814  &  84x29 *  & $4.4\times10^{-3}$ & $5.1\times10^{-2}$\\
             &             &            &        & 100x22  & $4.4\times10^{-3}$ & $5.1\times10^{-2}$\\
MRK 1498     & 16h28m04.0s & +51d46m31s & 0.0547 & 429x154 & $5.3\times10^{-8}$ & $4.6\times10^{-7}$\\
             &             &            &        & 583x137 & $4.4\times10^{-8}$ & $3.8\times10^{-7}$\\
B3 1636+418  & 16h37m53.4s & +41d46m01s & 0.867  & 57x52   & $1.1\times10^{-5}$ & $1.2\times10^{-4}$\\
             &             &            &        & 52x47   & $6.0\times10^{-5}$ & $6.3\times10^{-4}$\\
Hercules A *  & 16h51m08.1s & +04d59m33s & 0.154  & 200x67  & $2.5\times10^{-1}$ & $2.9$\\
B3 1701+423  & 17h02m55.9s & +42d17m49s & 0.476  & 120x89  & $6.2\times10^{-5}$ & $6.7\times10^{-4}$\\
             &             &            &        & 113x67  & $2.5\times10^{-5}$ & $2.6\times10^{-4}$\\
4C 34.47     & 17h23m20.8s & +34d17m58s & 0.206  & 92x88   & $1.7\times10^{-5}$ & $1.6\times10^{-4}$\\
             &             &            &        & 83x55   & $3.9\times10^{-5}$ & $3.8\times10^{-4}$\\
87GB 183438.3+620153 & 18h35m10.9s & +62d04m08s & 0.5194 & 69x26 & $8.0\times10^{-4}$ & $8.4\times10^{-3}$\\
             &             &            &        & 39x22   & $1.9\times10^{-3}$ & $2.0\times10^{-2}$\\
4C +74.26    & 20h42m37.3s & +75d08m02s & 0.104  & 773x193 & $5.8\times10^{-8}$ & $4.7\times10^{-7}$\\
 \hline
 \end{tabular}
 \end{center}
 \caption{\footnotesize{Full sample of RG lobes.
 }}
 \label{tab.1}
 \end{table*}
The quantity $\Delta I / I$ calculated at 150 GHz for the objects
we consider here takes values that are spread over a wide range
from $\sim 10^{-9}$ to $\sim 4 \cdot 10^{-1}$, depending on the
electron spectral index $\alpha$ and the value of the optical
depth $\tau_e$. The values of $\Delta I / I$ at 500 GHz for the
same objects are in the range $\sim 4 \times 10^{-9}-2.9$.
Figure \ref{fig.alphar_sz} shows the behavior of $\Delta I / I$ as a
function of $\alpha$ for the 21 RG lobes of Table \ref{tab.1}. The
larger the $\alpha_r$, the larger is the SZE brightness change
$\Delta I/I$. This is because in our assumptions we have
steep electron spectra with $\alpha>1$, the electron spectrum is
normalized to high-energy electrons (i.e., those that produce the
radio emission), and the value of $p_1=1$ is fixed. Therefore, the electron
density (see Eq.\ref{eq.norm}), and thus the optical depth
and the SZE, are mainly due to low-energy electrons.

We then estimated the ratio between the ICS flux and the
synchrotron flux calculated at 150 GHz (see Table \ref{tab.2}) in
order to assess the level of contamination of the SZE signal from
the synchrotron flux at the lowest frequency we consider in our
analysis. 
Table \ref{tab.2} reports the object name (col.1), angular size (col.2),
SZE signal at 150 GHz (col.3), calculated by multiplying the central SZE signal by the lobe angular
area and synchrotron flux at 150 GHz (col.4), derived  by extrapolation of
the low-frequency radio spectrum with the same spectral index.
The radio flux at 150 GHz was estimated by extrapolating the
synchrotron signal measured at low frequencies (from a few
hundreds MHz to a few GHz) by using the same spectral
shape at higher frequencies, up to hundreds of GHz. This
assumption implies therefore that the synchrotron flux evaluated
at 150 GHz is actually an upper limit, since there are several
arguments that suggest a steepening of the radio synchrotron
spectrum at higher and higher frequencies (see, e.g., Kardashev
1962).\\
An intrinsic hardening of the high-$\nu$ spectrum is
likely due to either the presence of a hot-spot or a compact
source with a flatter spectrum emerging at high-$\nu$. In general, however,
extended sources show spectra steepening at
high-$\nu$ (see, e.g., Laing \&  Peacock 1980).\\
The ICCMB flux was calculated, consistent with the
assumptions we use, by multiplying the SZE signal produced at the
center of the lobe by the lobe extension. This estimate thus yields
an order of magnitude value for the realistic ICCMB flux.
The flux value is an approximated value since, also for an
electron density and B-field that are constant within the lobe,
the SZE brightness is not completely uniform due to geometrical
projection effects.\\
We find, in agreement with our theoretical description, that the
synchrotron contamination of the SZE signal is only important for
RG lobes with very flat spectra and negligible for RG
lobes with steep spectra. This result indicates that the SZE is
best visible in RG lobes with steep electron spectra, because for
these objects there is the maximum SZE signal and the minimum
synchrotron contamination signal at high frequency.
\begin{table*}[htb]{}
\vspace{2cm}
\begin{center}
\begin{tabular}{|*{4}{c|}}
\hline Object name         &  Size (arcsec$^2$) & -$\Delta I_{SZ}$
(mJy) & $F_{sync}$ (mJy)\\ \hline CGCG 186-048 & 275x169 &
$3.4\times10^{-2}$ & 3.2\\
             & 388x141 & $9.2\times10^{-2}$ & 1.0\\
B2 1158+35   & 70x42   & 50 & 1.3\\
             & 61x42   & 61 & 2.0\\
3C 270       & 577x269 & 34 & 113\\
87GB 121815.5+635745 & 924x528 & $1.3\times10^{6}$ & 0.43\\
M 87         & 360x153 & $8.6\times10^{2}$ & 3.6$\times10^{3}$\\
3C 274.1     & 89x20 & $2.3\times10^{2}$ & 30\\
4C +69.15    & 822x414 & $5.3\times10^{-2}$ & 26\\
3C 292       & 63x21 &  29 &  15\\
             & 64x36 &  31 &  16\\
B2 1358+30C  & 384x264 & 8.9 & 3.6\\
             & 408x180 & $5.0\times10^{3}$ & 0.28\\
3C 294       & 29x12.5 & $3.4\times10^{2}$ & 2.0\\
PKS 1514+00  & 519x260 & $8.2\times10^{-4}$ & $4.8\times10^{2}$\\
GB1 1519+512 & 312x60  &   7.1 &   22\\
3C 326       & 684x267 &  11   &   22\\
             & 338x329 & 7.5 &  38\\
7C 1602+3739 & 84x29  &  72 &  0.23\\
             & 100x22 &  65 &  0.33\\
MRK 1498     & 429x154 & $2.4\times10^{-2}$ & 15\\
             & 583x137 & $2.8\times10^{-2}$ & 15\\
B3 1636+418  &  57x52  & 0.22 &  0.22\\
             &  52x47  & 0.99 &  0.47\\
Hercules A   & 200x67  & $2.3\times10^{4}$ & 253\\
B3 1701+423  & 120x89  & 4.5 & 1.1\\
             & 113x67  & 1.3 & 0.99\\
4C 34.47     & 92x88   & 0.93 & 9.5\\
             & 83x55   & 1.2  & 13\\
87GB 183438.3+620153 & 69x26 & 9.7 & 3.9\\
             & 39x22 &  11  &  2.2\\
4C +74.26    & 773x193 & $5.9\times10^{-2}$ & 94\\
 \hline
 \end{tabular}
 \end{center}
 \caption{\footnotesize{Comparison of SZE signal and synchrotron flux at 150 GHz for the 21 RG listed in Table \ref{tab.1}. 
 }}
 \label{tab.2}
 \end{table*}

\section{The SZE spectrum in selected RG lobes}

From the initial sample of 21 RGs listed in Table \ref{tab.1}, we
have selected a sub-set of seven RGs with extended lobes that
provide the most interesting cases to discuss the SZE in these
structures. These objects are denoted with an asterisk in
the list of Table \ref{tab.1} and are specifically Hercules A
(z=0.154), 87GB 121815.5+635745 (z=0.2), B2 1358+30C (z=0.206), 3C
274.1 (z=0.422), 3C 292 (z=0.71), 7C 1602+3739 (z=0.814), and 3C
294 (z=1.779). \footnote{The case of 3C 294 should be considered
with caution because the X-ray emission associated with this RG is
not co-spatial with the radio lobes (see Erlund et al. 2006).
However, this X-ray emission is likely a mixture of extended (i.e.,
associated with the lobes) and compact (i.e., associated with the
core) emission.}

For the radio lobes of this restricted set of RGs, we calculated the SZE spectra over
a wide frequency range up to $10^3$ GHz and the radial profile of the
SZE surface brightness at 150 GHz (at which the SZE signal has its
minimum). We complemented the SZE analysis by
calculating the multi-frequency ICCMB emission SED from $10^{14}$
to $10^{24}$ Hz, i.e., from the soft X-ray to the gamma-ray energy
range, which has relevance for the high-E observations of ICCMB in
these systems.

\begin{figure}[ht]
\begin{center}
 \epsfig{file=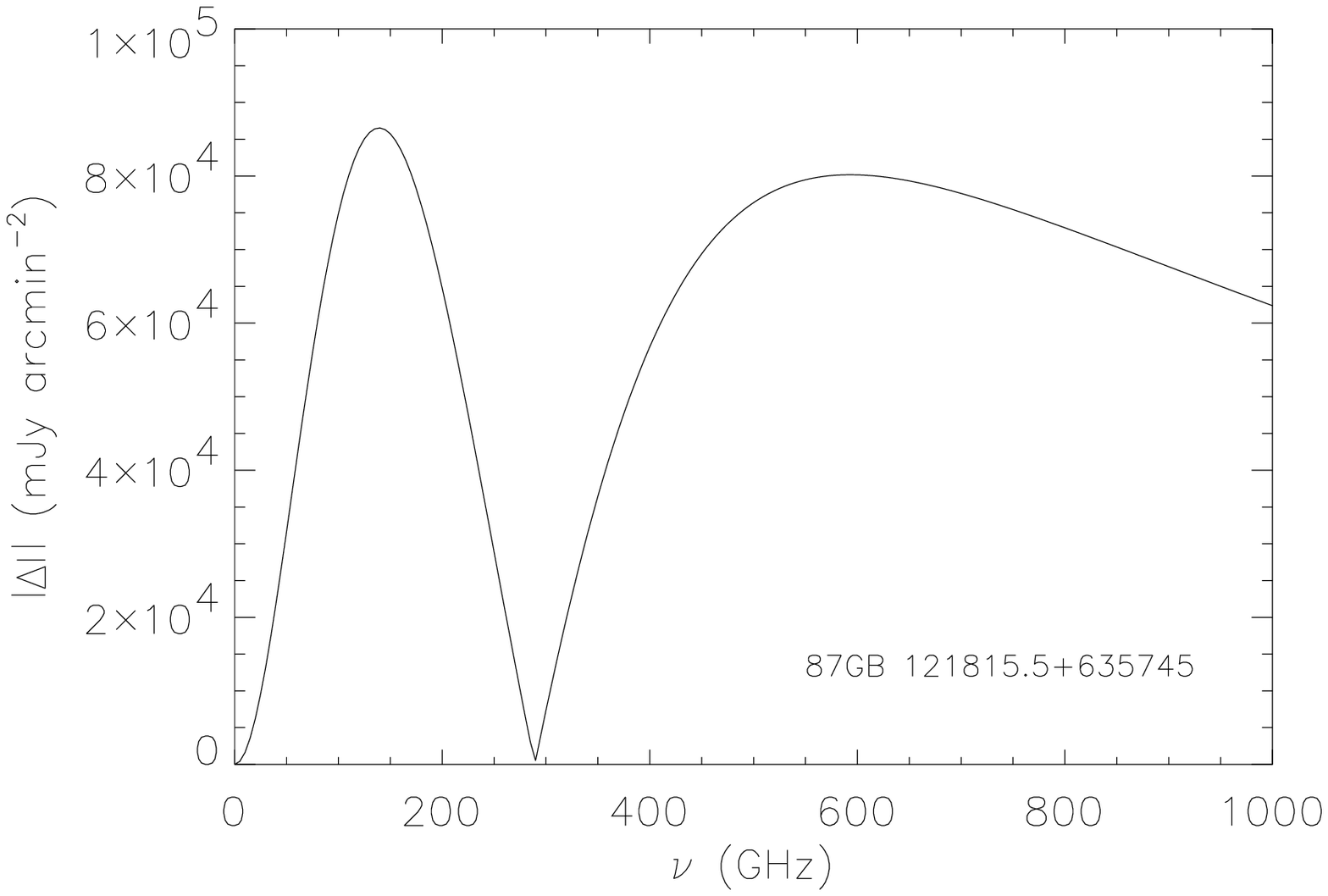,height=5.cm,width=9.cm,angle=0.0}
 \epsfig{file=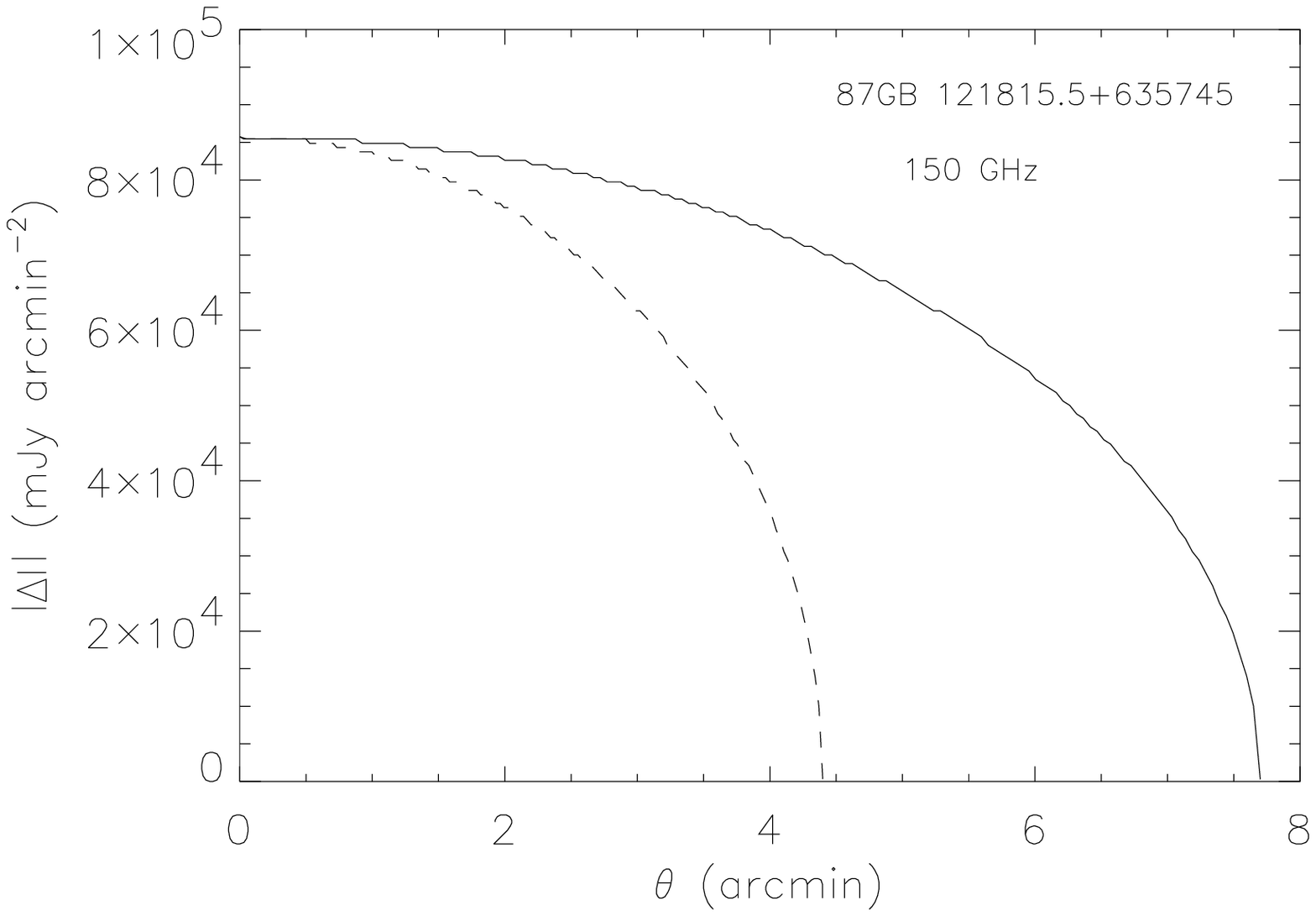,height=5.cm,width=9.cm,angle=0.0}
 \epsfig{file=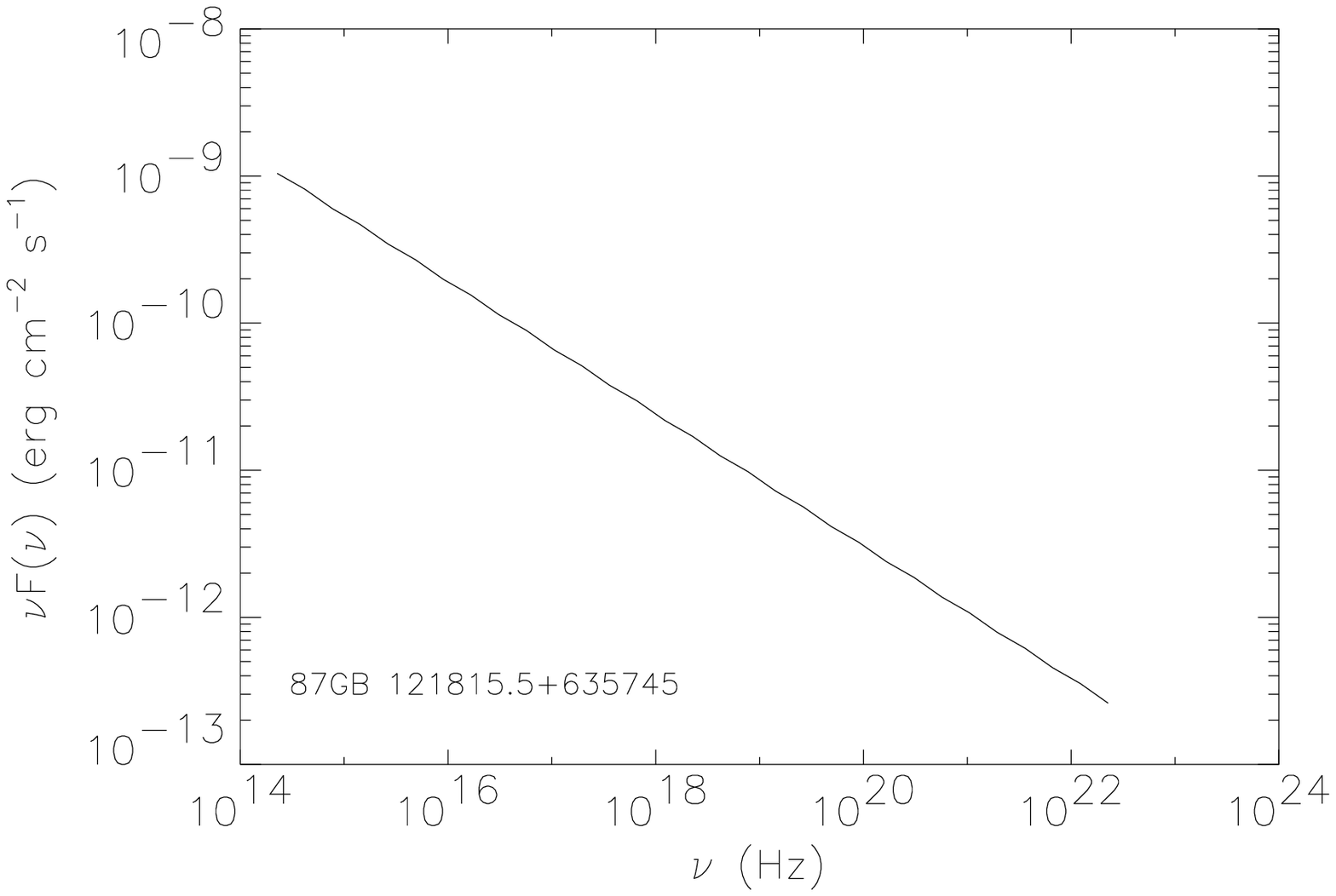,height=5.cm,width=9.cm,angle=0.0}
\end{center}
 \caption{\footnotesize{Upper panel: SZE spectrum at the center of the
 radio galaxy lobe 87GB 121815.5+635745. Mid panel: SZE signal profile at
 150 GHz calculated along the major axis (solid line) and along the minor
 XIS (dashed line). Lower panel: the ICS-on-CMB flux produced by
 the relativistic electron population in the radio lobes of this
 object.
 }}
 \label{fig.87gb12}
\end{figure}

\begin{figure}[ht]
\begin{center}
 \epsfig{file=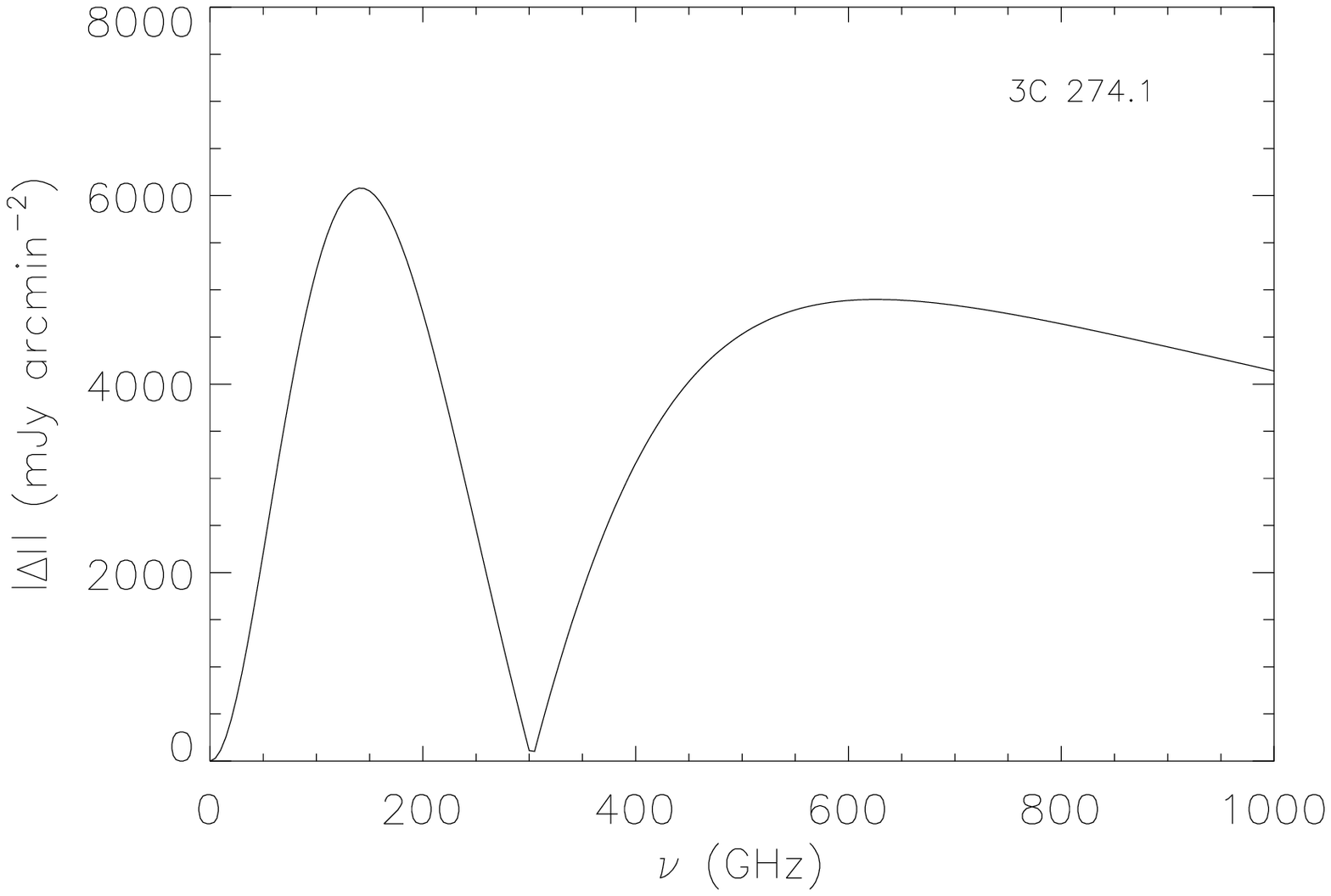,height=5.cm,width=9.cm,angle=0.0}
 \epsfig{file=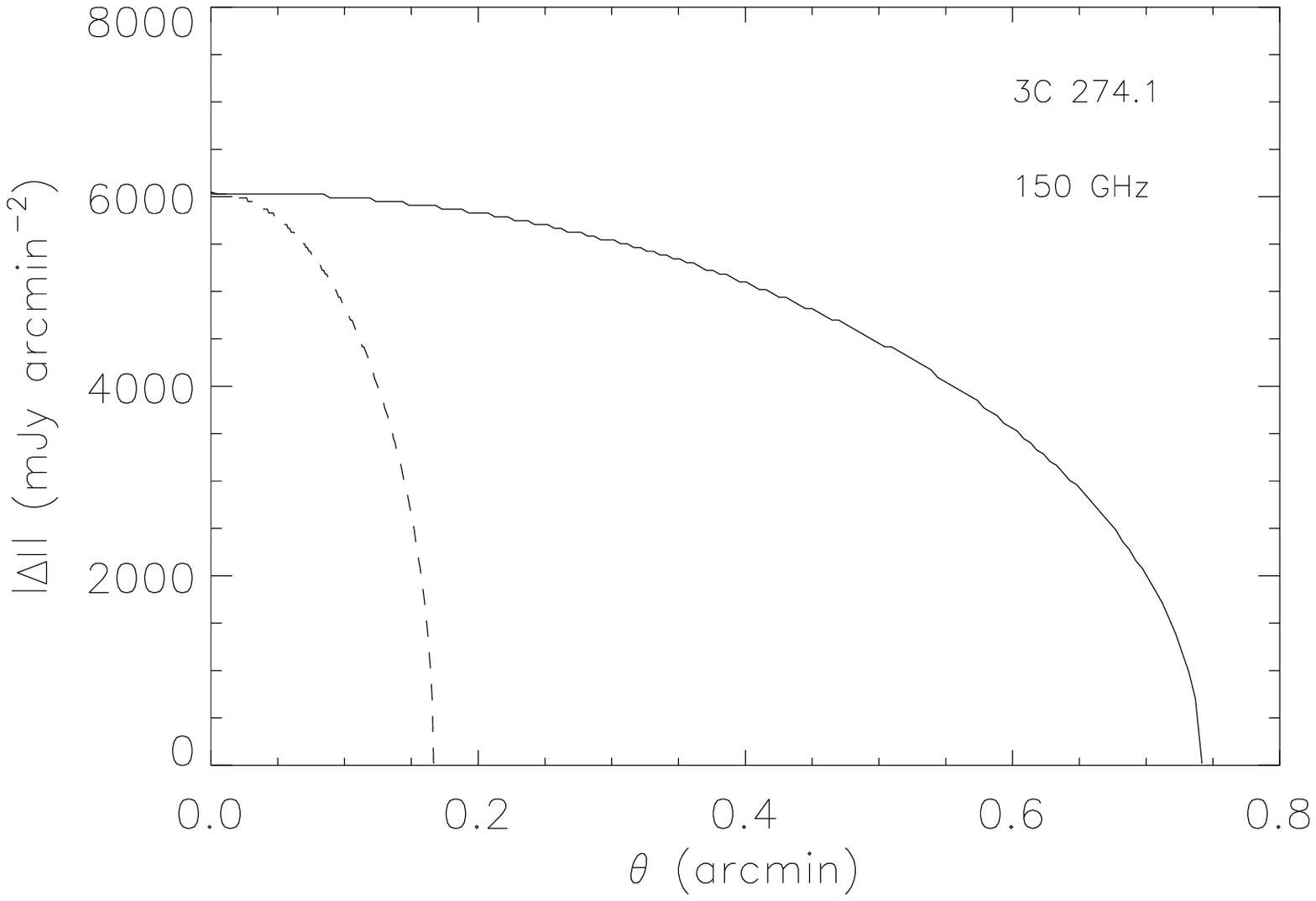,height=5.cm,width=9.cm,angle=0.0}
 \epsfig{file=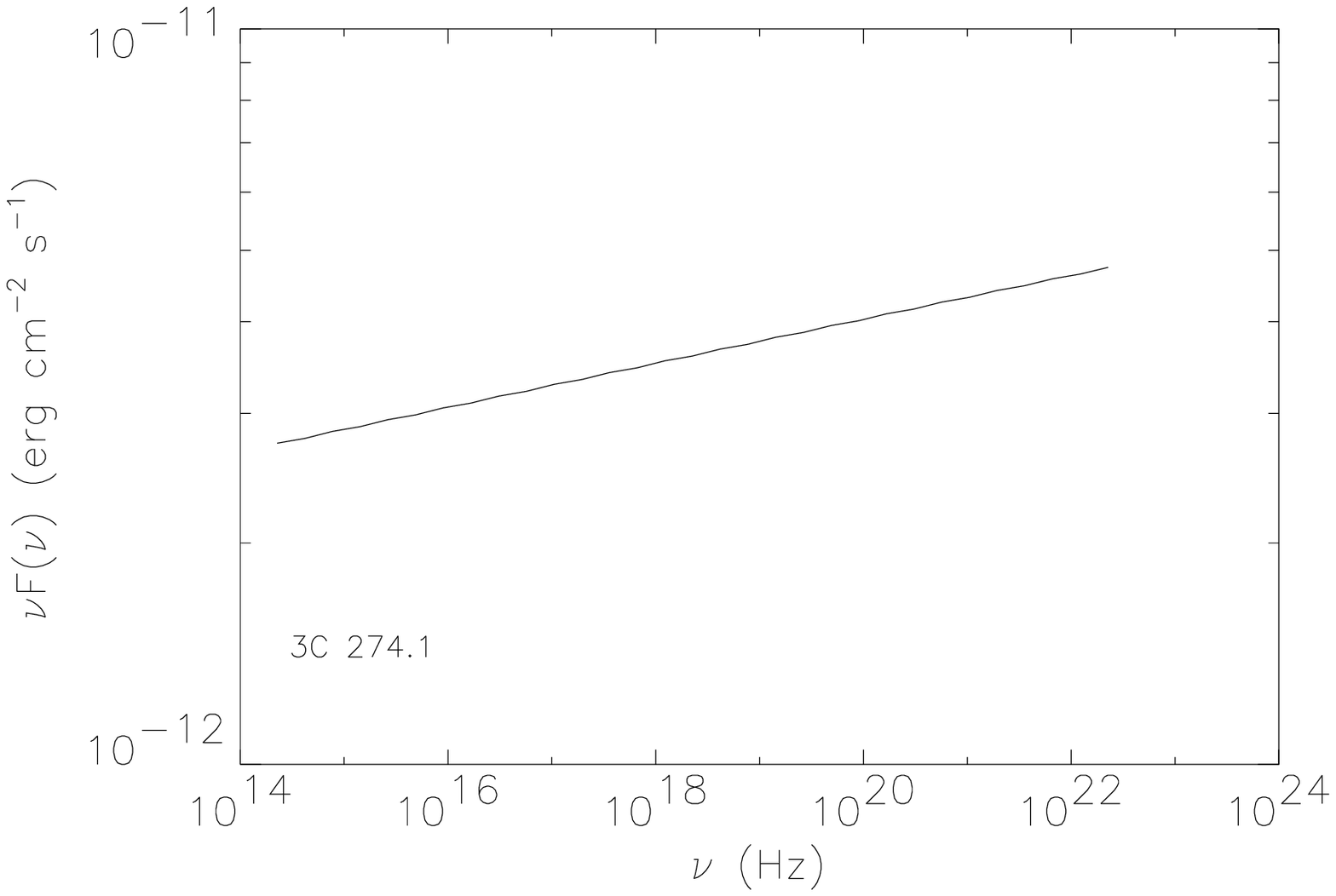,height=5.cm,width=9.cm,angle=0.0}
\end{center}
 \caption{\footnotesize{Same as Fig. \ref{fig.87gb12} but for 3C 274.1.
 }}
 \label{fig.3c2741}
\end{figure}

\begin{figure}[ht]
\begin{center}
 \epsfig{file=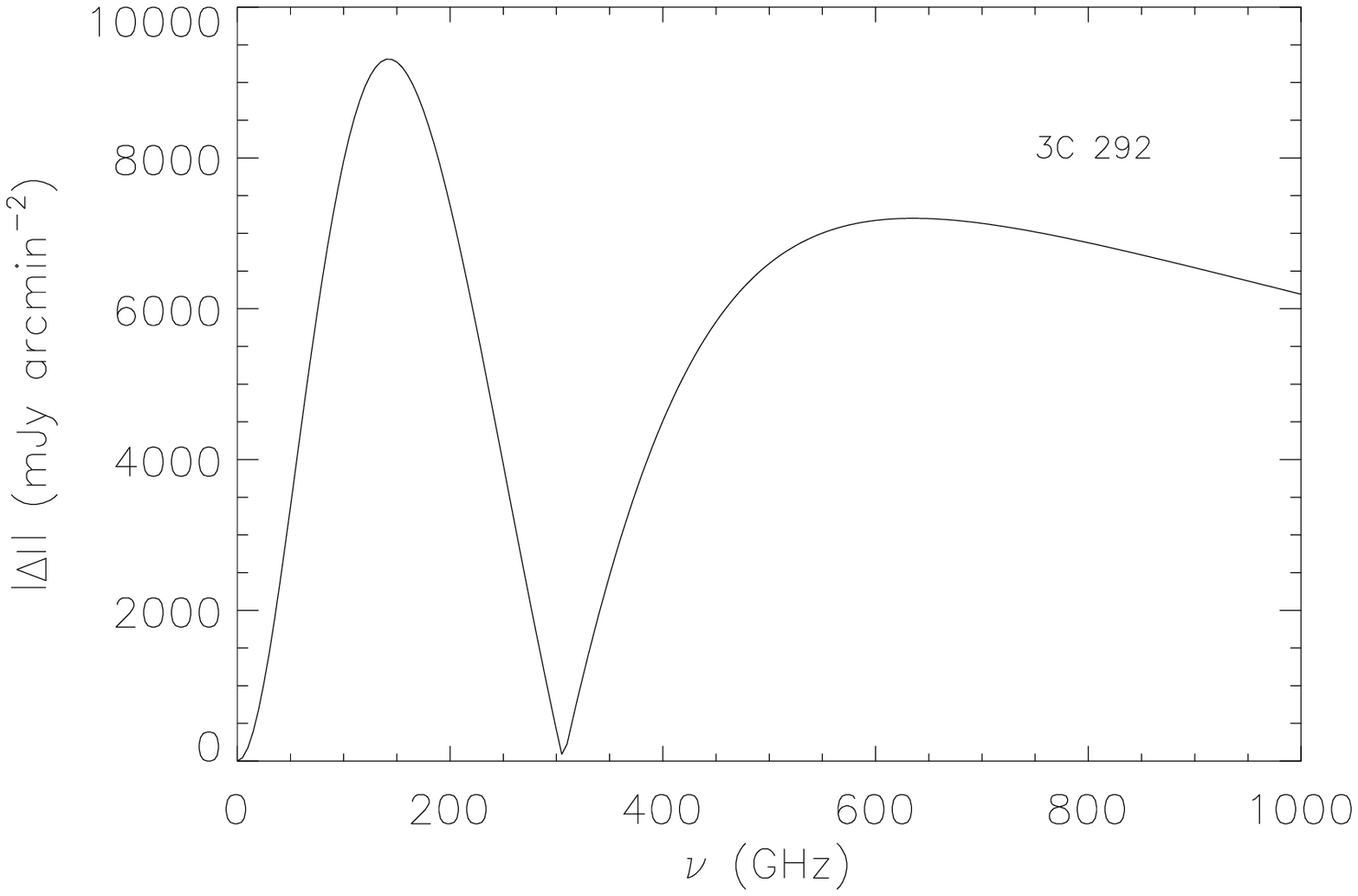,height=5.cm,width=9.cm,angle=0.0}
 \epsfig{file=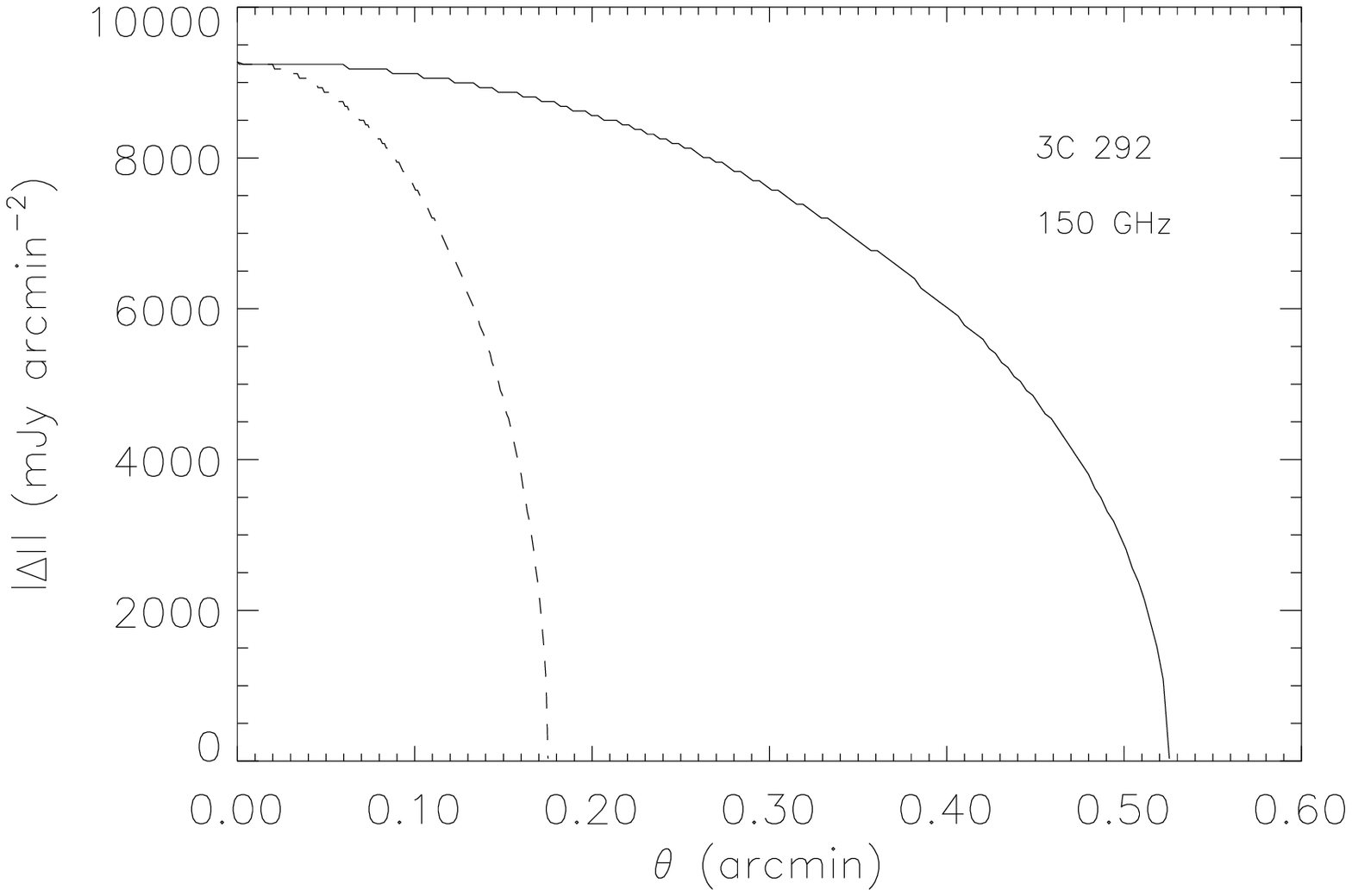,height=5.cm,width=9.cm,angle=0.0}
 \epsfig{file=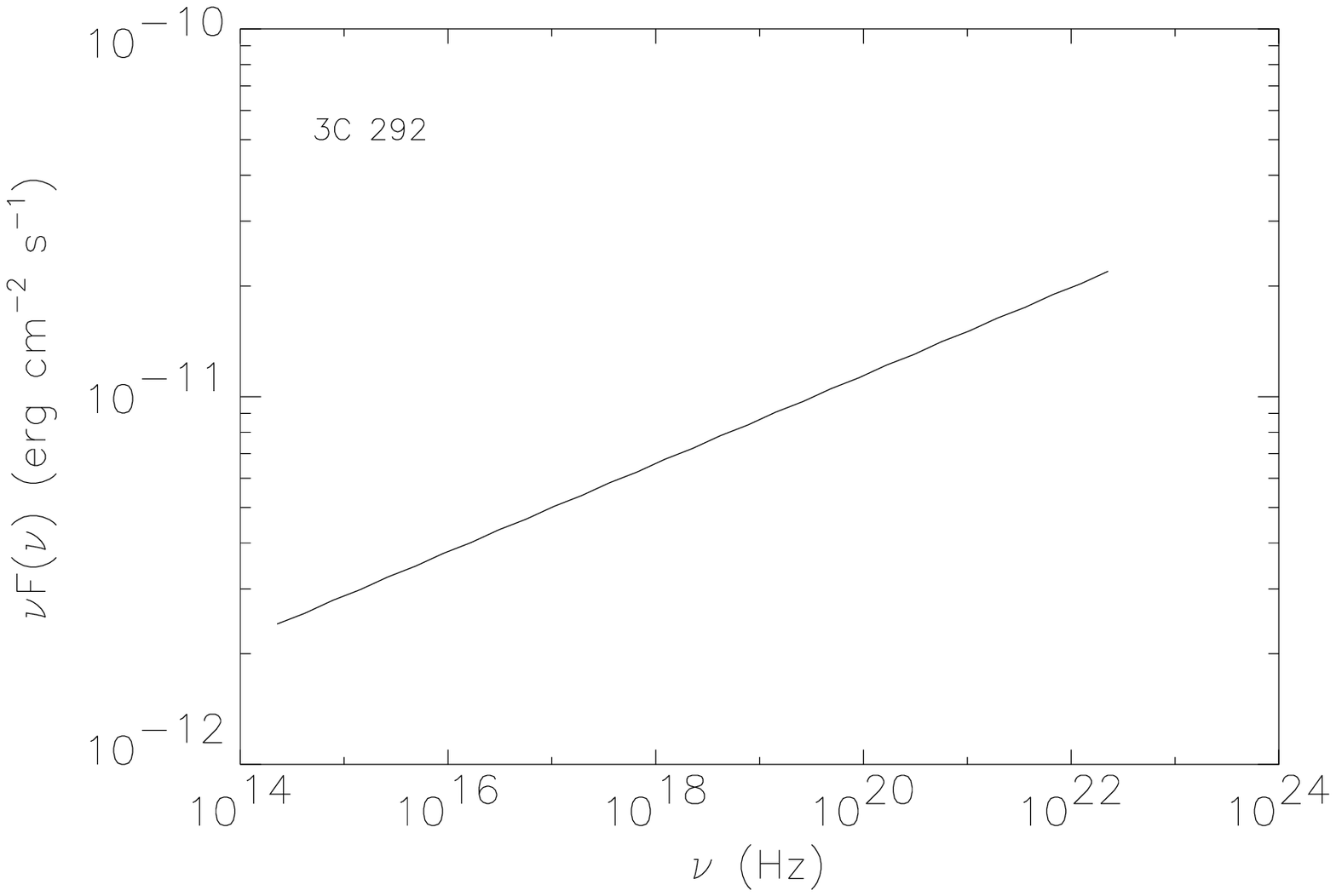,height=5.cm,width=9.cm,angle=0.0}
\end{center}
 \caption{\footnotesize{Same as Fig. \ref{fig.87gb12} but for 3C 292.
 }}
 \label{fig.3c292}
\end{figure}

\begin{figure}[ht]
\begin{center}
 \epsfig{file=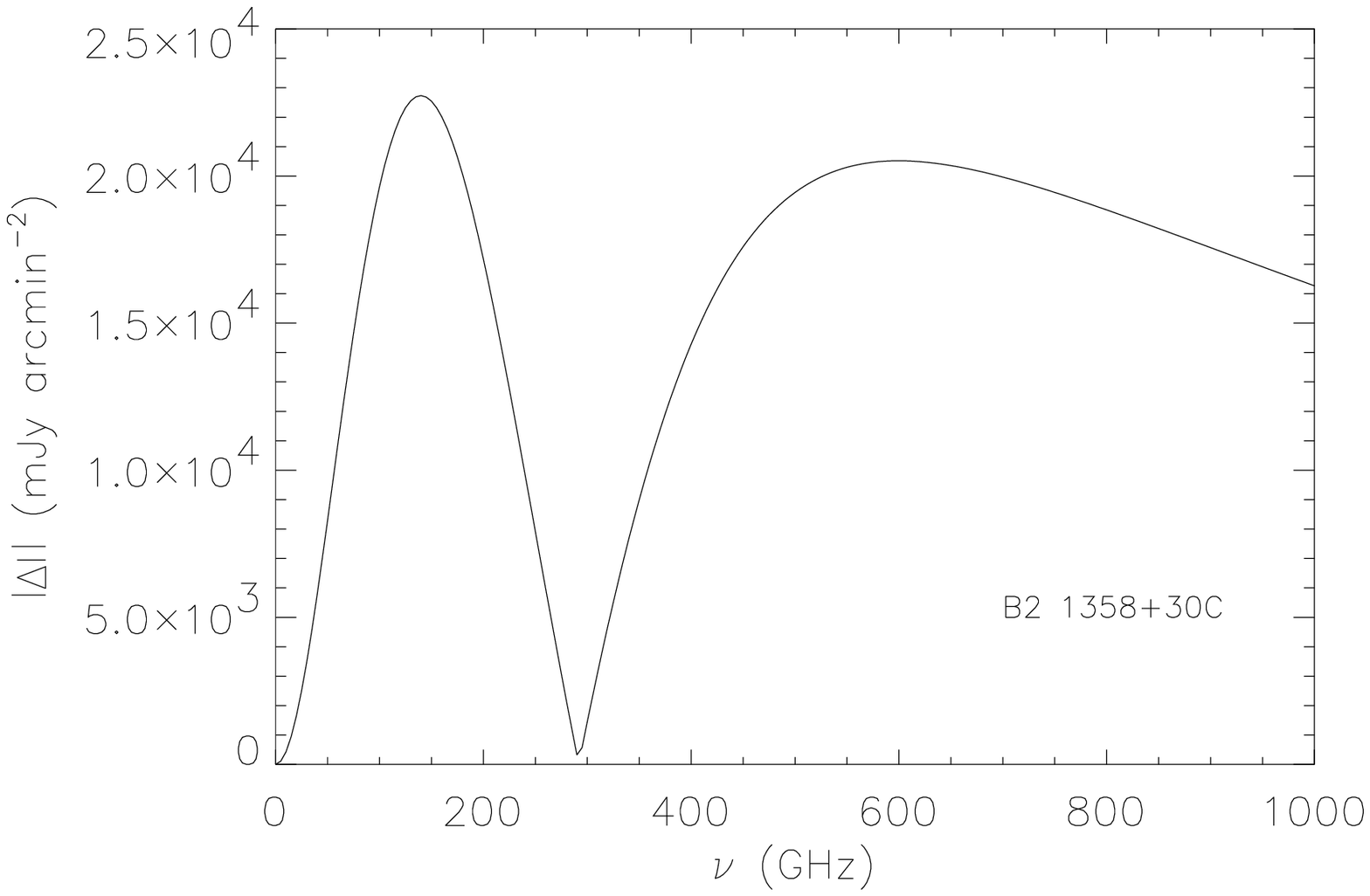,height=5.cm,width=9.cm,angle=0.0}
 \epsfig{file=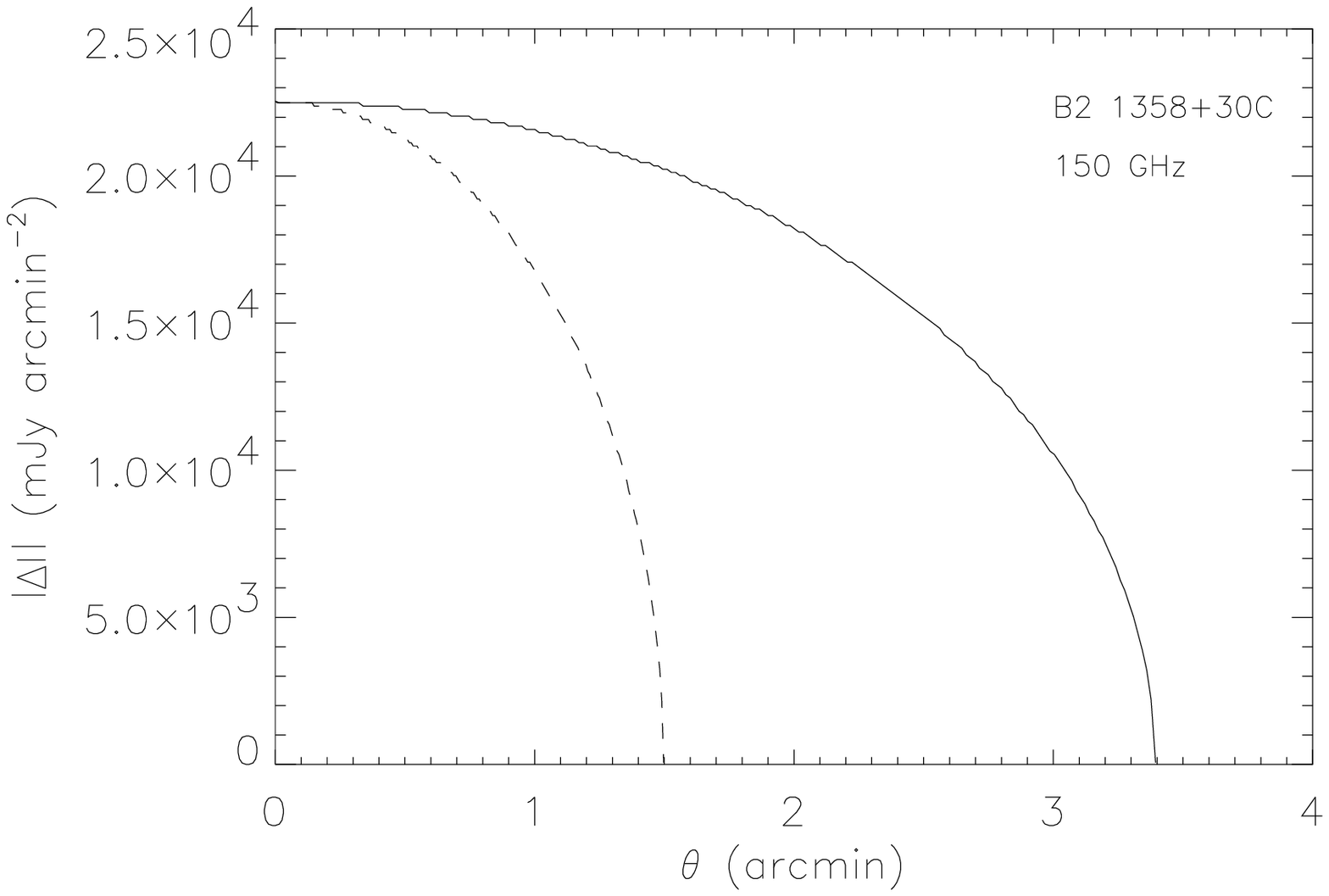,height=5.cm,width=9.cm,angle=0.0}
 \epsfig{file=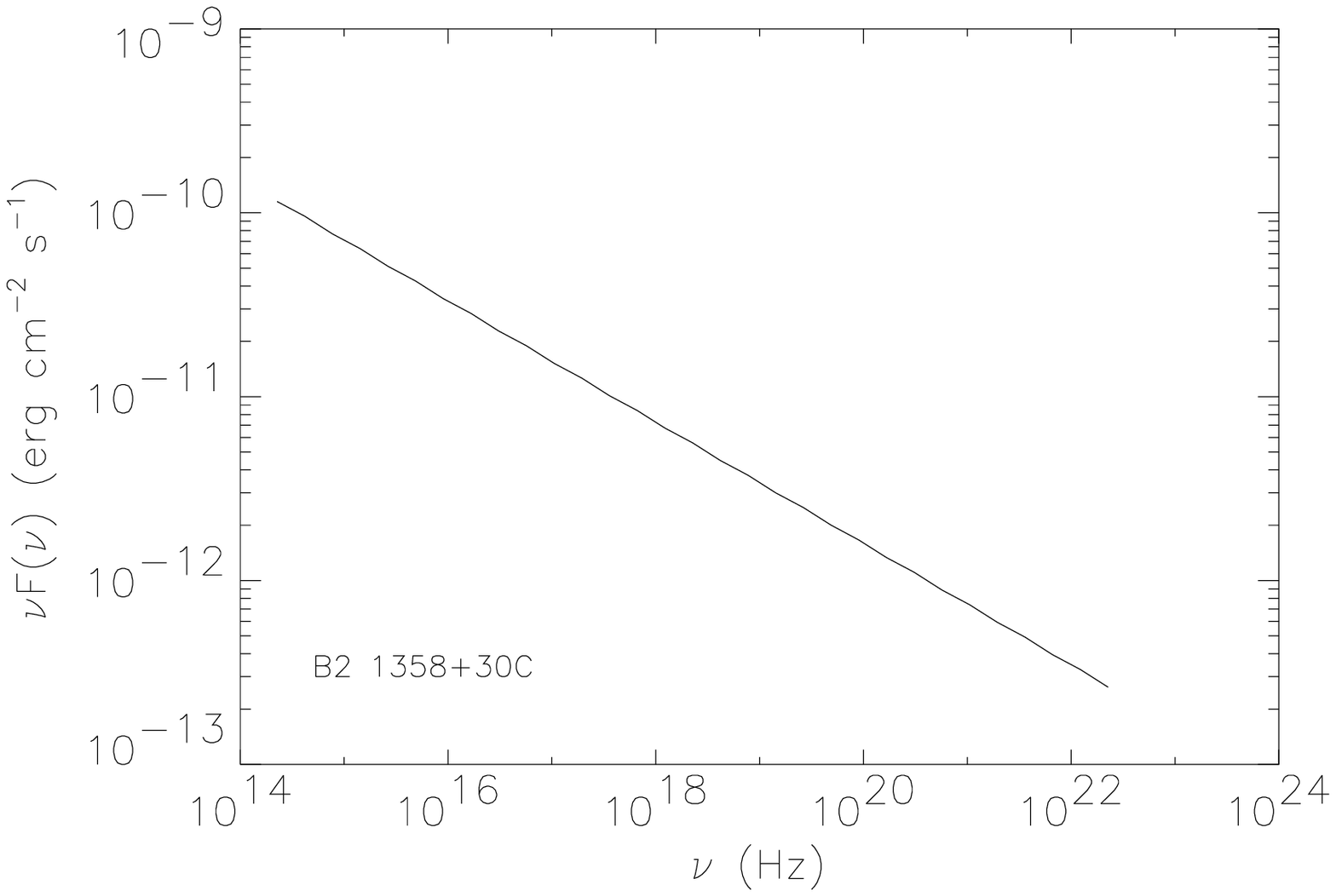,height=5.cm,width=9.cm,angle=0.0}
\end{center}
 \caption{\footnotesize{Same as Fig. \ref{fig.87gb12} but for B2 1358+30C.
 }}
 \label{fig.b2}
\end{figure}

\begin{figure}[ht]
\begin{center}
 \epsfig{file=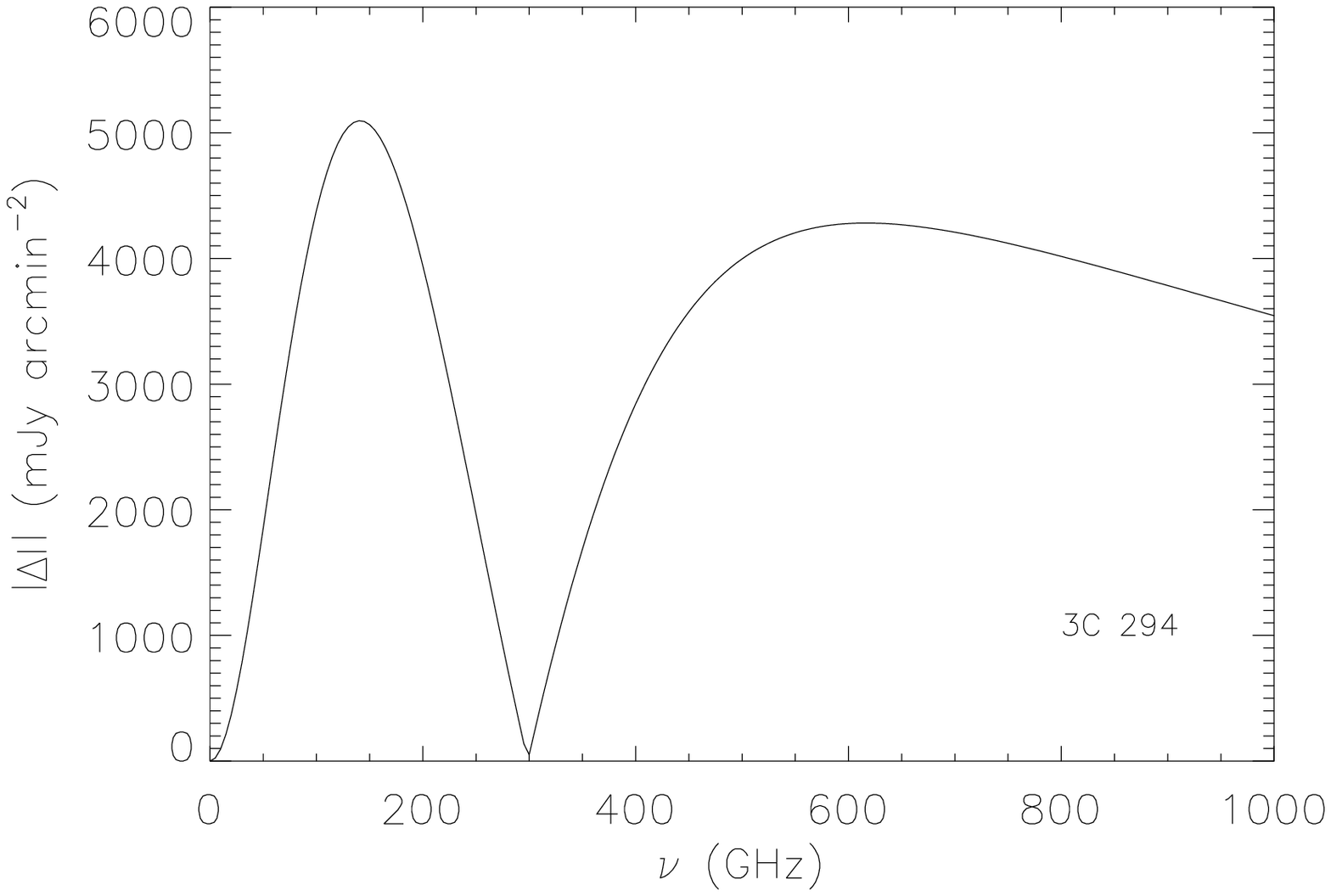,height=5.cm,width=9.cm,angle=0.0}
 \epsfig{file=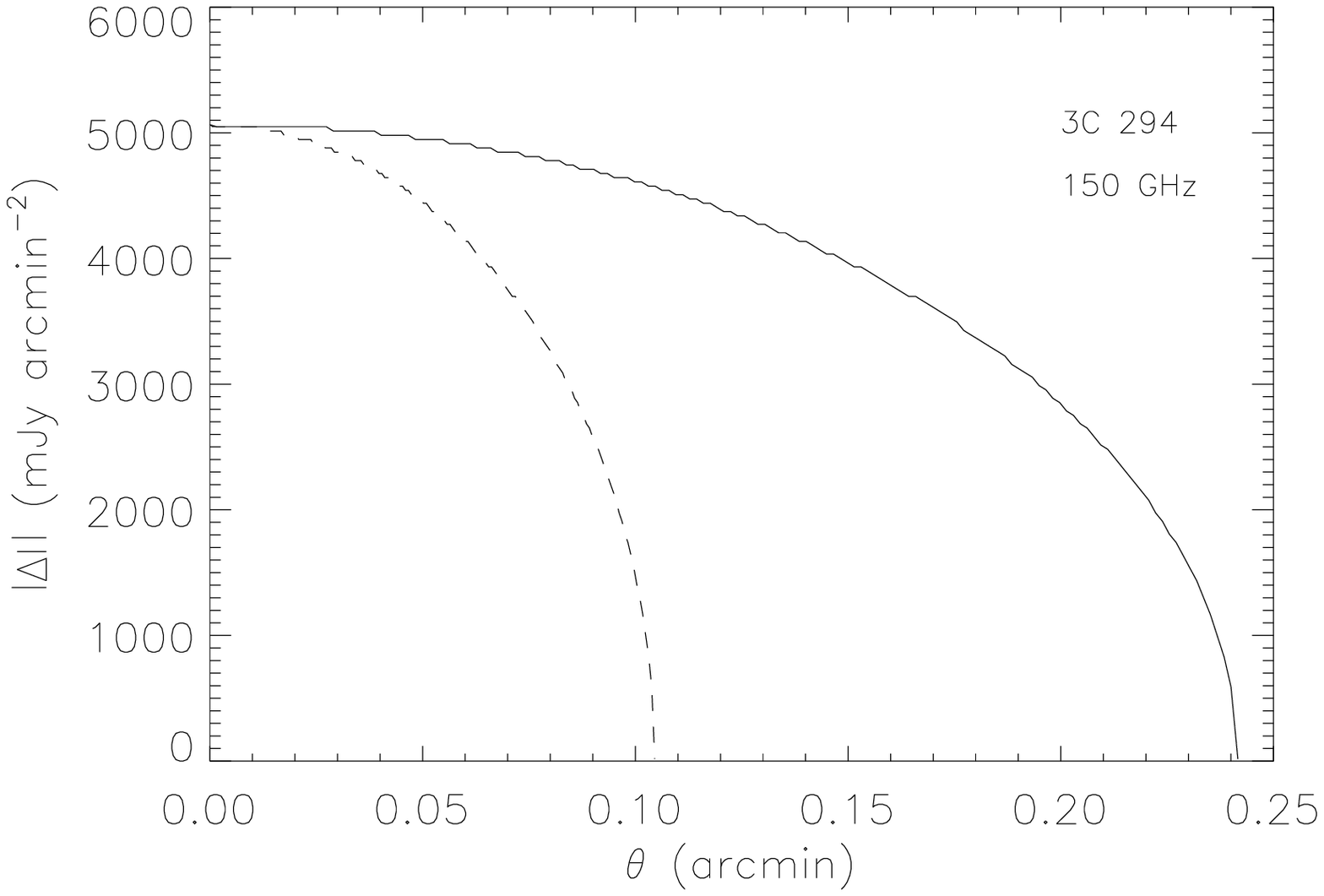,height=5.cm,width=9.cm,angle=0.0}
 \epsfig{file=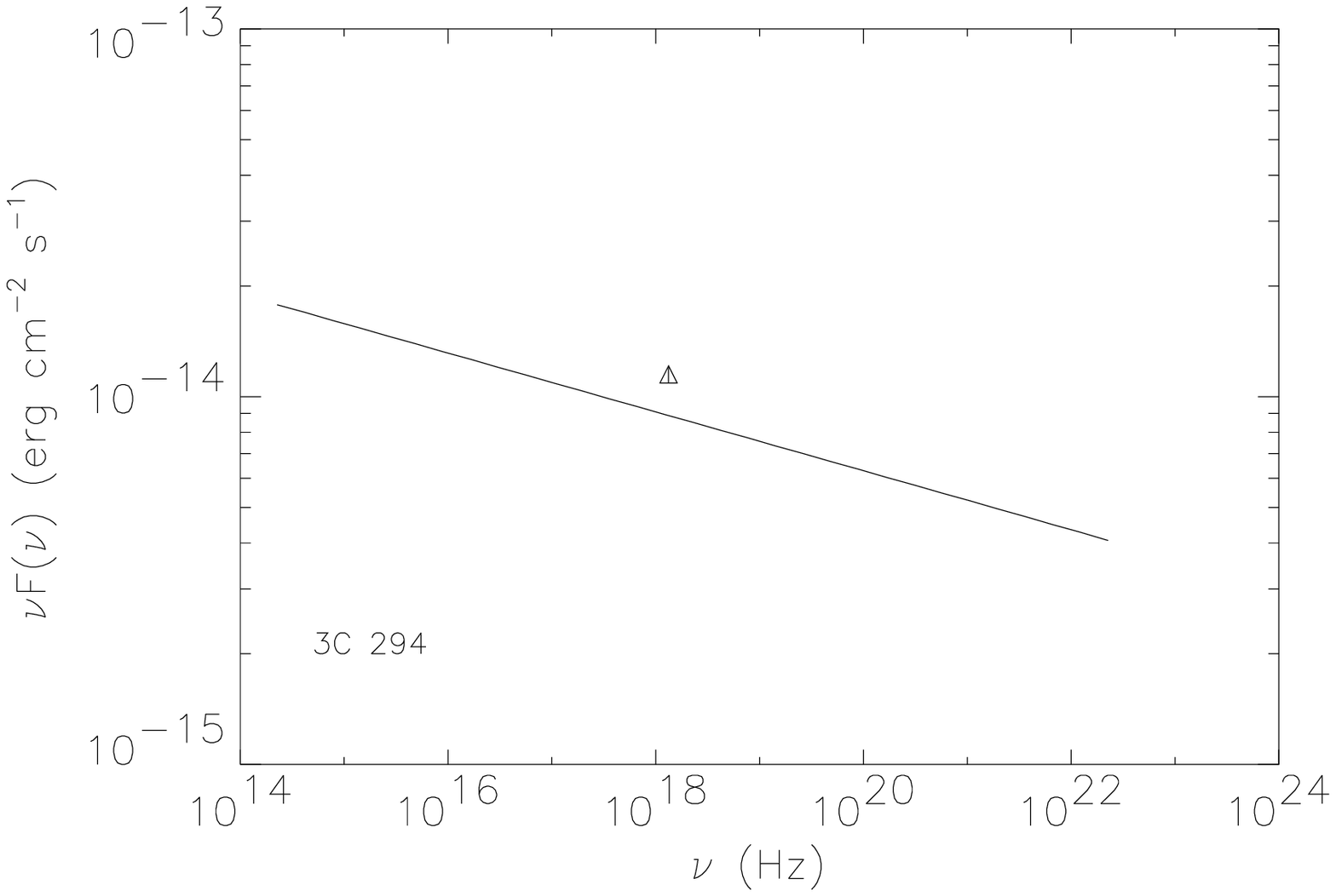,height=5.cm,width=9.cm,angle=0.0}
\end{center}
 \caption{\footnotesize{Same as Fig. \ref{fig.87gb12} but for 3C 294.
 }}
 \label{fig.3c294}
\end{figure}

\begin{figure}[ht]
\begin{center}
 \epsfig{file=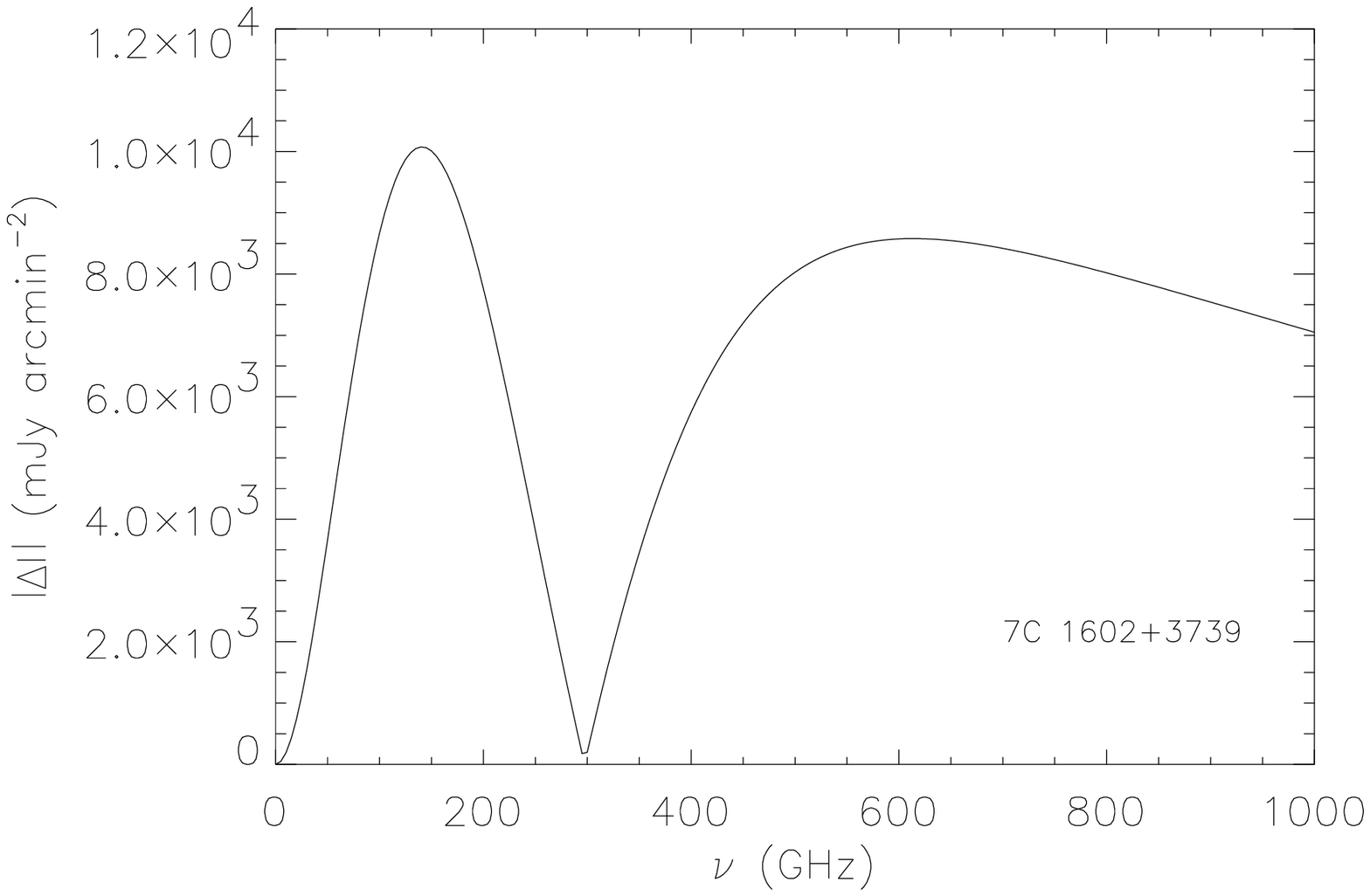,height=5.cm,width=9.cm,angle=0.0}
 \epsfig{file=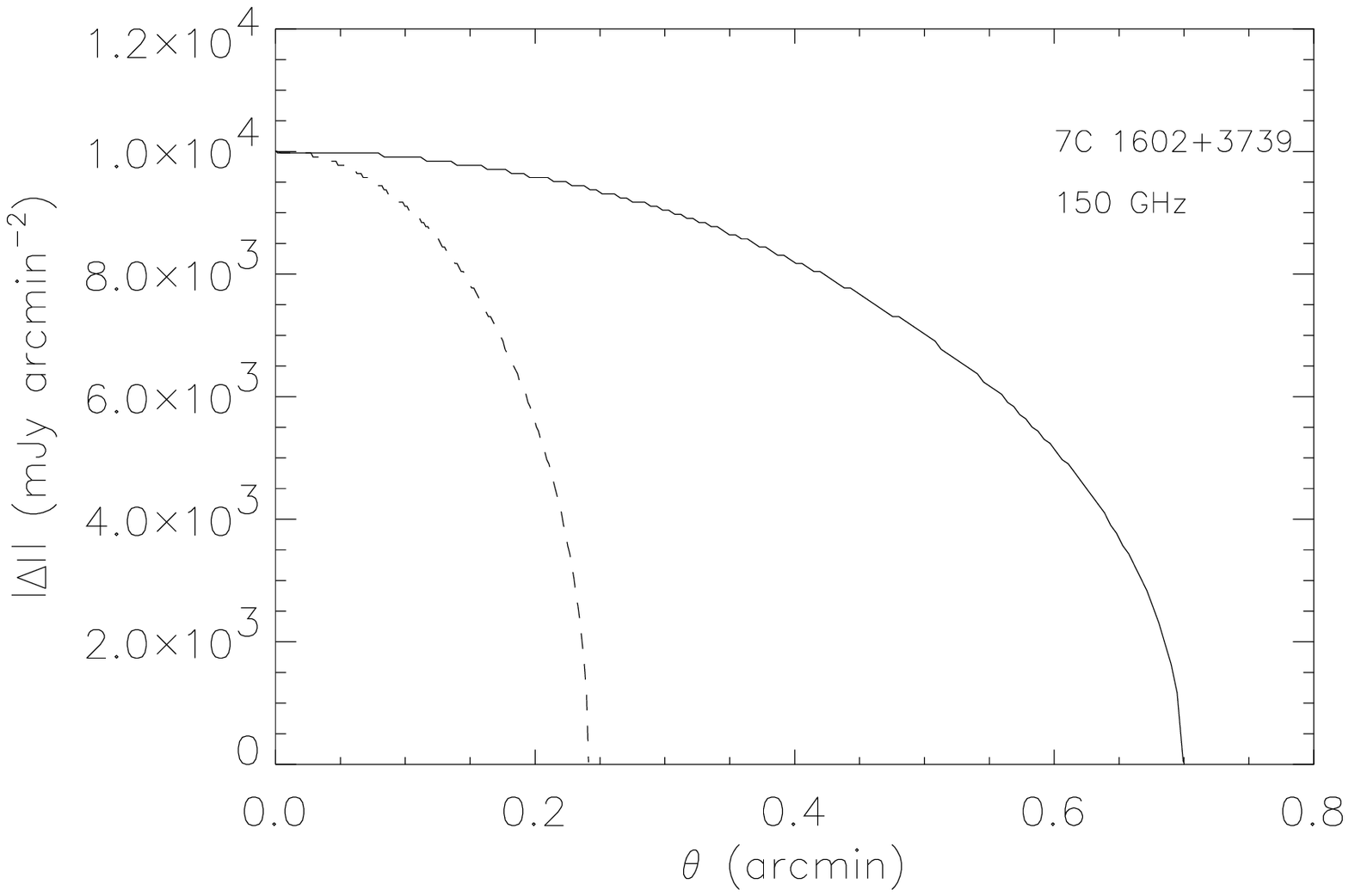,height=5.cm,width=9.cm,angle=0.0}
 \epsfig{file=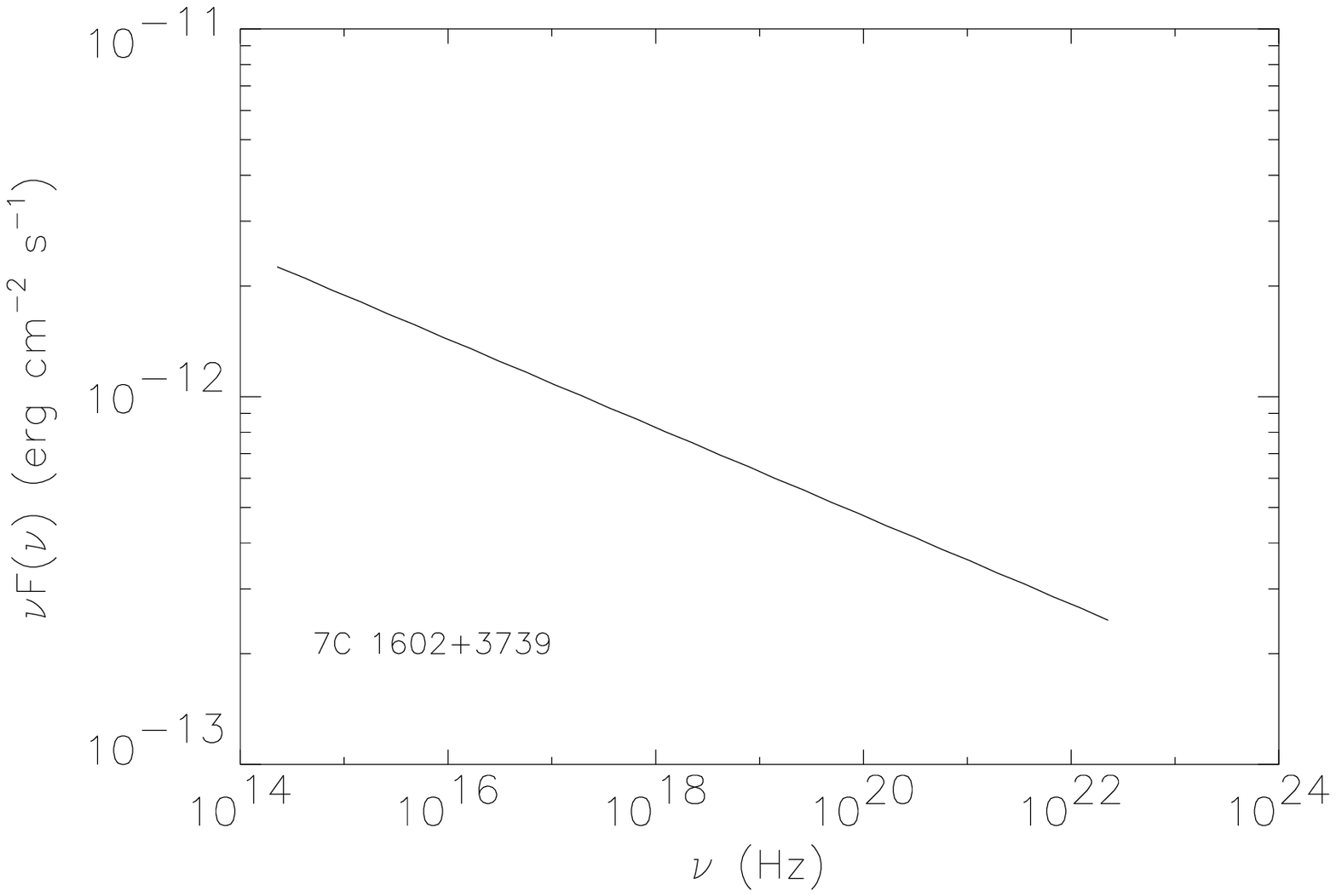,height=5.cm,width=9.cm,angle=0.0}
\end{center}
 \caption{\footnotesize{Same as Fig. \ref{fig.87gb12} but for 7C 1602+3739.
 }}
 \label{fig.7c}
\end{figure}

\begin{figure}[ht]
\begin{center}
 \epsfig{file=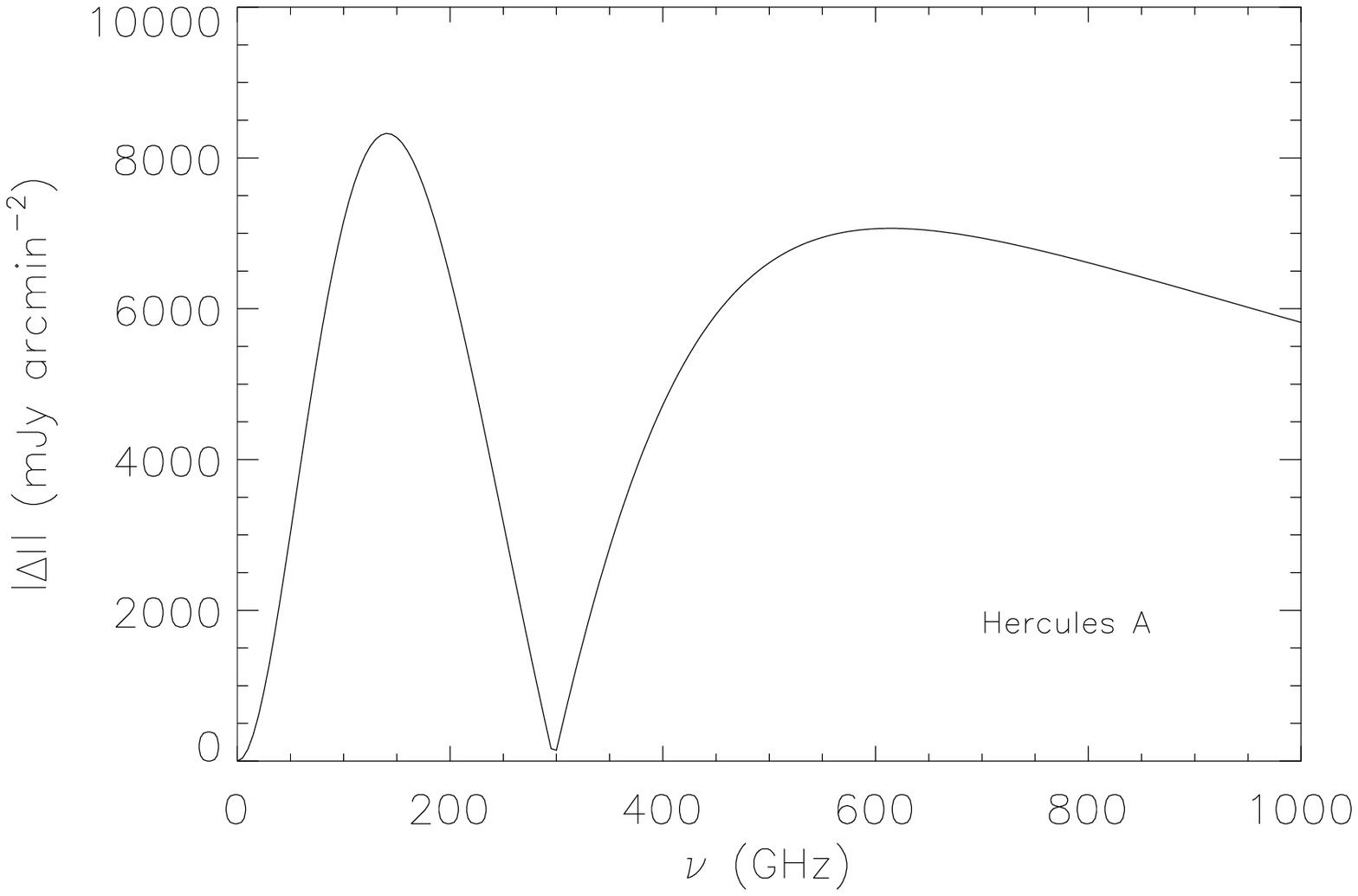,height=5.cm,width=9.cm,angle=0.0}
 \epsfig{file=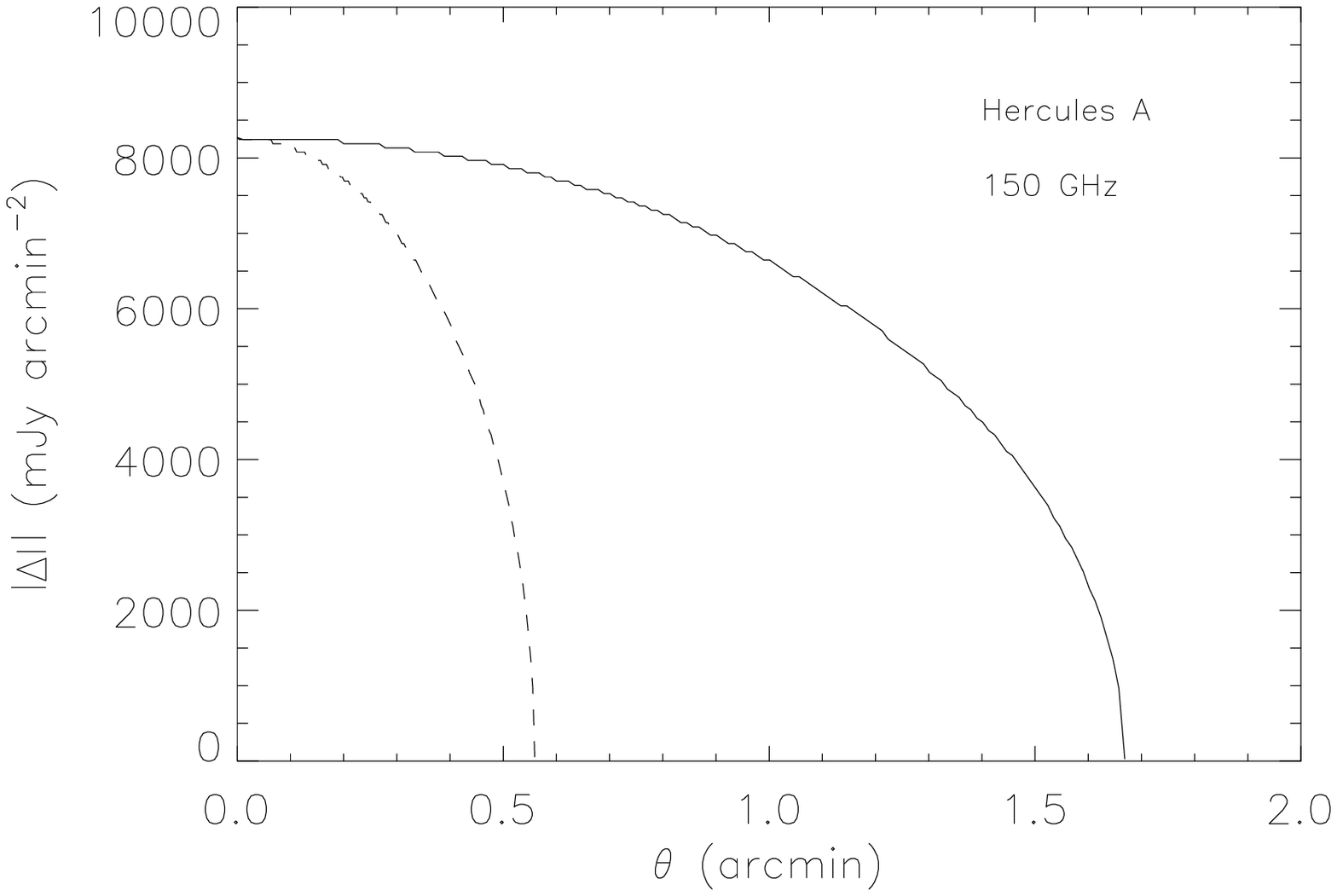,height=5.cm,width=9.cm,angle=0.0}
 \epsfig{file=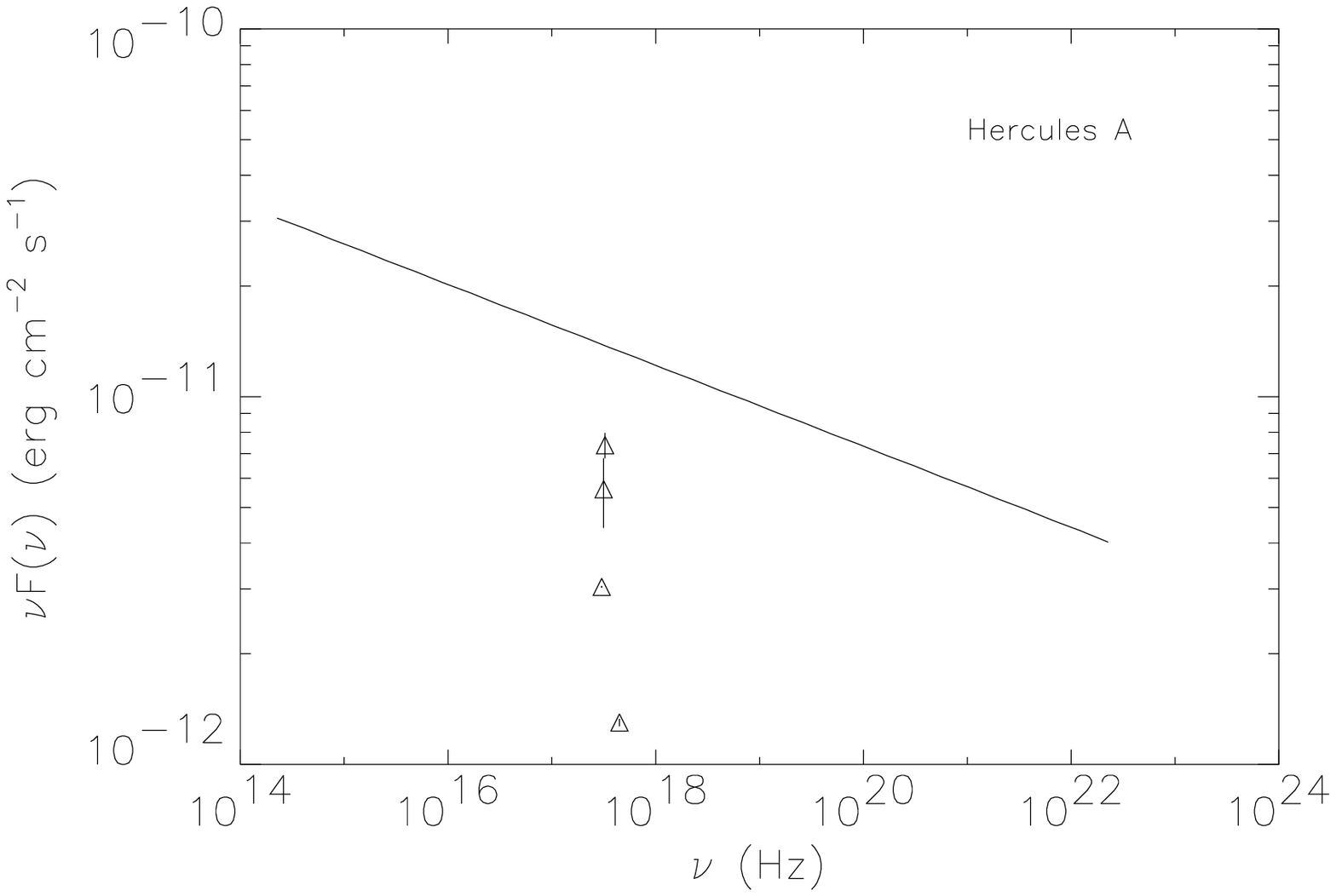,height=5.cm,width=9.cm,angle=0.0}
\end{center}
 \caption{\footnotesize{Same as Fig. \ref{fig.87gb12} but for Hercules A.
 }}
 \label{fig.hera}
\end{figure}

The electron spectra of these RG lobes are calculated assuming
$p_1 = 1$, $\alpha= 2 \alpha_r +1$ as obtained from the observed
values of $\alpha_r$. They are normalized to a fixed electron
density $k_0 = 2.6$ cm$^{-3}$. The values of the B-field,
estimated from the synchrotron flux with the previous electron
spectrum, are given in Table \ref{tab.6}. 
We report in this table the B-field values, assumed constant in the
emission region, as derived by radio data (Col.2) and assuming the same normalization for the electron spectrum
 (see Figs. \ref{fig.87gb12}--\ref{fig.hera}). Column 3 reports the
 uniform magnetic field $B^*$ derived from the combination of
 ICS X-ray data and radio synchrotron data for the three objects
 for which X-ray ICS data are available in the literature
 (see Figs. \ref{fig.3c292_normx}--\ref{fig.hera_normx}).
We stress here that we use the previous normalization as a reference value that recovers,
e.g., the observed synchrotron flux for Hercules A for a B-field
of 1 $\mu$G. We adopt this specific normalization because the
value of the B-field for Hercules A is the highest among the seven
objects we consider in detail. This normalization to a fixed
value of the B-field allows the dependence of the SZE on
the RG lobe spectral index to be studied, given a fixed normalization of the
electron spectrum.
\begin{table*}[htb]{}
\vspace{2cm}
\begin{center}
\begin{tabular}{|*{6}{c|}}
\hline Object name  & $B$ & $B^*$\\
             & $\mu$G & $\mu$G\\
\hline 87GB 121815.5+635745 & 0.21 & \\ 3C 274.1     & 0.31 & \\
3C 292       & 0.091 & 2.7 \\ B2 1358+30C  & 0.16 & \\ 3C 294 &
0.90 & 0.41\\ 7C 1602+3739 & 0.13 & \\ Hercules A   & 1.0 & 3.0\\
 \hline
 \end{tabular}
 \end{center}
 \caption{\footnotesize{Uniform magnetic field $B$ values for the restricted set of RG lobes. 
 }}
 \label{tab.6}
 \end{table*}

Figures \ref{fig.87gb12}--\ref{fig.hera} show the non-thermal SZE
spectrum at the center of the selected lobe of the seven RGs that we
discuss in detail, the spatial profile of the SZE brightness
change $\Delta I$ calculated along the major and minor axes of the
lobe, and the SED at high frequency
($10^{14}-10^{24}$ Hz) in a $Log (\nu) - Log (\nu F(\nu))$ plot,
produced by the electron ICS-on-CMB emission.

The SZE spectra calculated in this way distinctively show a
minimum at $\nu \sim 140$ GHz, a crossover frequency whose precise
value depends on the slope of the electron spectrum and on the
pressure/energy density of the electron population, and a maximum
that depends even more sensitively on the electron energy density
and on the electron spectral shape.
%
The frequency position of the minimum $\nu_{min}$, crossover
$\nu_{0}$, and maximum $\nu_{max}$ of the SZE spectra of the seven
objects that we study in detail are reported in Table \ref {tab.5}
together with the value of the electron spectral index $\alpha$.
For each one of these frequency values, the uncertainty is about $\pm2.5$ GHz.
The electron spectrum is assumed to be a single power-law with spectral index $\alpha$ and with $p_1=1$.
\begin{table*}[htb]{}
\vspace{2cm}
\begin{center}
\begin{tabular}{|*{6}{c|}}
\hline Object name      & $\alpha$ & $\nu_{min}$ & $\nu_{0}$ &
$\nu_{max}$\\ \hline 87GB 121815.5+635745 & 3.90 & 140 & 287.5 &
595\\ 3C 274.1 & 2.94 & 140 & 302.5 & 625\\ 3C 292       & 2.76 &
140 & 307.5 & 635\\ B2 1358+30C  & 3.66 & 140 & 292.5 & 600\\ 3C
294       & 3.16 & 140 & 297.5 & 615\\ 7C 1602+3739 & 3.24 & 140 &
297.5 & 615\\ Hercules A   & 3.22 & 140 & 297.5 & 615\\
 \hline
 \end{tabular}
 \end{center}
 \caption{\footnotesize{SZE spectral characteristic points for the restricted set of RG lobes.
 }}
 \label{tab.5}
 \end{table*}
\begin{table*}[htb]{}
\vspace{2cm}
\begin{center}
\begin{tabular}{|*{6}{c|}}
\hline Object name         &  $\alpha$ & $-\Delta I(150 \mbox{
GHz})$ & $\Delta I(500 \mbox{ GHz})$ & $S_{sync}(150 \mbox{ GHz})$
& $S_{sync}(500 \mbox{ GHz})$\\
             &      & mJy arcmin$^{-2}$           & mJy arcmin$^{-2}$ & mJy arcmin$^{-2}$           & mJy arcmin$^{-2}$          \\
\hline 87GB 121815.5+635745 & 3.90 & $8.55\times10^{4}$ &
$7.66\times10^{4}$ & 0.13 & 0.02\\ 3C 274.1     & 2.94 &
$6.05\times10^3$ & $4.53\times10^3$ & 487 & 152\\ 3C 292       &
2.76 & $9.27\times10^3$ & $6.59\times10^3$ & 679 & 236\\ B2
1358+30C  & 3.66 & $2.25\times10^4$ & $1.95\times10^4$ &
$5.8\times10^{-2}$ & $1.2\times10^{-2}$\\ 3C 294       & 3.16 &
$5.07\times10^3$ & $4.00\times10^3$ & 277 & 76\\ 7C 1602+3739 &
3.24 & $9.99\times10^3$ & $8.01\times10^3$ & 3.4 & 0.88\\ Hercules
A   & 3.22 & $8.27\times10^3$ & $6.61\times10^3$ & 243 & 64\\
 \hline
 \end{tabular}
 \end{center}
 \caption{\footnotesize{SZE signal and radio flux at 150 and 500 GHz for the restricted set of RG lobes.
 }}
 \label{tab.4}
 \end{table*}

\begin{figure}[ht]
\begin{center}
 \epsfig{file=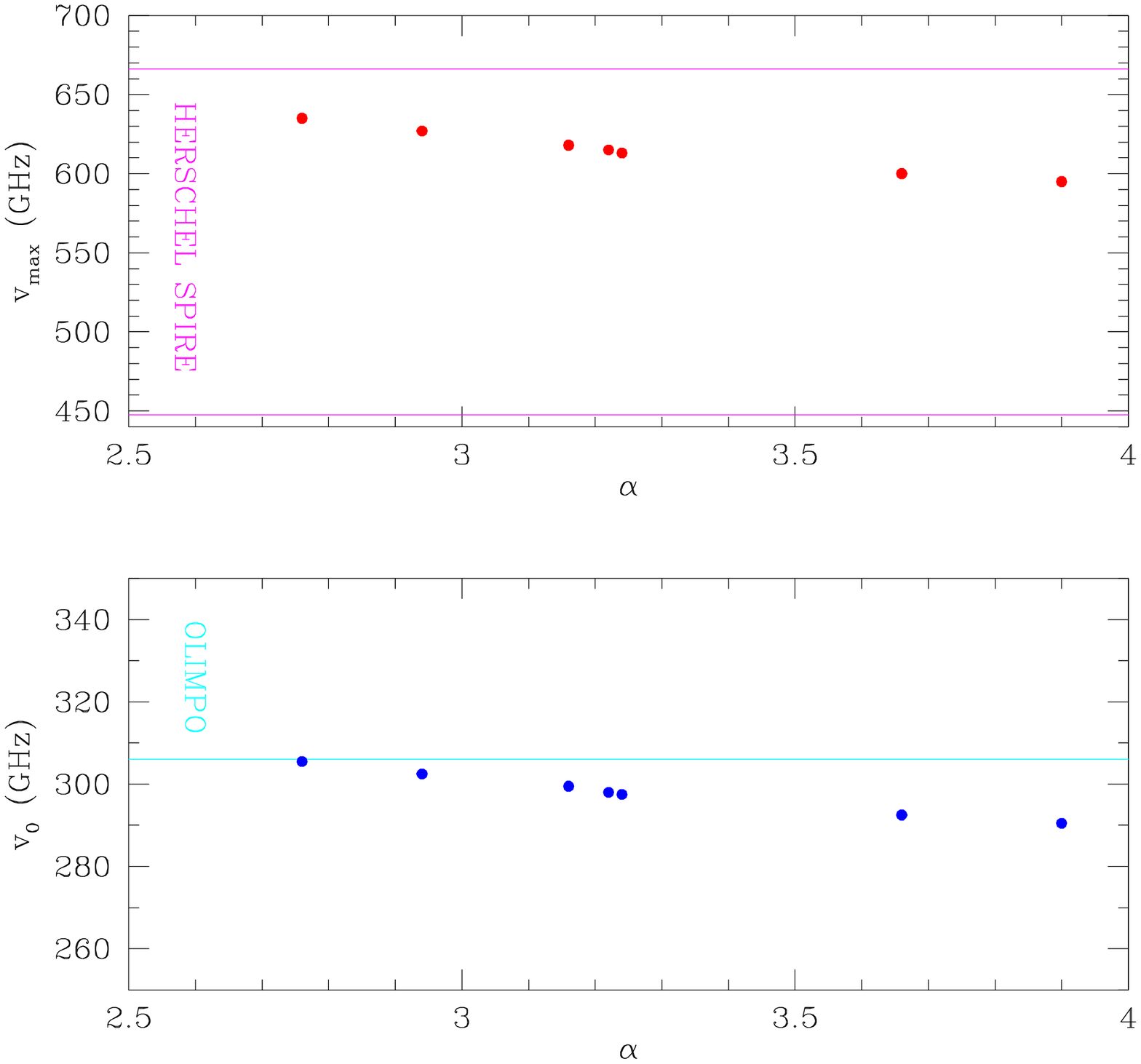,height=9.cm,width=9.cm,angle=0.0}
\end{center}
 \caption{\footnotesize{Correlation of the frequency position
 of the maximum $\nu_{max}$ (upper panel) and of the crossover $\nu_0$ (lower panel)
 of the non-thermal SZE from RG lobes as a function of the
 electron spectrum index $\alpha$ for the seven objects we study here.
 The lowest frequency Herschel-SPIRE band ($\approx 447-666$ GHz, as enclosed by the magenta lines) is
 shown for comparison.
 }
 }
 \label{fig.numax_s}
\end{figure}

We plot in Fig.\ref{fig.numax_s} the position of the
characteristic frequencies $\nu_{0}$ and $\nu_{max}$ as a function
of the slope $\alpha$ of the electron spectrum. The analytical
formulae with a polynomial of power 2 that fit the frequency
locations of $\nu_{max}$ and $\nu_0$ as a function of $\alpha$ are
\begin{equation}
 \nu_{max} = 13.35 \times \alpha^2 - 124.19 \times \alpha + 877.13
\end{equation}
and
\begin{equation}
 \nu_{0} = 4.35 \times \alpha^2 - 42.11 \times \alpha + 389.07 \; ,
\end{equation}
(frequencies are given in GHz) and are valid in the range
$\alpha=2.5 - 4$ probed by the RG lobes we consider here.

\begin{figure}[ht]
\begin{center}
 \epsfig{file=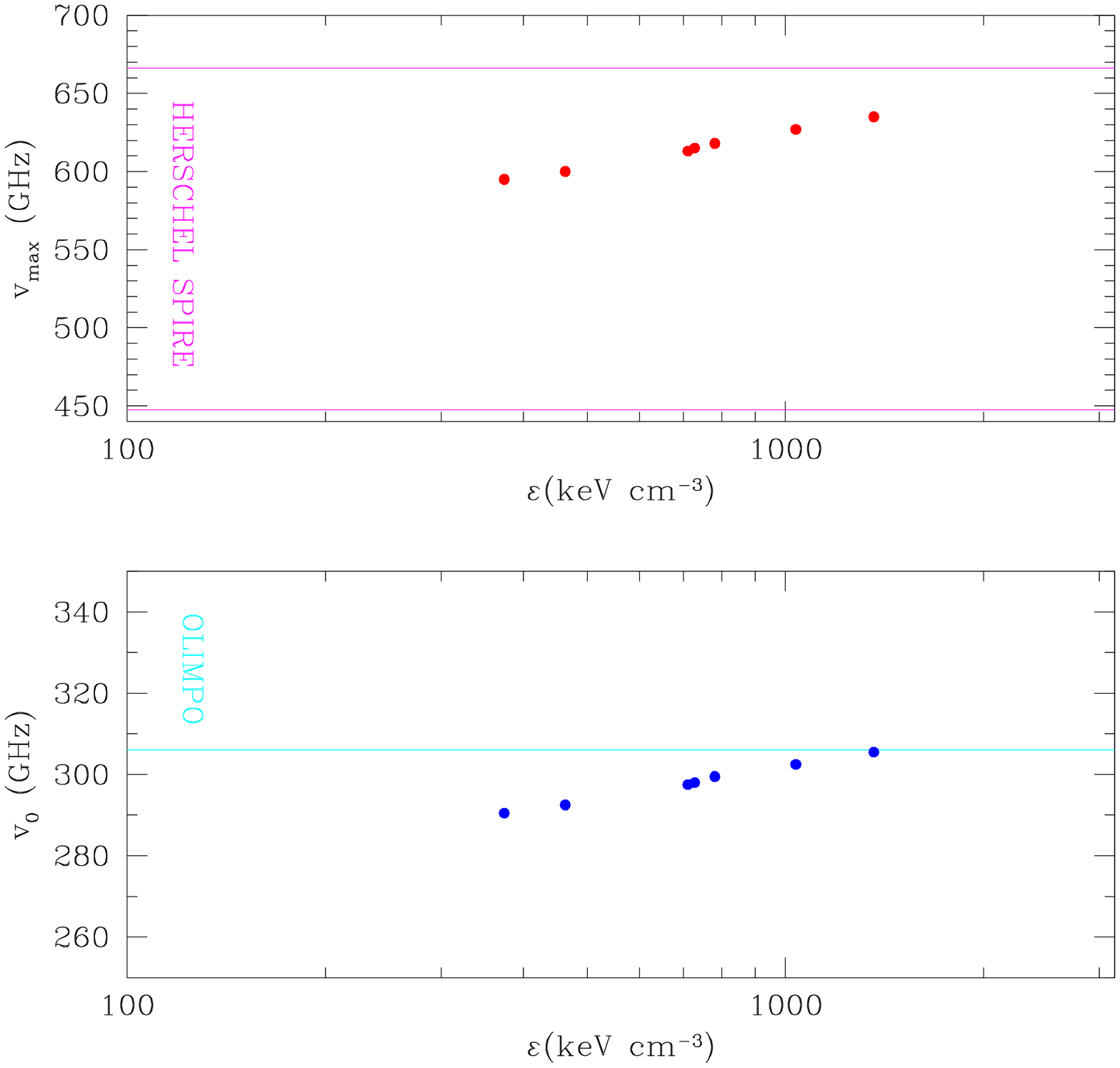,height=9.cm,width=9.cm,angle=0.0}
\end{center}
 \caption{\footnotesize{Correlation of the frequency position
 of the maximum $\nu_{max}$ (upper panel) and of the crossover $\nu_0$ (lower panel)
 of the non-thermal SZE from RG lobes as a function of the
 electron energy density at the center of the radio lobes for the seven objects
 we study here.
 The lowest frequency Herschel-SPIRE band ($\approx 447-666$ GHz, as enclosed by the
 magenta lines) is shown for comparison.
 }}
 \label{fig.numax_e}
\end{figure}

A similar correlation is also found, consistent with the
theoretical results described in Sect. 2.1, between $\nu_0$ and
$\nu_{max}$ and the energy density ${\cal E}$ of the electron
population in the radio lobes (see Fig.\ref{fig.numax_e}). The
analytical formulae with a polynomial of power 2 that fit the
frequency locations of $\nu_{max}$ and $\nu_0$ as a function of
${\cal E}$ are
\begin{equation}
 \nu_{max} = -2.49 \cdot 10^{-5} \times {\cal E}^2 +0.0840 \times {\cal E}+566.78
\end{equation}
and
\begin{equation}
 \nu_{0} = -1.00\cdot 10^{-5} \times {\cal E}^2 +0.0326 \times {\cal E}+279.64 \; ,
\end{equation}
where the energy densities are given in keV cm$^{-3}$ and frequencies in
GHz.
An analogous correlation with the pressure of the electron
population is expected because $P = {\cal E} /3$.

The values of $\nu_0$ and $\nu_{max}$ decrease 
 with
increasing values of $\alpha$ and they increase with increasing value of ${\cal E}$.
This is because the ICS of more
energetic electron populations (which have lower values of
$\alpha$, i.e., harder spectra) provide on average a larger frequency
increase to the CMB photons (see Colafrancesco et al. 2003,
Colafrancesco 2008, Colafrancesco \& Marchegiani 2010 for
details).
The presence and the specific value of a possible high-energy
cutoff of the electron spectrum do not change the previous
results appreciably for steep electron spectra because they do not
affect the value of the total pressure and energy of the electron
population.

Table \ref{tab.4} reports, for the same objects we consider in the
previous figures, the values of the central synchrotron brightness
$S_{sync}$ at 150 and 500 GHz (evaluated under the assumptions
previously described) that we compare with the central SZE
brightness change $\Delta I$ evaluated at the same frequencies.
The values of the brightness are computed using a uniform spatial distribution of the electrons within the assumed ellipsoidal geometry, a uniform magnetic field,
 and we normalized the electron spectrum with $k_0=2.6$ cm$^{-3}$.
 We also assume a power-law electron spectrum with a spectral
 index $\alpha$ indicated in Table \ref{tab.4} and a minimum momentum $p_1=1$.
These results allow the cases to be identified in which the synchrotron
emission either dominates over the SZE signal or
provides an appreciable correction to the SZE signal.
We find, in agreement with the results of Colafrancesco (2008), that
for RG lobes with steep spectra the synchrotron emission provides
a negligible contamination of the SZE signal at both 150 and 500
GHz. However, in lobes with flat spectra the synchrotron emission
becomes more important. In some cases (e.g., for 3C 292,
which is the flattest spectra RG lobes we consider with
$\alpha=2.76$), it dominates over the SZE signal even at high
frequencies. There are also cases (like 3C 274.1 with an
intermediate value of $\alpha=2.94$) in which the flat spectrum of
the electron population provides a substantial contamination
to the SZE signal at low frequency (150 GHz), but not at higher
frequency (500 GHz).\\
We stress again that we obtained these results by
extrapolating at 150 and 500 GHz the electron spectral index
measured at lower radio frequencies (of a few GHz). Therefore,
the predicted synchrotron flux in the radio lobes must be
considered as an upper limit to the realistic synchrotron signal
at high frequencies, which is likely to be steeper due to electron
losses and aging (e.g., Longair 1993).

\begin{figure}[ht]
\begin{center}
 \epsfig{file=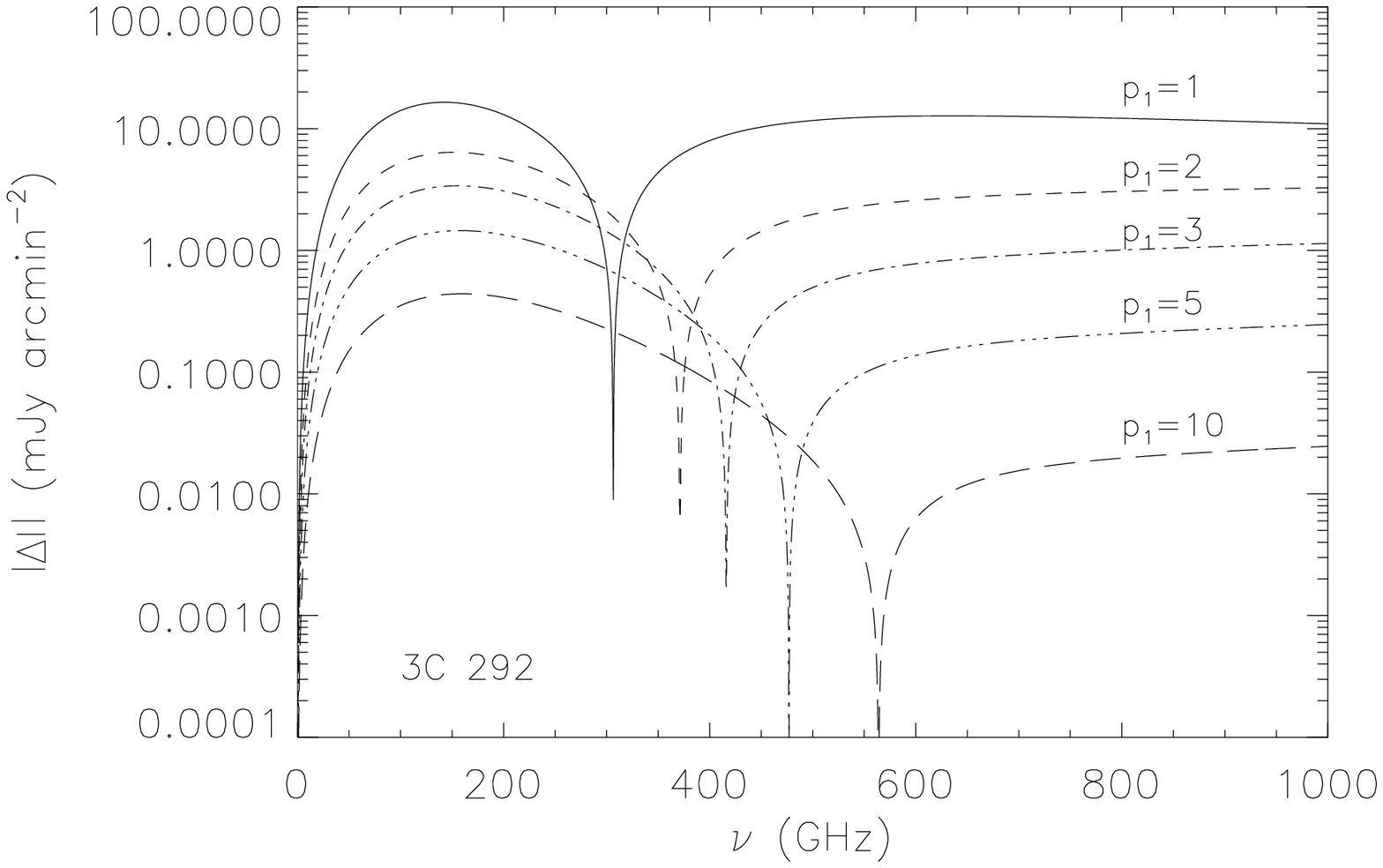,height=6.cm,width=9.cm,angle=0.0}
\end{center}
 \caption{\footnotesize{SZE spectrum of 3C 292 for a fixed
 value of $\alpha = 2.76$ and for values of $p_1=$ 1 (solid), 2 (dashes), 3 (dot-dashes),
 5 (3 dots-dashes) and 10 (long dashes).
 }}
 \label{fig.3c294_p1}
\end{figure}

We find that the SZE spectral shape also depends on the minimum
momentum $p_1$ (see also Colafrancesco \& Marchegiani 2010, 2011
for a discussion of this effect). However, since the value of
$p_1$ is unknown, but likely close to its minimum possible value,
we assumed in the previous calculations a unique value
$p_1=1$ for the electron spectrum of all the RG lobes.
Because we used an electron population with a
power-law spectrum and with a minimum momentum $p_1=1$, the
electrons that mostly contribute to the SZE signal are those with
low momenta, i.e., with $p\sim 1$. For this reason the crossover
frequency of the non-thermal SZE is found at relatively low
frequencies, $\nu_0 \sim 300$ GHz, for all the objects we consider.
Increasing the minimum value of $p_1$ has the effect of moving the
values of $\nu_{min}$, $\nu_0$ and $\nu_{max}$ toward higher
frequencies because the electronic population
is more populated by higher energy electrons, which thus yield
larger frequency up-shifts to the CMB photons. We show this effect
for the reference case of the RG 3C 292 in Fig.\ref{fig.3c294_p1}.
The fit to the frequency dependence of the main spectral
signatures of the SZE (i.e., $\nu_{min}$, $\nu_{0}$, and
$\nu_{max}$) for a fixed value of the electron spectral slope
$\alpha=2.76$ (as for the case of 3C 292) are
\begin{equation}
\nu_{min}=-0.431 \times p_1^2 + 6.43 \times p_1 + 138 \; ,
\end{equation}
\begin{equation}
\nu_0= -2.97 \times p_1^2 + 60.5 \times p_1 + 256 \; ,
\end{equation}
and
\begin{equation}
\nu_{max}= 403 \times p_1^2 - 73.5 \times p_1 + 256 \; .
\end{equation}
The frequency values of $\nu_{min}, \nu_{0}$, and
$\nu_{max}$ are given in GHz. The previous fits are
valid in the range $p1=1 \div 10$, based on the curves reported in
Fig.11. In this case, the minimum of the SZE also 
increases its location in frequency with increasing values of
$p_1$, even if its dependence on $p_1$ is milder than for the
other two spectral signatures $\nu_0$ and $\nu_{max}$. This effect
is due to the fact that the minimum of the SZE is found in the
Rayleigh-Jeans parts of the CMB (upscattered and
unscattered) spectrum, whose difference is weakly sensitive to the
larger distortions of the CMB spectrum produced by the ICCMB
process for high-energy electrons.
The dependence of $\nu_{min}$ from $p_1$ remains unchanged by
varying $\alpha$. The same holds for the dependence of $\nu_0$
from $p_1$, but with a shift in the initial value; the dependence
of $\nu_{max}$ on $p_1$ is instead steeper (flatter) for lower
(larger) values of $\alpha$.

\section{The multi-frequency ICS spectrum of RG lobes}

In order to establish a more robust normalization of the SZE
signals predicted for the RG lobes, we calculate the
multi-frequency spectrum of the ICCMB emission in the energy bands
where observations are already performed or can be
performed, i.e., the soft X-ray band (SXR: 2-10 keV), the hard
X-ray band (HXR: 20-80 keV) and the gamma-ray band (GRB: 0.1-100
GeV).
Table \ref{tab.3} reports the ICCMB flux produced by the
relativistic electrons filling the radio lobes in the  SXR, HXR,
and GRB energy bands, together with the electron spectral index
$\alpha$ for each one of the 21 RG lobes listed in
Table \ref{tab.1}. These ICS fluxes are calculated with a
fixed B-field of 1 $\mu$G analogous to the SZE signal calculated
in Table \ref{tab.1}.
\begin{table*}[htb]{}
\vspace{2cm}
\begin{center}
\begin{tabular}{|*{5}{c|}}
\hline Object name         &  $\alpha$ & $F(2-10 \mbox{ keV})$ &
$F(20-80 \mbox{ keV})$ & $F(0.1-100 \mbox{ GeV})$\\
                                  &              & erg cm$^{-2}$ s$^{-1}$ &
erg cm$^{-2}$ s$^{-1}$ & erg cm$^{-2}$ s$^{-1}$ \\
 \hline 
CGCG 186-048 & 2.48 & $6.6\times10^{-15}$ & $9.9\times10^{-15}$ &
$2.1\times10^{-13}$\\
             & 2.66 & $5.0\times10^{-15}$ & $6.3\times10^{-15}$ & $5.9\times10^{-14}$\\
B2 1158+35   & 3.00 & $5.7\times10^{-14}$ & $4.9\times10^{-14}$ &
$1.0\times10^{-13}$\\
             & 3.06 & $4.7\times10^{-14}$ & $3.8\times10^{-14}$ & $5.9\times10^{-14}$\\
3C 270       & 2.78 & $9.9\times10^{-13}$ & $1.1\times10^{-12}$ &
$5.9\times10^{-12}$\\ 87GB 121815.5+635745 & 3.90 &
$8.3\times10^{-14}$ & $2.7\times10^{-14}$ & $1.0\times10^{-15}$\\
M 87         & 2.76 & $2.9\times10^{-11}$ & $3.3\times10^{-11}$ &
$2.0\times10^{-10}$\\ 3C 274.1     & 2.94 & $5.7\times10^{-13}$ &
$5.2\times10^{-13}$ & $1.4\times10^{-12}$\\ 4C +69.15    & 2.32 &
$2.5\times10^{-14}$ & $4.4\times10^{-14}$ & $1.9\times10^{-12}$\\
3C 292       & 2.76 & $1.2\times10^{-13}$ & $1.3\times10^{-13}$ &
$8.0\times10^{-13}$\\
             & 2.76 & $1.3\times10^{-13}$ & $1.4\times10^{-13}$ & $8.8\times10^{-13}$\\
B2 1358+30C  & 2.90 & $5.6\times10^{-14}$ & $5.3\times10^{-14}$ &
$1.7\times10^{-13}$\\
             & 3.66 & $1.7\times10^{-13}$ & $7.2\times10^{-14}$ & $7.9\times10^{-15}$\\
3C 294       & 3.16 & $1.3\times10^{-14}$ & $9.5\times10^{-15}$ &
$9.5\times10^{-15}$\\ PKS 1514+00  & 1.72 & $2.4\times10^{-14}$ &
$8.2\times10^{-14}$ & $4.9\times10^{-11}$\\ GB1 1519+512 & 2.68 &
$1.2\times10^{-13}$ & $1.5\times10^{-13}$ & $1.3\times10^{-12}$\\
3C 326       & 2.80 & $2.2\times10^{-13}$ & $2.3\times10^{-13}$ &
$1.2\times10^{-12}$\\
             & 2.72 & $2.6\times10^{-13}$ & $3.0\times10^{-13}$ & $2.2\times10^{-12}$\\
7C 1602+3739 & 3.24 & $1.9\times10^{-14}$ & $1.3\times10^{-14}$ &
$8.8\times10^{-15}$\\
             & 3.20 & $2.2\times10^{-14}$ & $1.5\times10^{-14}$ & $1.3\times10^{-14}$\\
MRK 1498     & 2.32 & $1.5\times10^{-14}$ & $2.7\times10^{-14}$ &
$1.1\times10^{-12}$\\
             & 2.32 & $1.5\times10^{-14}$ & $2.7\times10^{-14}$ & $1.1\times10^{-12}$\\
B3 1636+418  & 2.78 & $2.0\times10^{-15}$ & $2.1\times10^{-15}$ &
$1.2\times10^{-14}$\\
             & 2.84 & $5.7\times10^{-15}$ & $5.9\times10^{-15}$ & $2.5\times10^{-14}$\\
Hercules A   & 3.22 & $1.9\times10^{-11}$ & $1.3\times10^{-11}$ &
$9.9\times10^{-12}$\\ B3 1701+423  & 2.94 & $2.1\times10^{-14}$ &
$1.9\times10^{-14}$ & $5.2\times10^{-14}$\\
             & 2.84 & $5.6\times10^{-15}$ & $5.1\times10^{-15}$ & $1.4\times10^{-14}$\\
4C 34.47     & 2.62 & $3.9\times10^{-14}$ & $5.1\times10^{-14}$ &
$5.7\times10^{-13}$\\
             & 2.62 & $5.2\times10^{-14}$ & $6.7\times10^{-14}$ & $7.5\times10^{-13}$\\
87GB 183438.3+620153 & 2.82 & $4.2\times10^{-14}$ &
$4.4\times10^{-14}$ & $2.0\times10^{-13}$\\
             & 2.88 & $3.2\times10^{-14}$ & $3.1\times10^{-14}$ & $1.1\times10^{-13}$\\
4C +74.26    & 2.22 & $5.4\times10^{-14}$ & $1.1\times10^{-13}$ &
$7.2\times10^{-12}$\\
 \hline
 \end{tabular}
 \end{center}
 \caption{\footnotesize{ICCMB flux in the SXR, HXR, GRB for the full set of RG lobes in Table \ref{tab.1}.
 }}
 \label{tab.3}
 \end{table*}

\begin{figure}[ht]
\begin{center}
 \epsfig{file=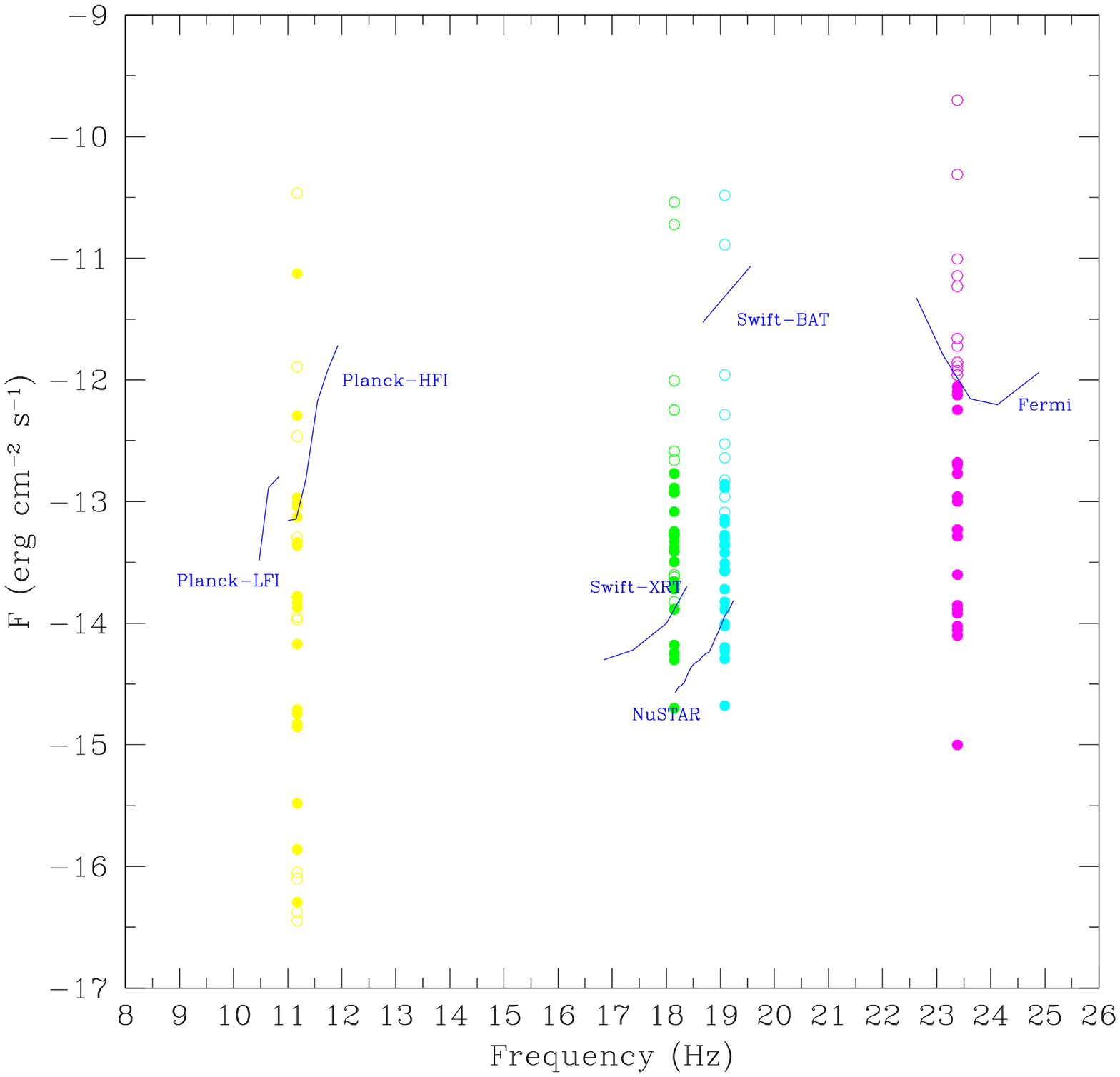,height=8.cm,width=9.cm,angle=0.0}
\end{center}
 \caption{\footnotesize{ICS flux predicted for the 21 RG lobes in Table \ref{tab.1}
 in the SXR, HXR, and gamma-ray bands is compared to the
 instrumental sensitivity of Swift-XRT, Swift-BAT, and Fermi-LAT. Open
 dot mark the RG lobes with $F_{ICS}(1 GeV)$ larger than the
 Fermi-LAT sensitivity. The black diamonds mark the seven RGs for which we perform
 a detailed study. We plot the gamma-ray flux at 1
 GeV and the X-ray flux at the centers of the
 2-10 keV and 20-80 keV energy bands. The SZE flux is calculated at 150 GHz.
 }}
 \label{fig.ics_multi}
\end{figure}

The predicted ICS flux from about half of the RG lobes in
Table \ref{tab.1} overcomes the Fermi-LAT 5yr sensitivity limit at 1
GeV (see Fig.\ref{fig.ics_multi}).
This therefore requires renormalizing the electron
spectrum of the RG lobes at the electron energy corresponding to a
1 GeV emitted photon by ICCMB, i.e., at $E_e = (1 GeV / 8
keV)^{1/2} \approx 354$ GeV. The electron spectrum normalization
for these RG lobes is thus an upper limit on their actual
spectrum, given the fact that it is set at the value of the
Fermi-LAT sensitivity.
The synchrotron frequency corresponding to this energy is $\approx
1.99 \cdot 10^{3}$ GHz, above the frequency range probed by
available space experiments.\\
The electrons in these RG lobes can therefore have either a
power-law spectrum extending beyond this energy but renormalized
to the Fermi-LAT upper limit or a cutoff at this energy in order
to be consistent with the multi-frequency limits.
After this renormalization, several RG lobes have SXR and
HXR flux still detectable with operative instruments (e.g.,
Chandra, XMM-Newton, Swift-XRT) and with upcoming HXR
instruments (e.g., NuSTAR).\\
The SZE observations for the RG lobes will thus be
relevant in determining their high-energy features.

As previously discussed, the ICCMB spectra of the seven RG lobes
reported in Figs. \ref{fig.87gb12}--\ref{fig.hera} for which we
perform a detailed analysis are normalized instead to a
fixed electron density $k_0 = 2.6$ cm$^{-3}$. The relative
values of the B-field that fit the radio observations are given
in Table \ref{tab.6} (we used the previous
normalization as a reference value that provides the observed
synchrotron flux for Hercules A for a B-field of 1 $\mu$G).

\begin{figure}[ht]
\begin{center}
 \epsfig{file=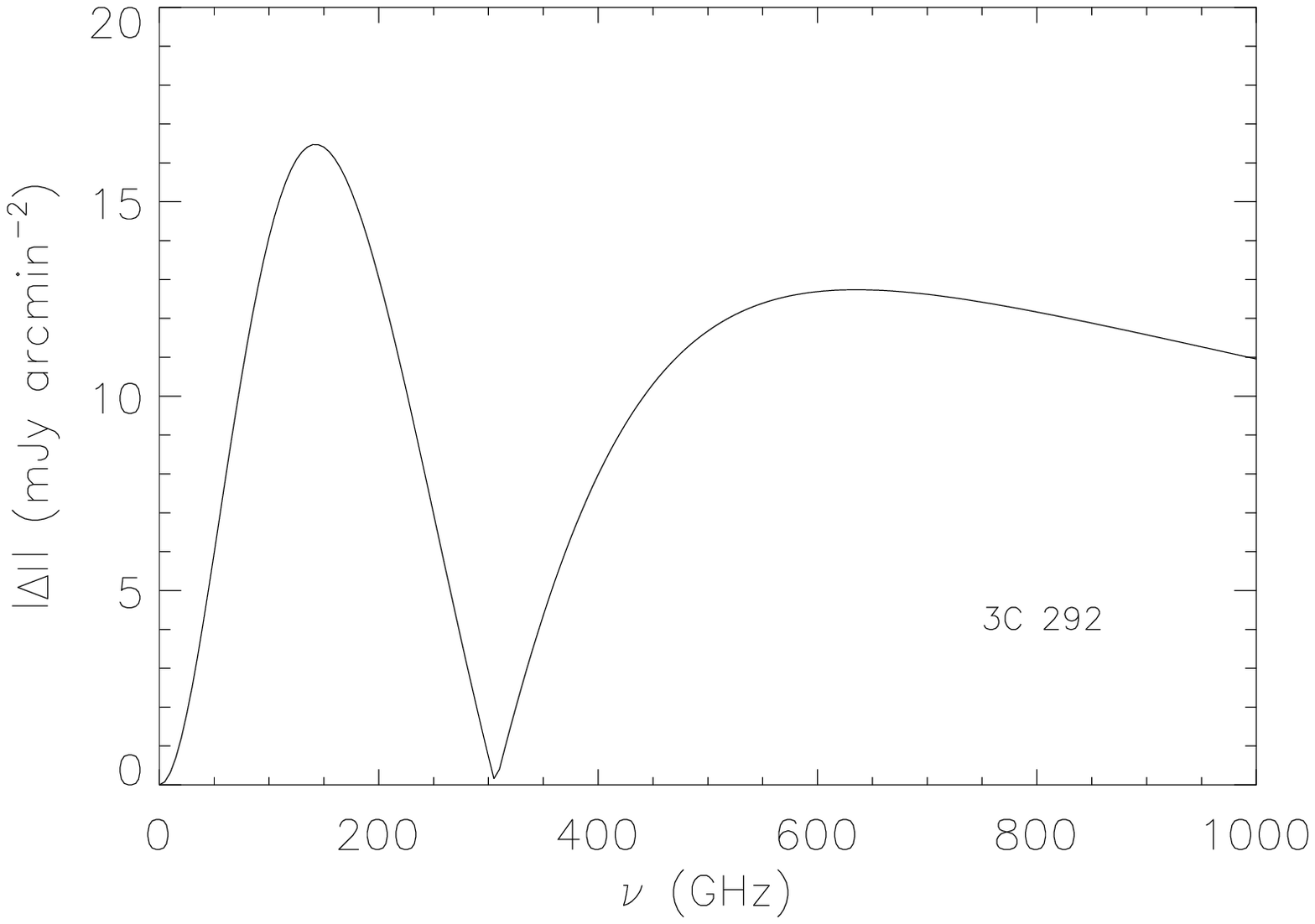,height=5.cm,width=9.cm,angle=0.0}
 \epsfig{file=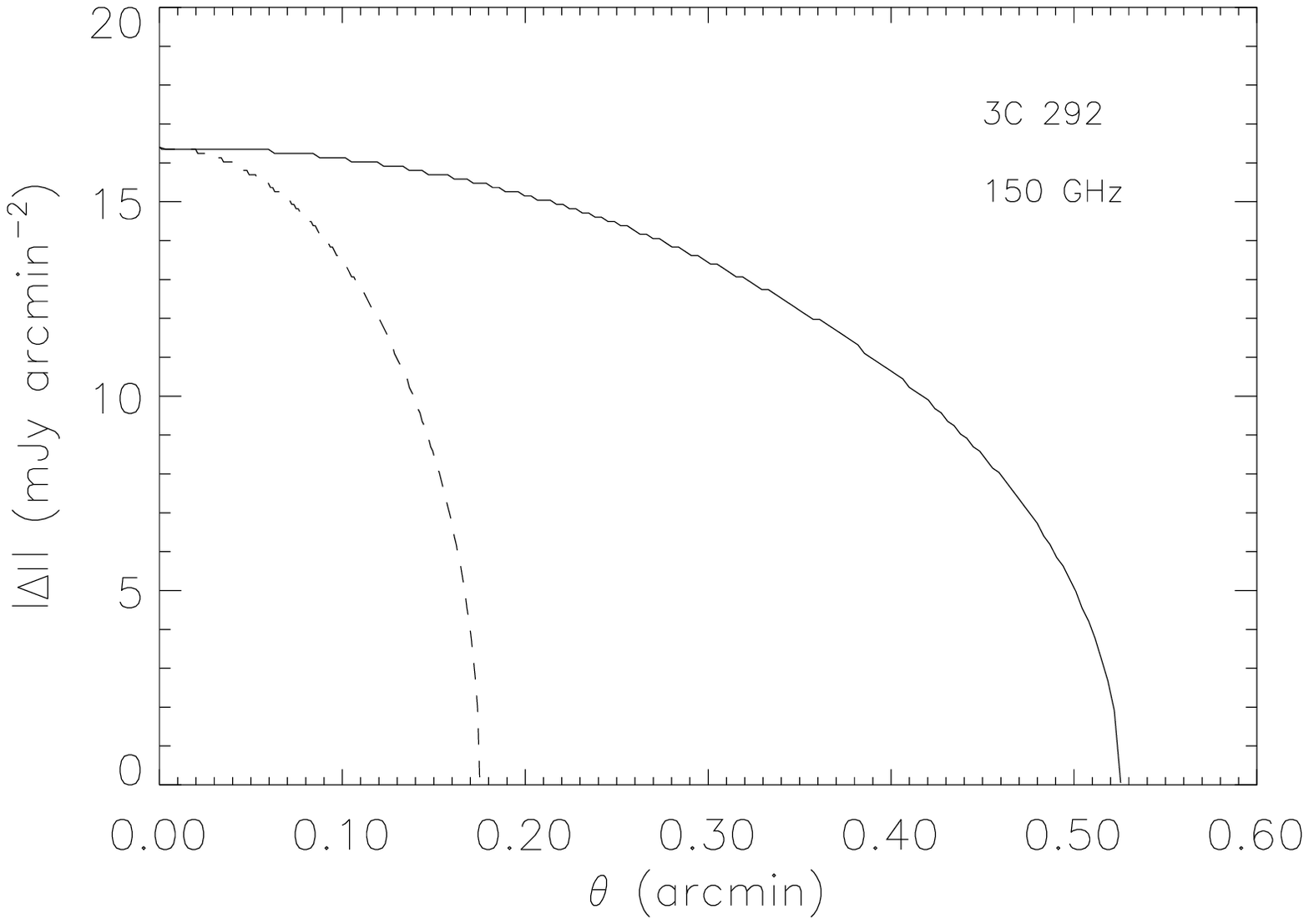,height=5.cm,width=9.cm,angle=0.0}
 \epsfig{file=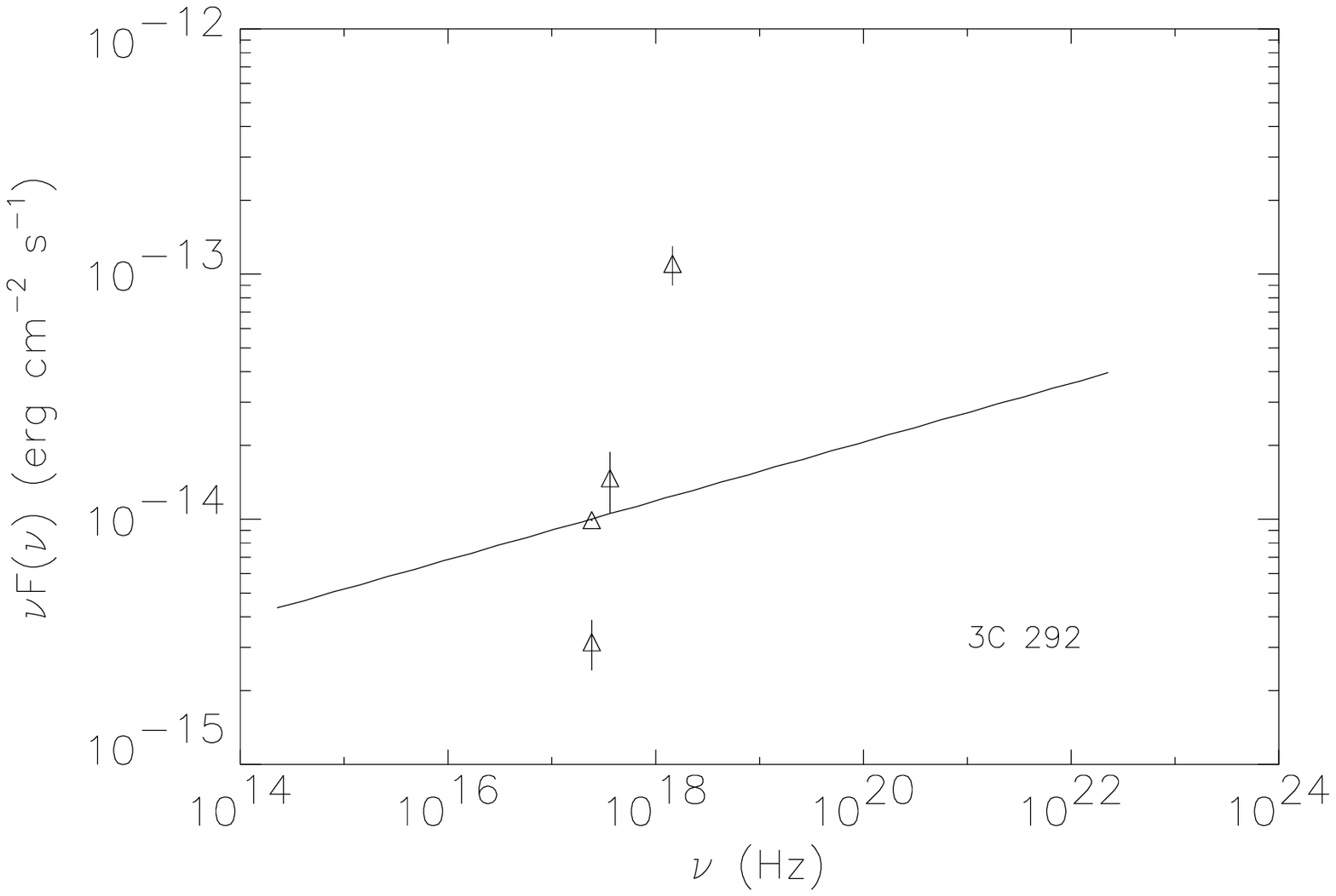,height=5.cm,width=9.cm,angle=0.0}
\end{center}
 \caption{\footnotesize{Same as Fig. \ref{fig.87gb12} but for 3C 292
and with normalization to X-ray data. The ICS X-ray emission model
is normalized to the X-ray flux from the RG lobe (Croston et al.
2005).
 }}
 \label{fig.3c292_normx}
\end{figure}

\begin{figure}[ht]
\begin{center}
 \epsfig{file=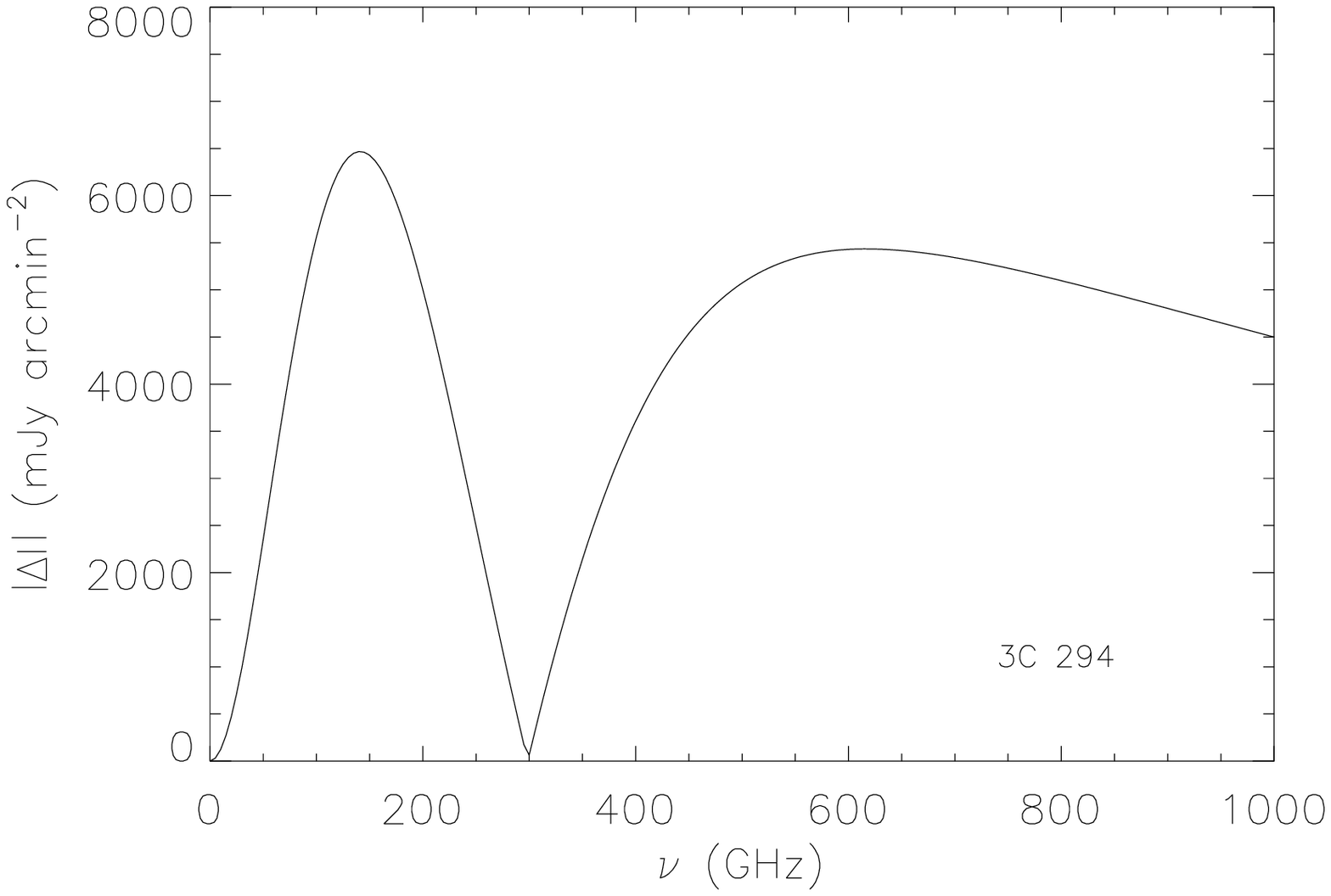,height=5.cm,width=9.cm,angle=0.0}
 \epsfig{file=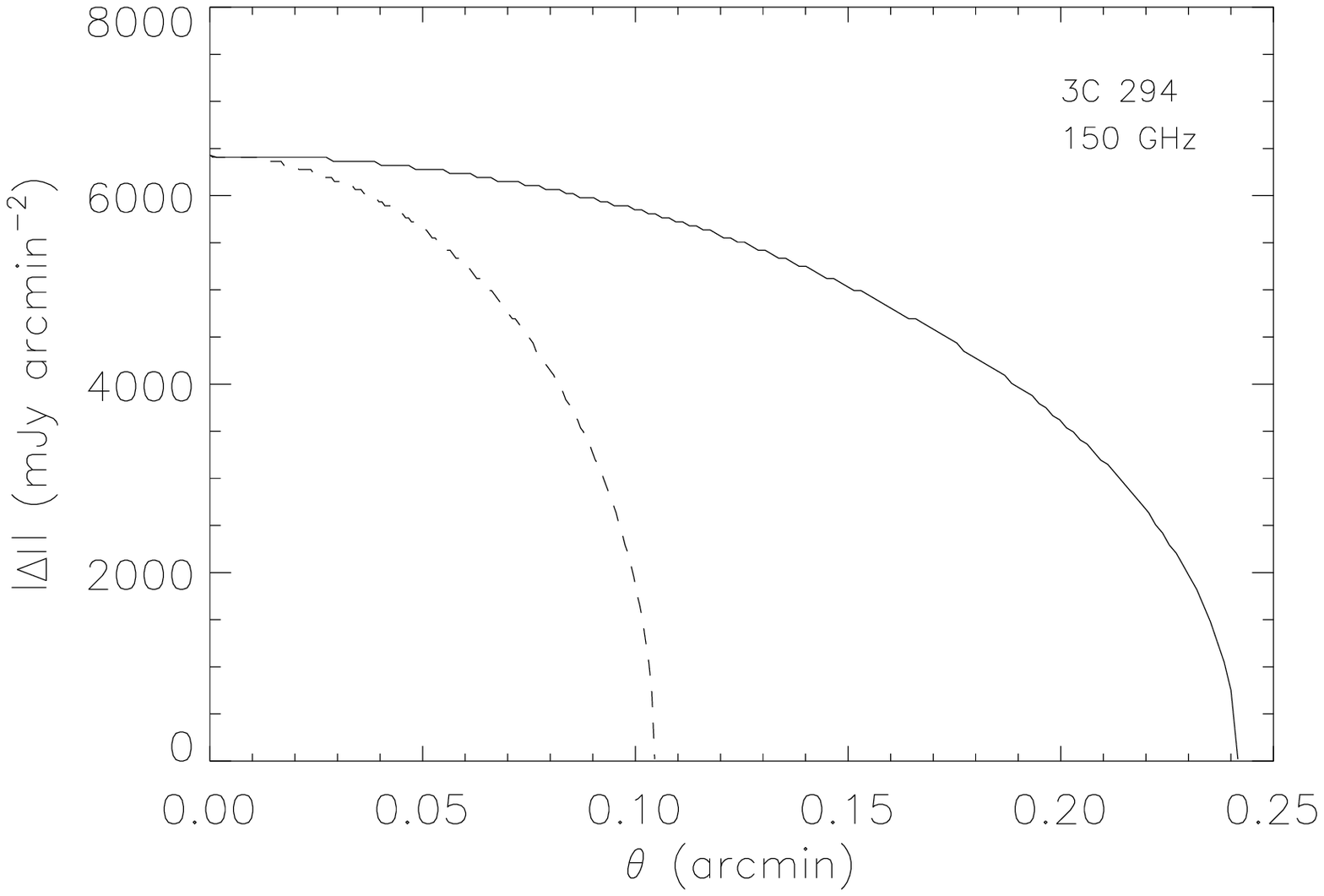,height=5.cm,width=9.cm,angle=0.0}
 \epsfig{file=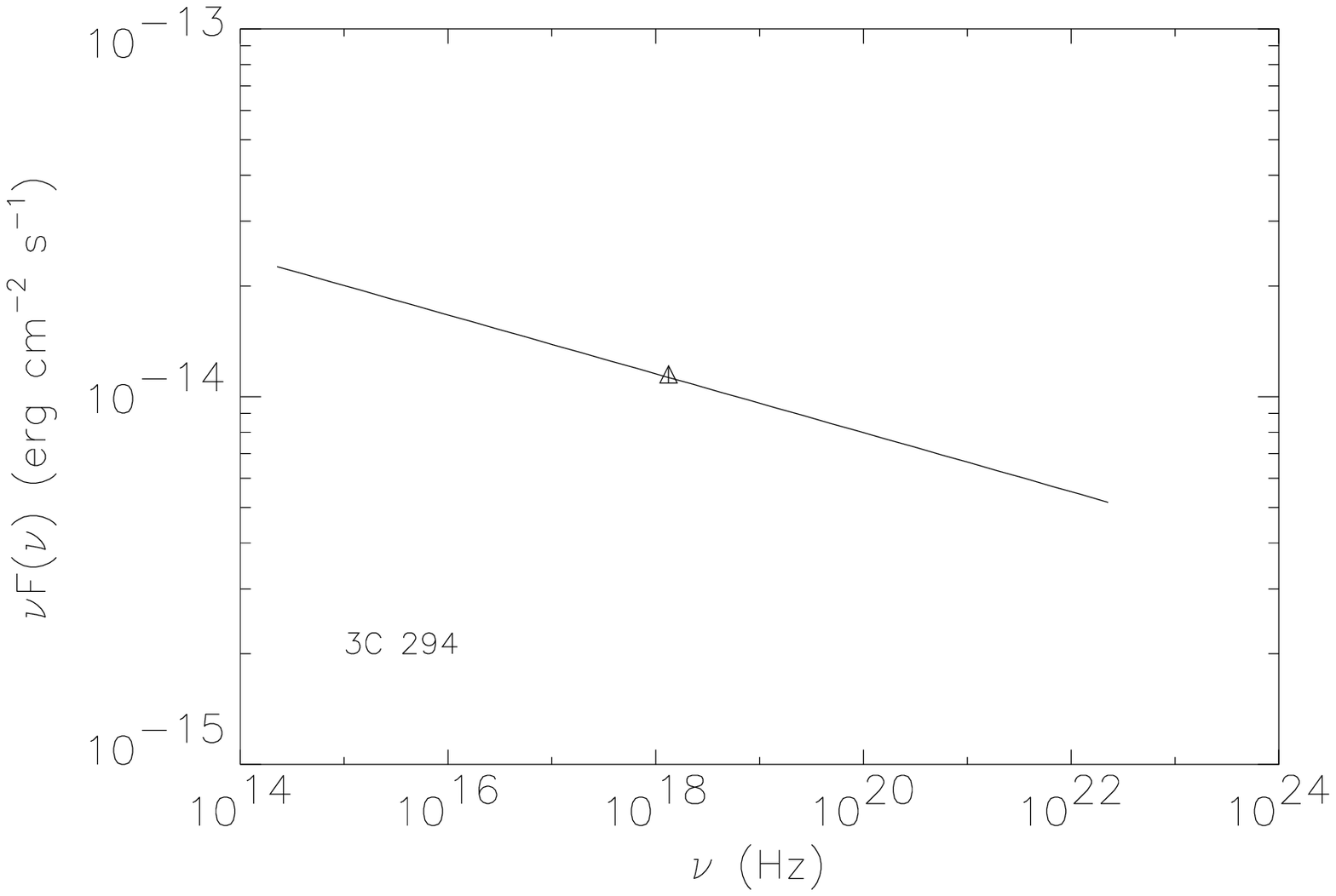,height=5.cm,width=9.cm,angle=0.0}
\end{center}
 \caption{\footnotesize{Same as Fig. \ref{fig.87gb12} but for 3C 294
and with normalization to X-ray Chandra data (Gilmour et al.
2009).
 }}
 \label{fig.3c294_normx}
\end{figure}

\begin{figure}[ht]
\begin{center}
 \epsfig{file=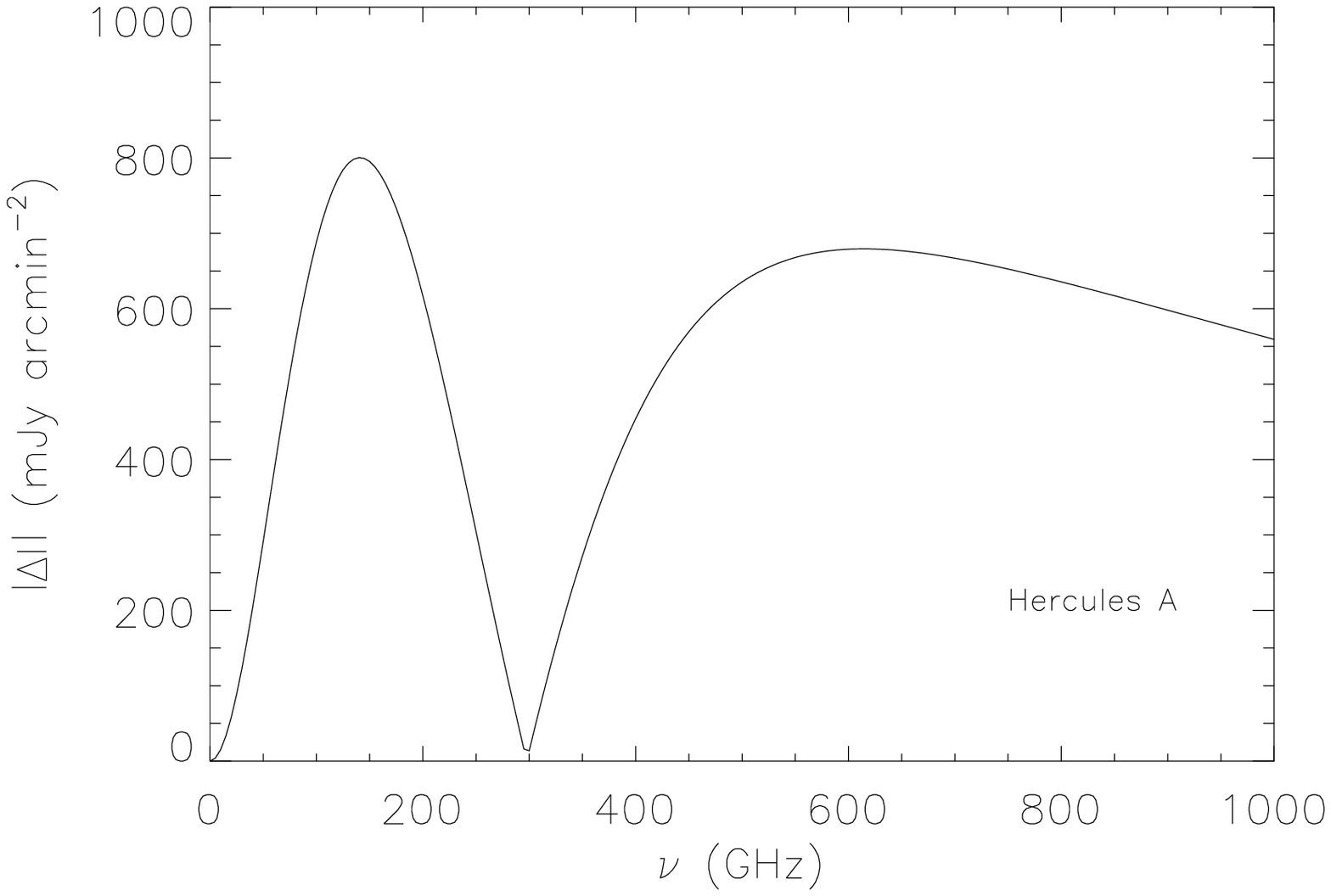,height=5.cm,width=9.cm,angle=0.0}
 \epsfig{file=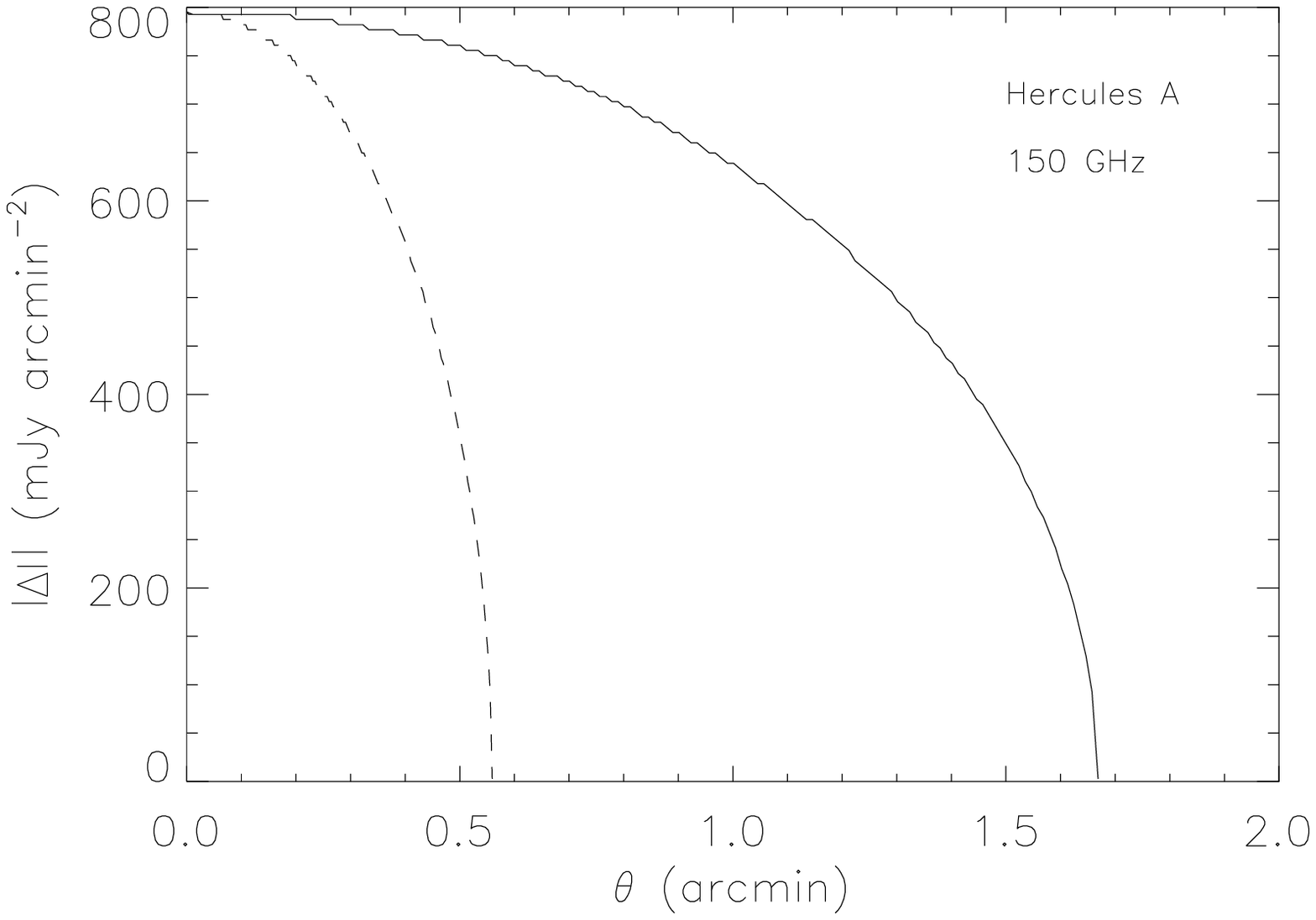,height=5.cm,width=9.cm,angle=0.0}
 \epsfig{file=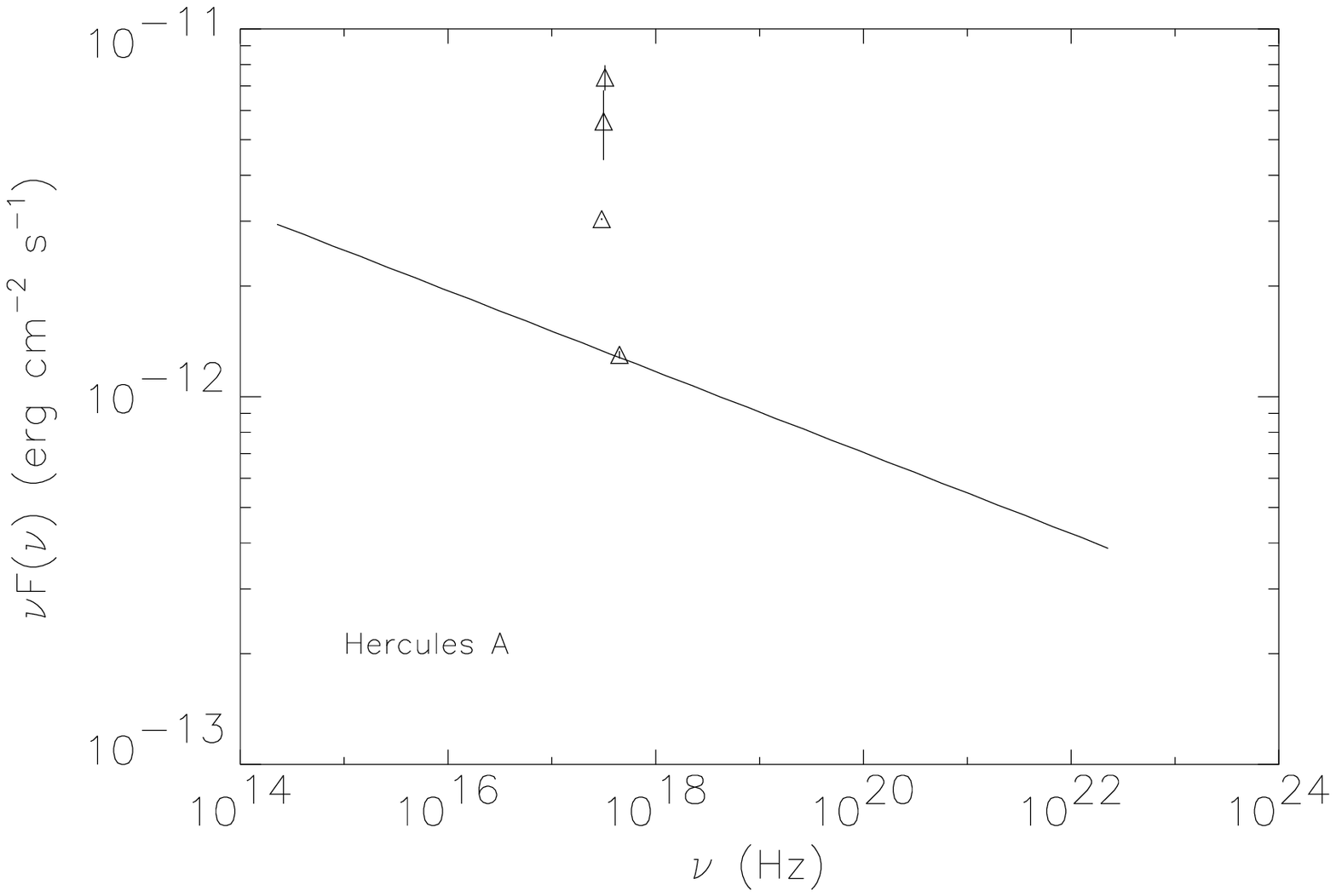,height=5.cm,width=9.cm,angle=0.0}
\end{center}
 \caption{\footnotesize{Same as Fig. \ref{fig.87gb12} but for Hercules A
and with normalization to the lowest value of the X-ray flux at 2
keV (Feigelson \& Berg 1983).
 }}
 \label{fig.hera_normx}
\end{figure}

For three of the RGs listed in Table \ref{tab.3} there are
high-energy observations available that allow the
degeneracy between the relativistic electron density and the
magnetic field present in synchrotron radio observations
to be broken. As a result,
the electron spectrum normalization and the RG
lobe magnetic field can be determined 
by combining radio and X-ray observations.
Here, we assume that the X-ray flux is due to to ICCMB emission, as
in the case of other RGs (see discussion in Sect.1).
Figures \ref{fig.3c292_normx}--\ref{fig.hera_normx} show the ICS
SEDs normalized to the available X-ray data that yield the
following electron spectrum normalization: $k_0= 4.6 \cdot
10^{-3}$ cm$^{-3}$ for 3C 292, $k_0= 3.3$ cm$^{-3}$ for 3C 294, and
$k_0= 2.5 \cdot 10^{-1}$ cm$^{-3}$ for Hercules A. We note
(see Erlund et al. 2006 for a detailed analysis) that the
X-ray emission associated with 3C 294 is offset in angle compared
with the radio emission, so that the normalization $k_0$ for this
object should be considered as an upper limit.
In addition, the lobes of Hercules A need to have an electron
spectrum with a normalization that makes the X-ray and gamma-ray
ICCMB emission consistent with the Fermi-LAT limit. Interestingly,
the Fermi limit is consistent with a normalization of the ICCMB
flux at the lowest observed X-ray flux for Hercules A (see
Fig.\ref{fig.hera_normx}).

Given these normalizations for the electron density in these RG
lobes, we can estimate the value of the (uniform) B-field in the
same lobes by combining the X-ray (ICCMB) flux and the radio
synchrotron flux to obtain: $B=2.7$ $\mu$G for 3C 292, $B=0.41$
$\mu$G for 3C 294, and $B=3.0$ $\mu$G for Hercules A (see Table
\ref{tab.6}).
The B-field estimated for 3C 294 is likely larger than $B=0.41$
$\mu$G because the X-ray flux is an upper limit to
the X-ray ICCMB flux from the RG lobe.
The B-field estimated for Hercules A is also probably larger than
$B=3.0$ $\mu$G because the X-ray flux used to
normalize the ICS emission is likely an upper limit to the X-ray
ICS flux from the RG lobe.
We emphasize that these estimates of the B-field rely upon the
extrapolation of a unique electron spectrum from the energies at
which the electrons produce ICS X-ray emission, i.e., $E_e\approx$
a few tenths of GeV to a GeV, to those at which electrons produce
synchrotron emission, i.e., $E_e \approx$ several to many GeV
(see Sect.1 and the discussion in Colafrancesco \&
Marchegiani 2011).

The SZE spectra and surface brightness distribution of the three
RG lobes calculated using the previous re-normalized values of
$k_0$ are shown in
Figs.\ref{fig.3c292_normx}--\ref{fig.hera_normx}.
In this context, however, observations of the SZE in the
lobes of radio galaxies can provide a better and more reliable
estimate of the absolute normalization of the electron spectrum
and of the RG lobe magnetic field. This is because the SZE is a measure of
the integrated electron spectrum and hence less sensitive to
the extrapolation of the electron spectrum when comparing the
electrons responsible for the synchrotron and X-ray ICCMB
emissions.

In the next section, we discuss the optimal observational strategy
to determine the full spectrum of the SZE from RG lobes in order
to derive the physical parameters of the electron population
contained in these systems and to further determine the full
physical structure of the RG lobes.

\clearpage

\section{Observational strategy}

\begin{figure}[ht]
\begin{center}
 \epsfig{file=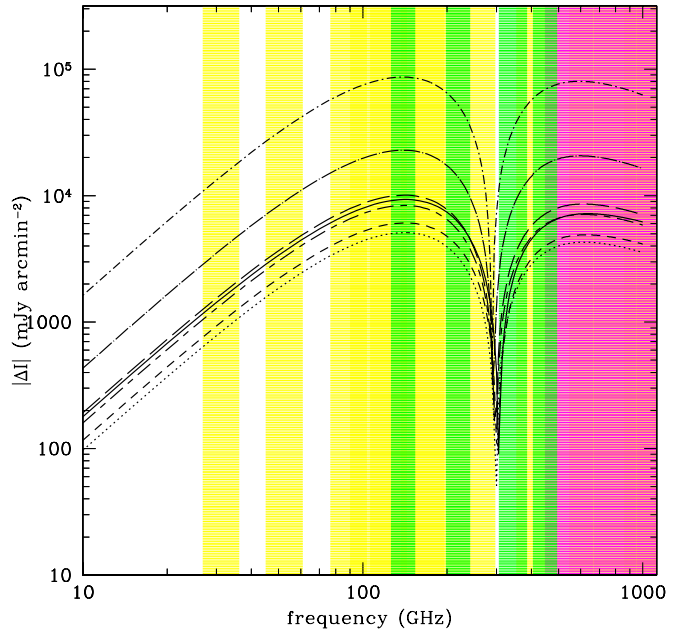,height=9.cm,width=9.cm,angle=0.0}
\end{center}
 \caption{\footnotesize{Superposition of the SZE spectrum for the seven
 objects we consider in our detailed analysis of the non-thermal SZE from RG
 lobes: 3C 292 (solid), 3C 294 (dots), 3C 274.1 (short dashes), 7C
 1602+3739 (long dashes), 87GB 121815+635745 (dot-dashes), B2
 1358+30C (dot-long dashes), and Hercules A (short-long dashes).
 The frequency coverage of Planck (yellow), OLIMPO (green), and
 Herschel-SPIRE (magenta) are also shown for comparison.
 }}
 \label{fig.szall}
\end{figure}

In this section, we discuss a possible observational strategy to
study the SZE from the RG lobes. The determination of the
full properties of the SZE spectrum from the relativistic
electrons populating the RG lobes requires combining observations
in the microwave, mm, and sub-mm frequency ranges.\\
Figure \ref{fig.szall} shows the SZE spectra evaluated for the seven
RG lobes (under the same assumption  used in Figs. 2-8) superposed
to the frequency bands in which Planck (yellow), OLIMPO (green),
and Herschel-SPIRE (magenta) operate. The optimal
frequency channels to study the non-thermal SZE from RG lobes are
at $\sim 150$ GHz and $\simgt 500$ GHz, where there is
optimal coverage of these instruments.\\
The Planck-HFI instrument has various frequency bands of interest
for studying this SZE signal. These are the bands B2 at $143 \pm
30$ GHz, B3 at $217 \pm 3$ GHz, B4 at $353 \pm 28$ GHz, and B5 at
$545 \pm 31$ GHz. Actually, the Planck-HFI instrument covers the whole
frequency range $\sim 100-857$ GHz, with NE$\Delta T_{CMB}$ from
$\sim 60 \mu K_{CMB} \sqrt{s}$ to $\sim 40 m K_{CMB} \sqrt{s}$
(see, e.g., The Planck Core Team 2011). However, the highest frequency
bands are heavily affected by instrumental and confusion noise.\\
The SPIRE instrument on board Herschel (see, e.g., Griffin
et al. 2010) is a FTS spectrometer with three bands (from 600 to
1200 GHz). It is therefore well suited to observe the SZE from
RG lobes, especially  the high-$\nu$ (positive) region of the SZE,
which contains crucial astrophysical information on the electron
plasma.\\
OLIMPO is a balloon-borne experiment carrying a four-band
bolometer array that is sensitive to four frequency bands 
with a band width of $\sim 10 \%$: 150,
220, 340, and 450 GHz (see Masi et al. 2005, 2008, Nati et al. 2007).
The OLIMPO instrument is being upgraded, including a differential
Fourier-Transform spectrometer similar to the one proposed for the
SAGACE space mission project (de Bernardis et al. 2010). The two
input ports of a Martin-Puplett interferometer are located in
symmetrical positions on the focal plane of the telescope. In this
way, the instrument measures the difference of the spectra from
couples of pixels (symmetrical with respect to the meridian
passing through the center of the focal plane). The resolution of
this instrument is 1.8 GHz, and spectra are obtained
simultaneously for the four bands of OLIMPO. With such an
improvement, the instrument is converted effectively in a 180-band
instrument.

The combination of these three experiments allows
the relevant parts of the RG lobes SZE to be covered, from the frequency region
of its minimum (at $\nu \sim 140$ GHz) to the maximum of the SZE
(at $\nu \sim 500-600$ GHz). It also probes the tail of the
SZE spectrum up to 1200 GHz (see Fig.\ref{fig.szall}), even though
the 857 and 1200 GHz channels are quite susceptible to foreground
contamination. This experimental combination allows
all the parameters of the non-thermal SZE in RG lobes to be determined: the
electron optical depth (mostly from low-frequency data) and the
energetics and pressure of the electron population (mostly from
high-frequency data).
Planck can cover most of the RG lobe SZE spectrum with good
sensitivity, especially at $\nu \simlt 450$ GHz, but has a spatial
resolution of several arcmin that does not allow a morphological
analysis of the SZE from RG lobes. Such a morphological analysis
is more suitable with OLIMPO and Herschel-SPIRE.
At low and intermediate frequencies, the OLIMPO observations are
crucial to determine the minimum of the SZE that provides a
measure of the electron optical depth quite independent from the
electron energetics (see Fig.\ref{fig.3c294_p1}).
At high frequencies, the Herschel-SPIRE observations are
crucial to  provide information on the energetics and total
pressure of the electron population because: i) the SZE spectrum
peaks at $\sim 600$ GHz (see Figs.\ref{fig.numax_s},
\ref{fig.numax_e}, \ref{fig.szall}) and extends in the SPIRE
frequency band, so that three high-quality data points can be
obtained to describe the smooth shape  of the SZE spectrum; ii)
the SZE brightness extends for several arcmins and depends
only on the spatial distribution of the electrons. The spatial
resolution of the Herschel-SPIRE instrument ($36.6 ''$ at 600 GHz,
$24.9''$ at 857 GHz, and $18.1''$ at 1200 GHz) can also map the
spatial distribution of the SZE and determine the distribution of
the electron population and its total energy content, a
goal that is not easily feasible with both radio (because of the
degeneracy between electrons and B-field) and X-ray or
gamma-ray (because only high-E electrons contribute)
observations.

Detailed observations of RG lobes in the microwave and mm
frequency range have not been obtained so far. Such observations
are very relevant to set constraints on the SZE in a frequency
region that is marginally affected by synchrotron emission for the
steep-spectrum RG lobes, which are the best candidates for a
detection of their SZE. We analyze in the following the WMAP 7yr
data to set constraints on the SZE from the seven RG lobes we
consider in Table 5 at frequencies of 41, 70, and 94 GHz.

\subsection{WMAP constraints}

The WMAP 7yr data in the direction of the RG lobes can set
limits on the amplitude of the SZE at $\mu$waves and hence on
the value of the minimum momentum $p_1$ (or the momentum break
$p_b$ for a double power-law spectrum) of the electron spectrum in
the RG lobes. In turn, it can set limits on the relevant quantities that depend
on it, namely, the electron optical depth, pressure, and energy
density.
We then use these limits on the electron spectrum to derive more
reliable predictions for the shape and amplitude of the SZE
spectrum of these RG lobes observable with  Planck, OLIMPO, and
Herschel-SPIRE.

\begin{figure}[ht]
\begin{center}
\vbox{
 \hbox{
 \epsfig{file=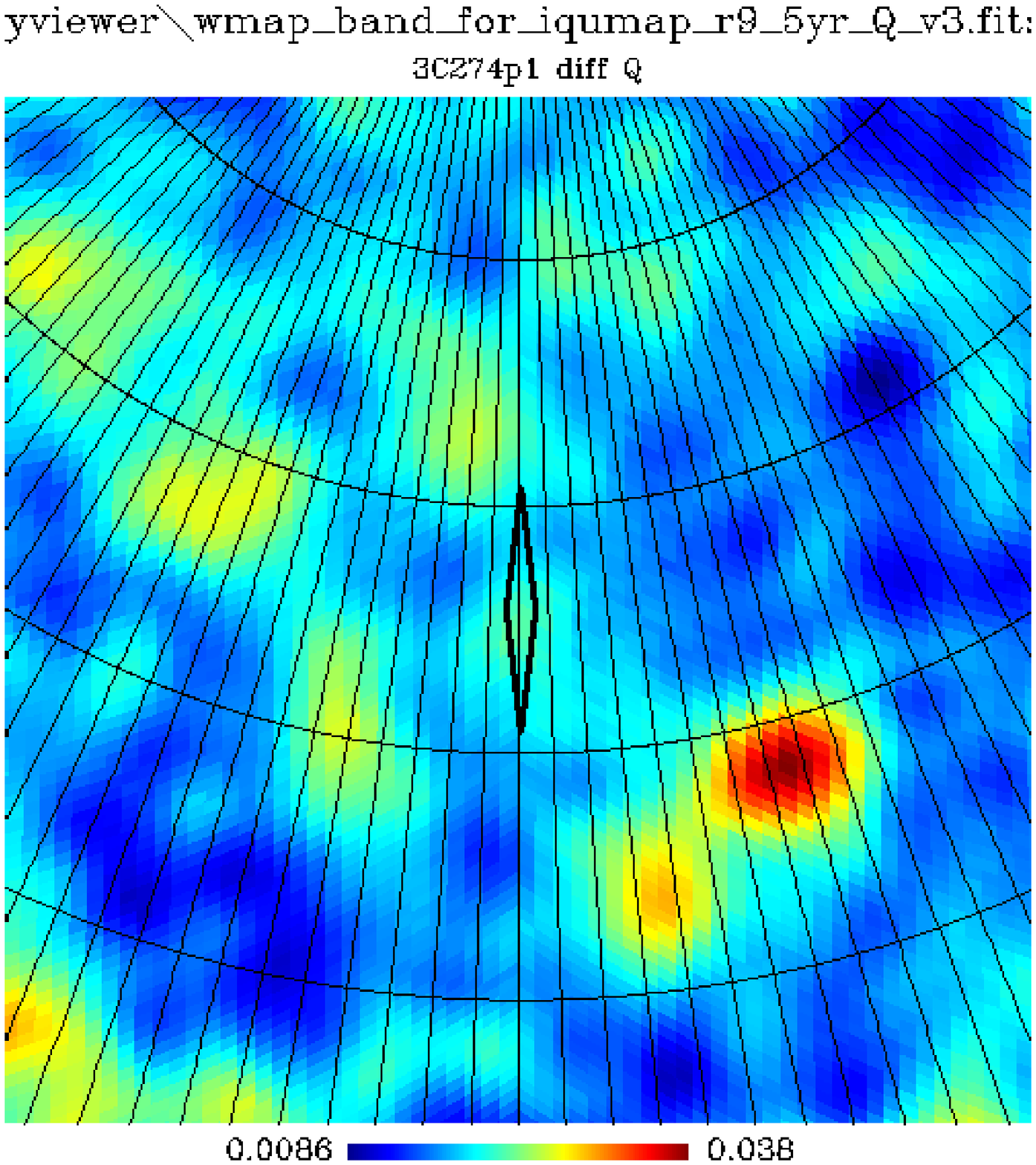,height=3.5cm,angle=0.0}
 \epsfig{file=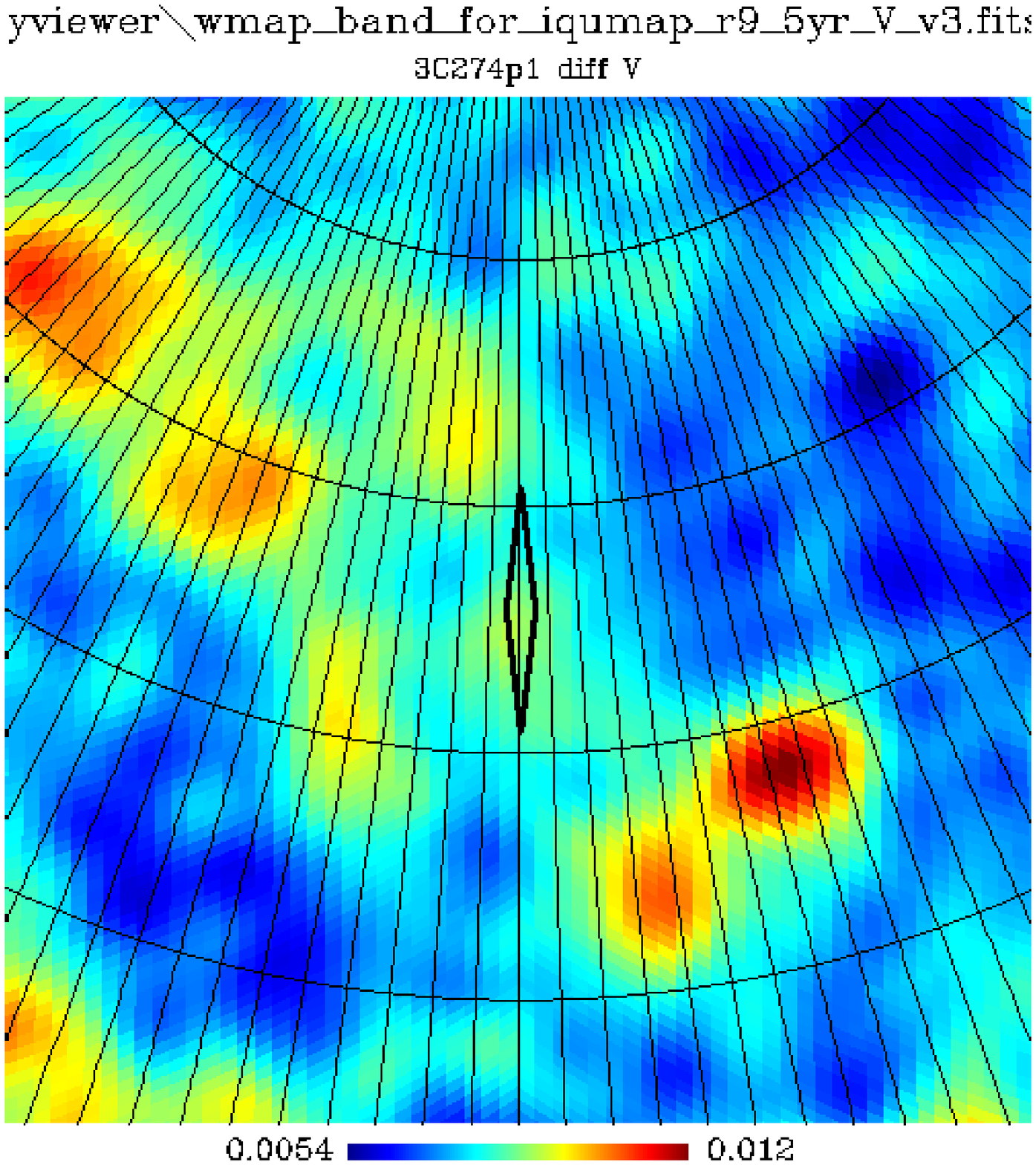,height=3.5cm,angle=0.0}
 \epsfig{file=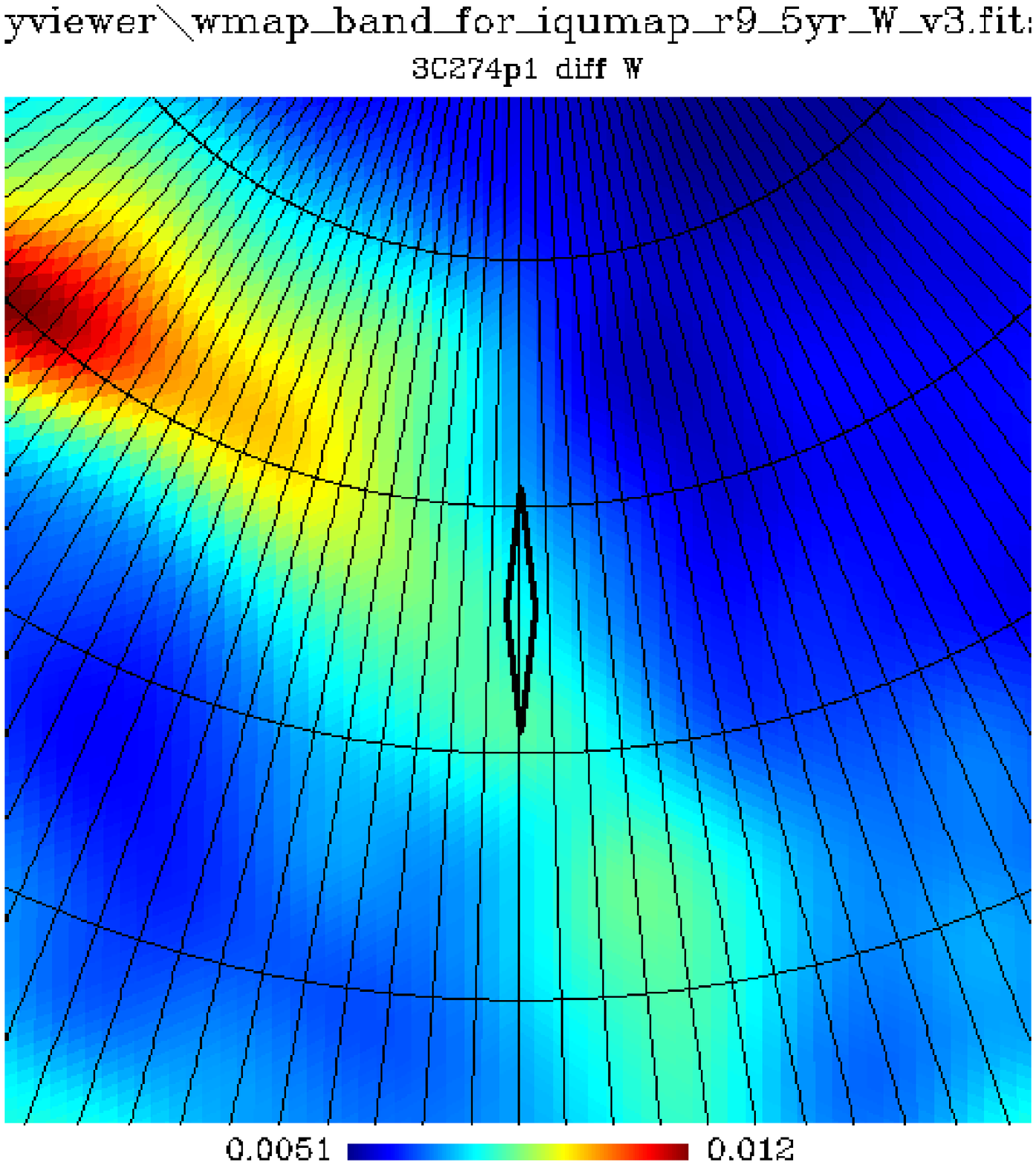,height=3.5cm,angle=0.0}
 }
 \hbox{
 \epsfig{file=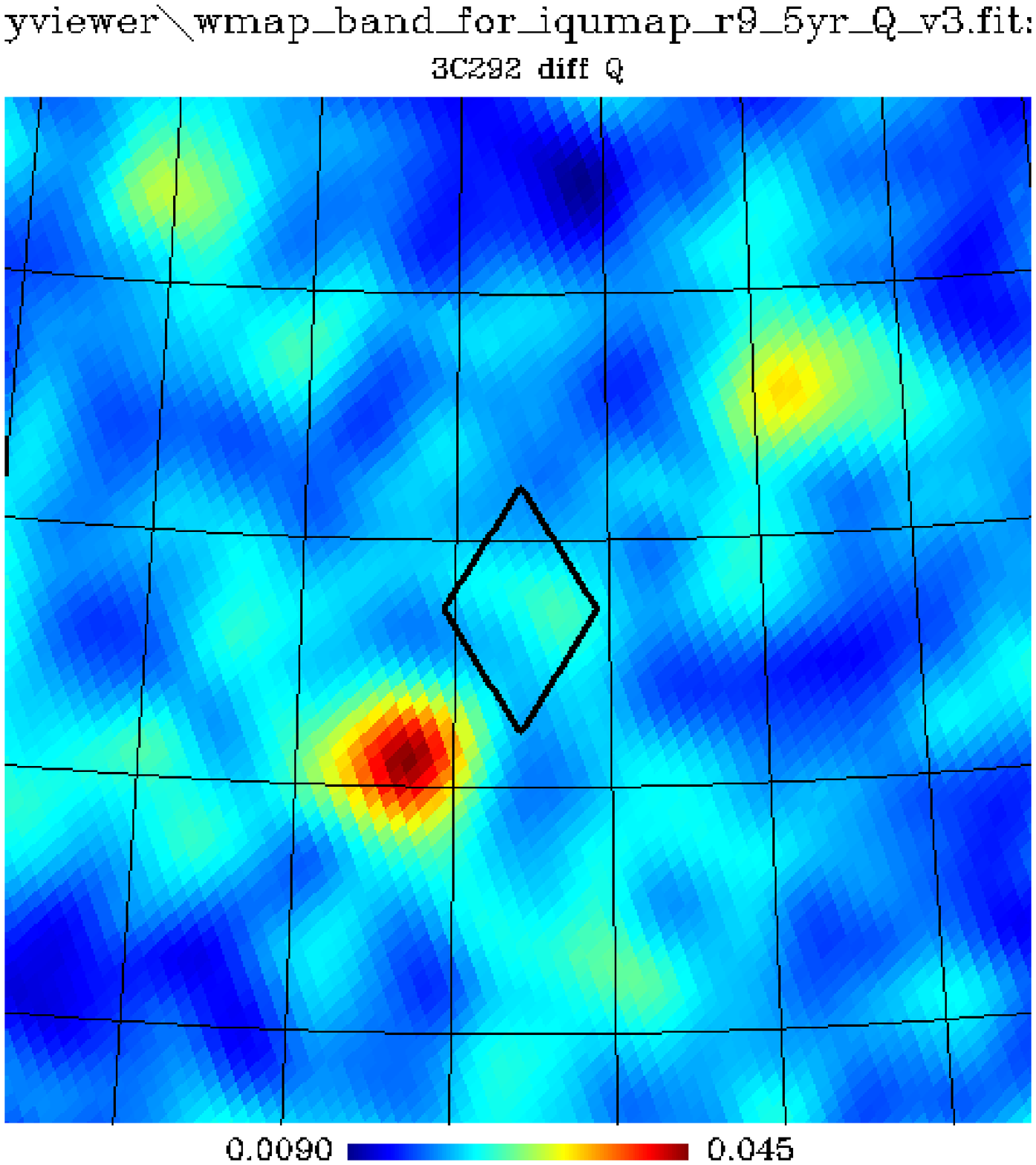,height=3.5cm,angle=0.0}
 \epsfig{file=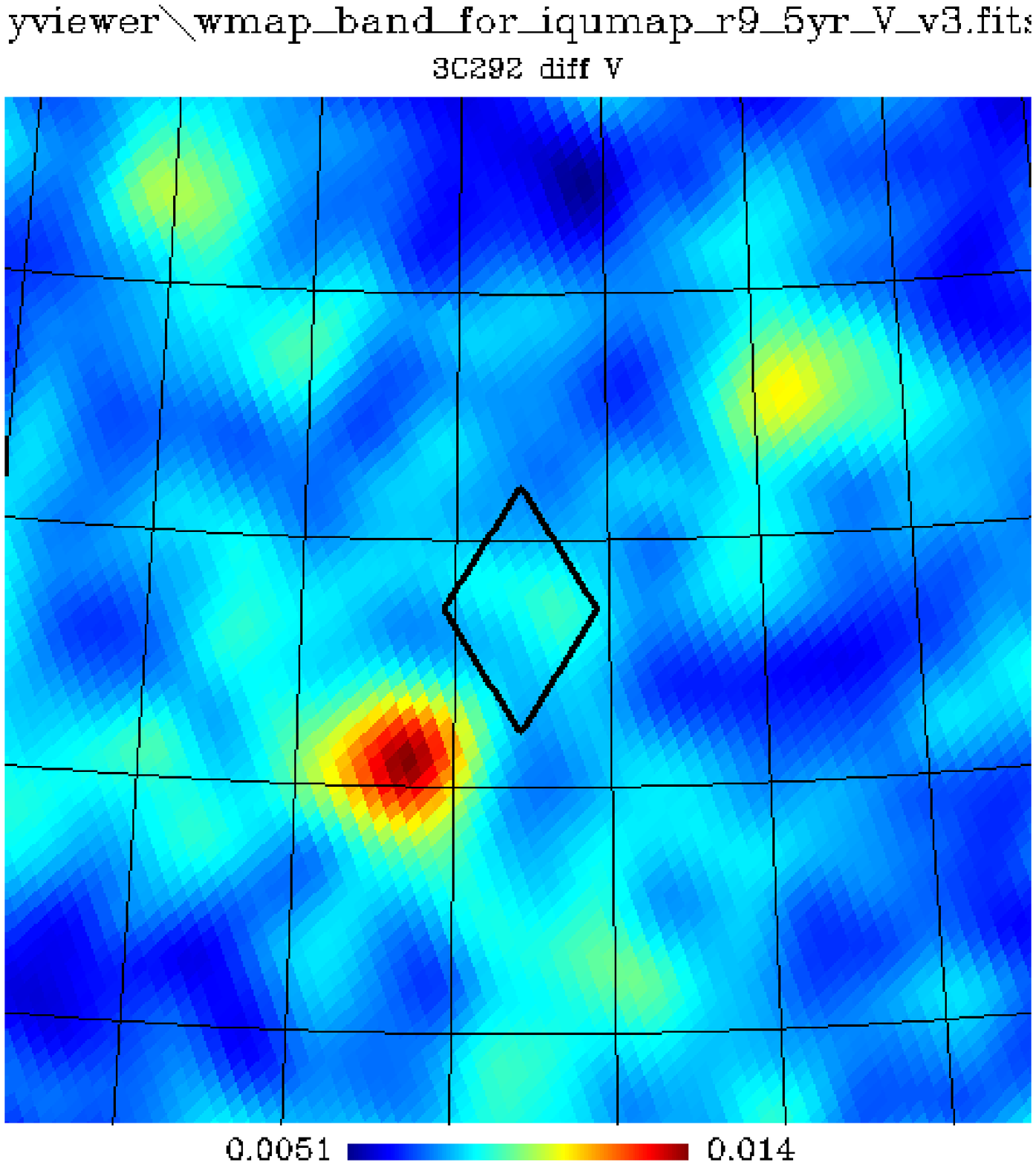,height=3.5cm,angle=0.0}
 \epsfig{file=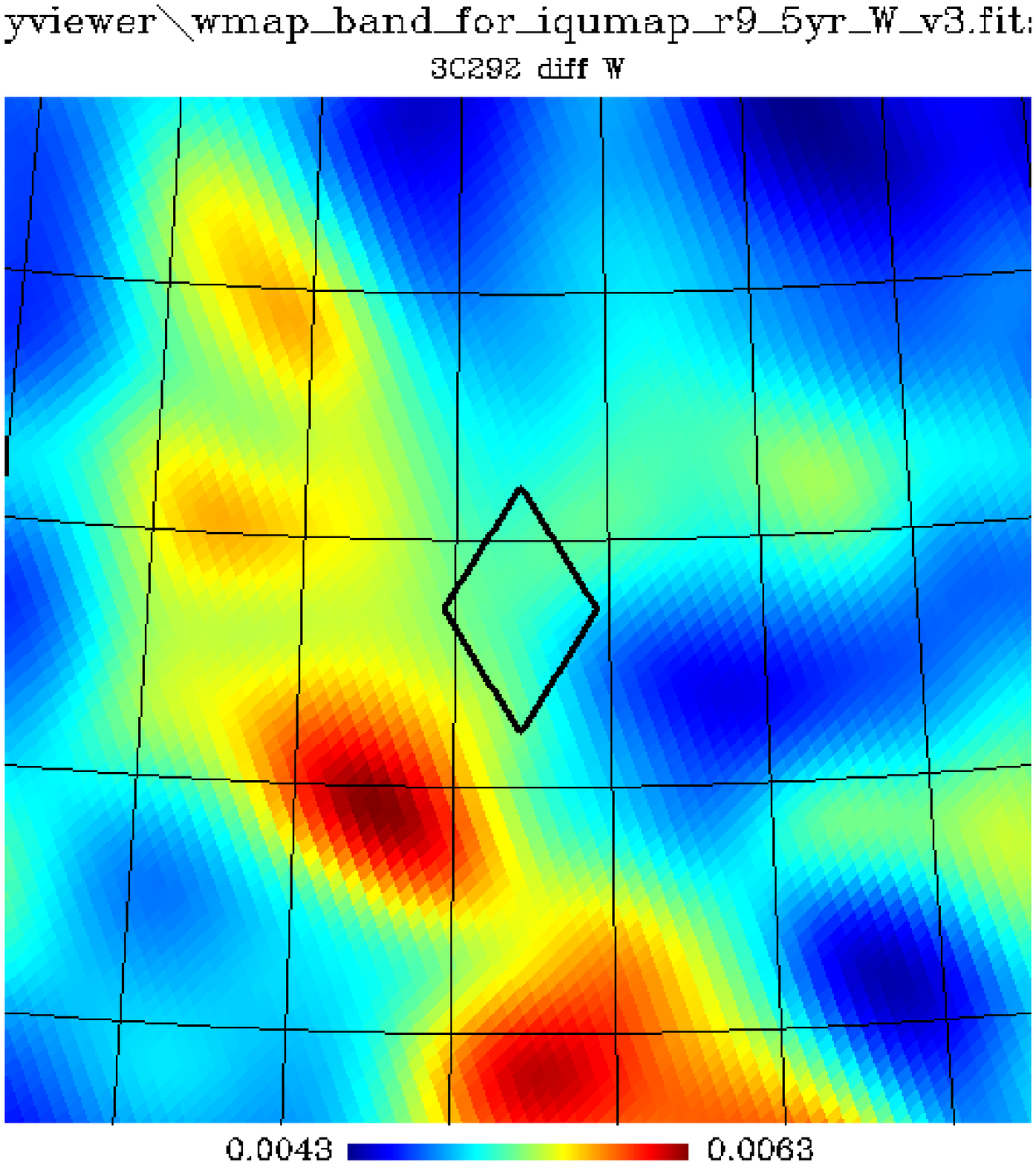,height=3.5cm,angle=0.0}
 }
 \hbox{
 \epsfig{file=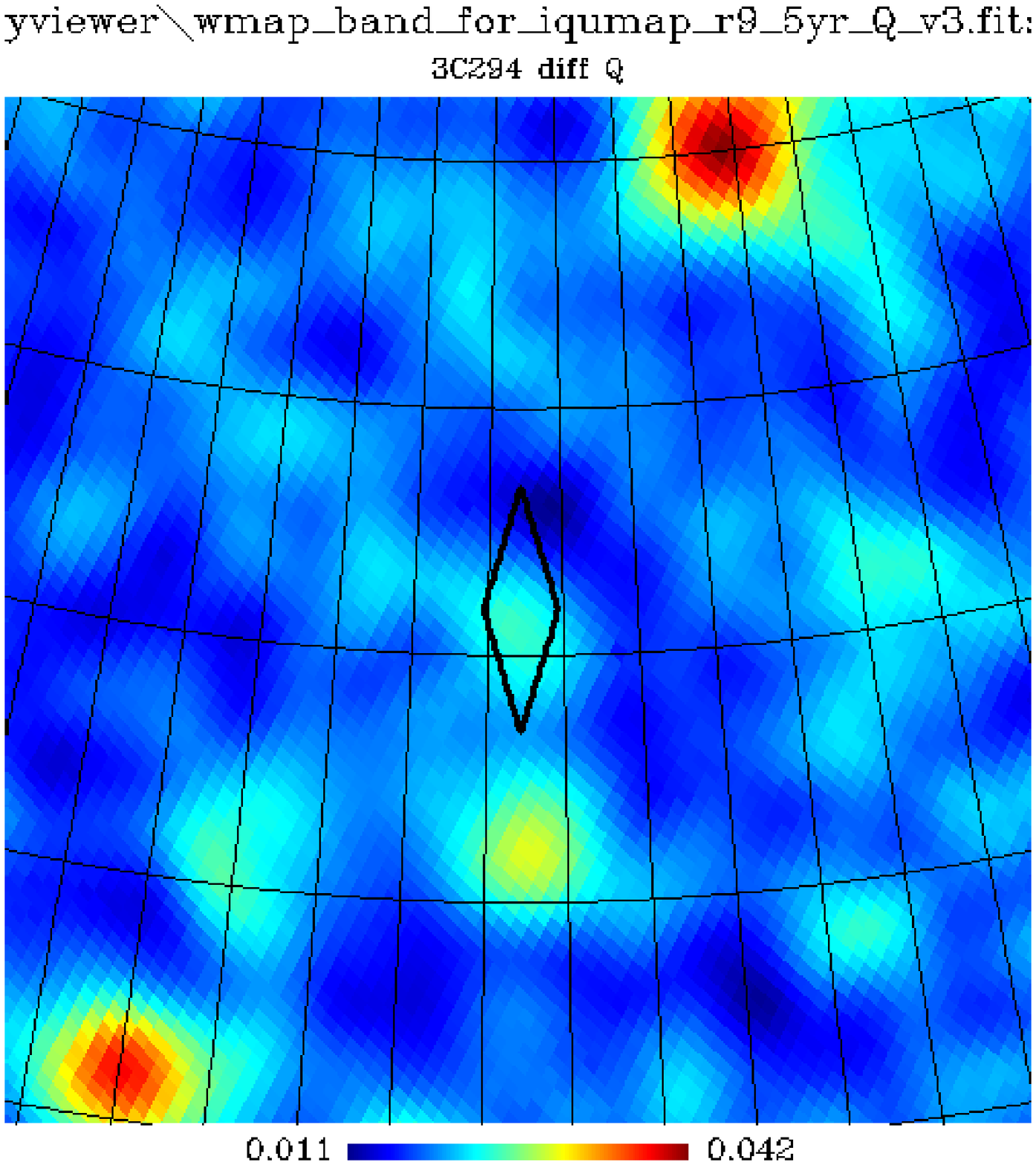,height=3.5cm,angle=0.0}
 \epsfig{file=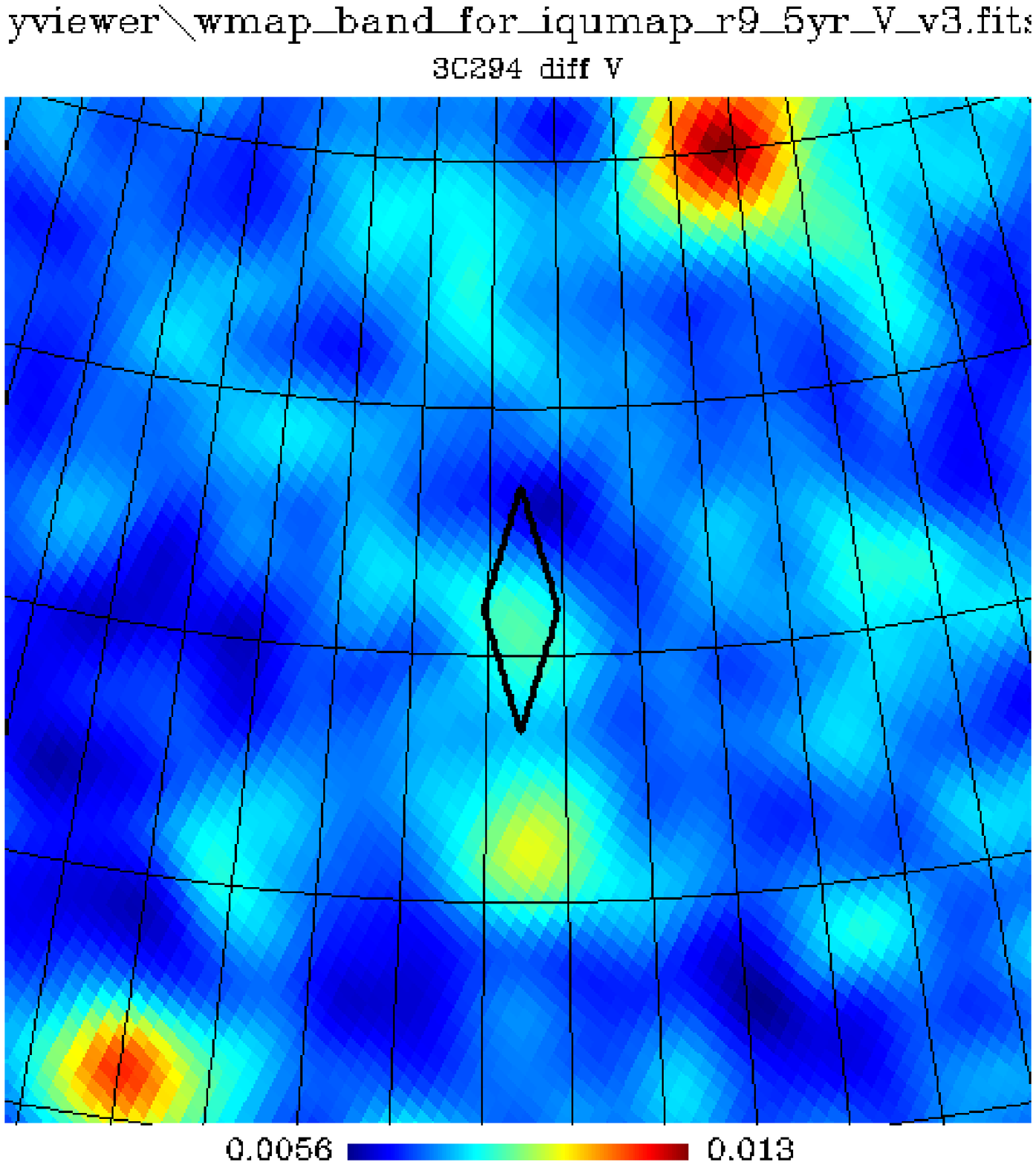,height=3.5cm,angle=0.0}
 \epsfig{file=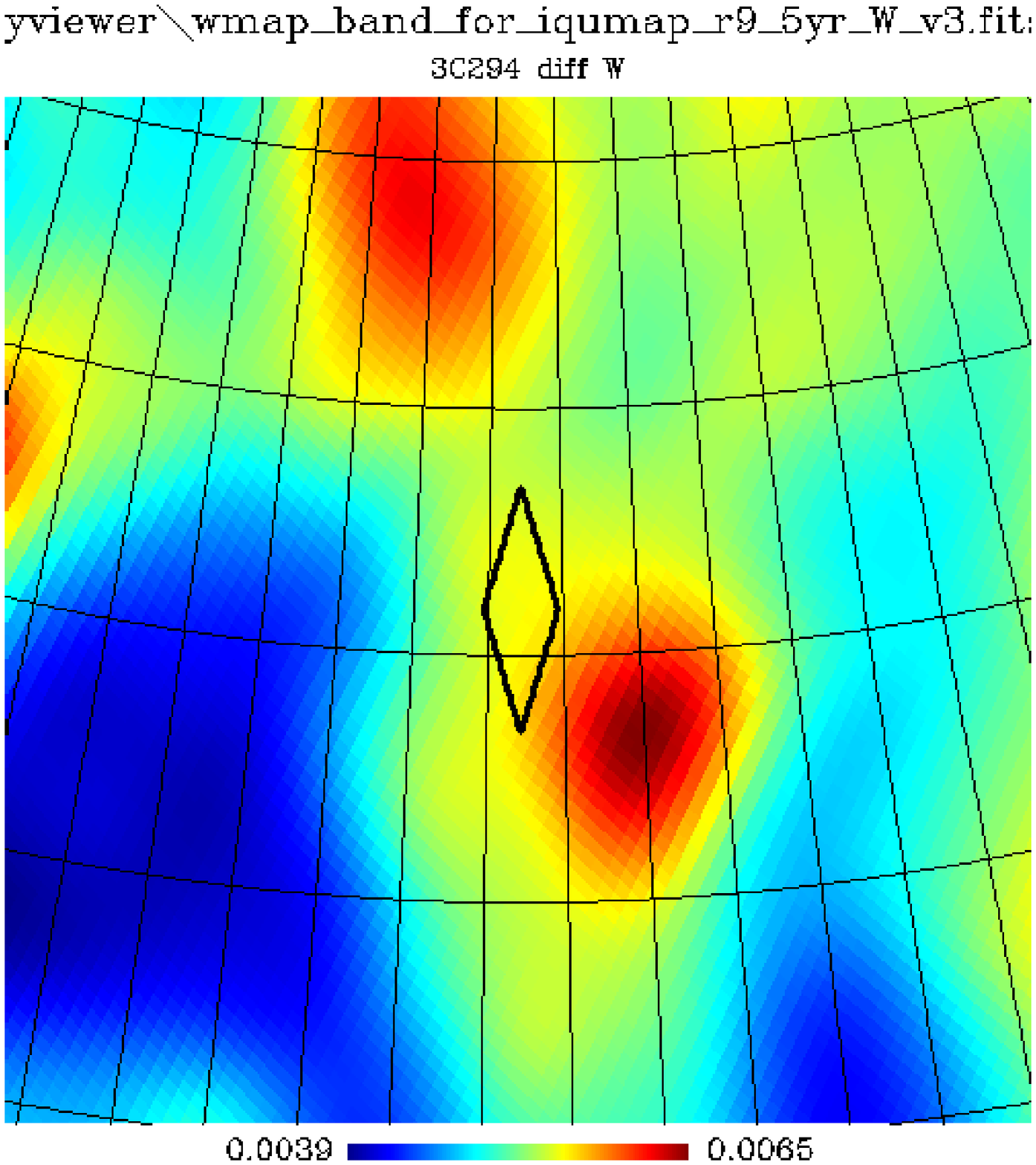,height=3.5cm,angle=0.0}
 }
 \hbox{
 \epsfig{file=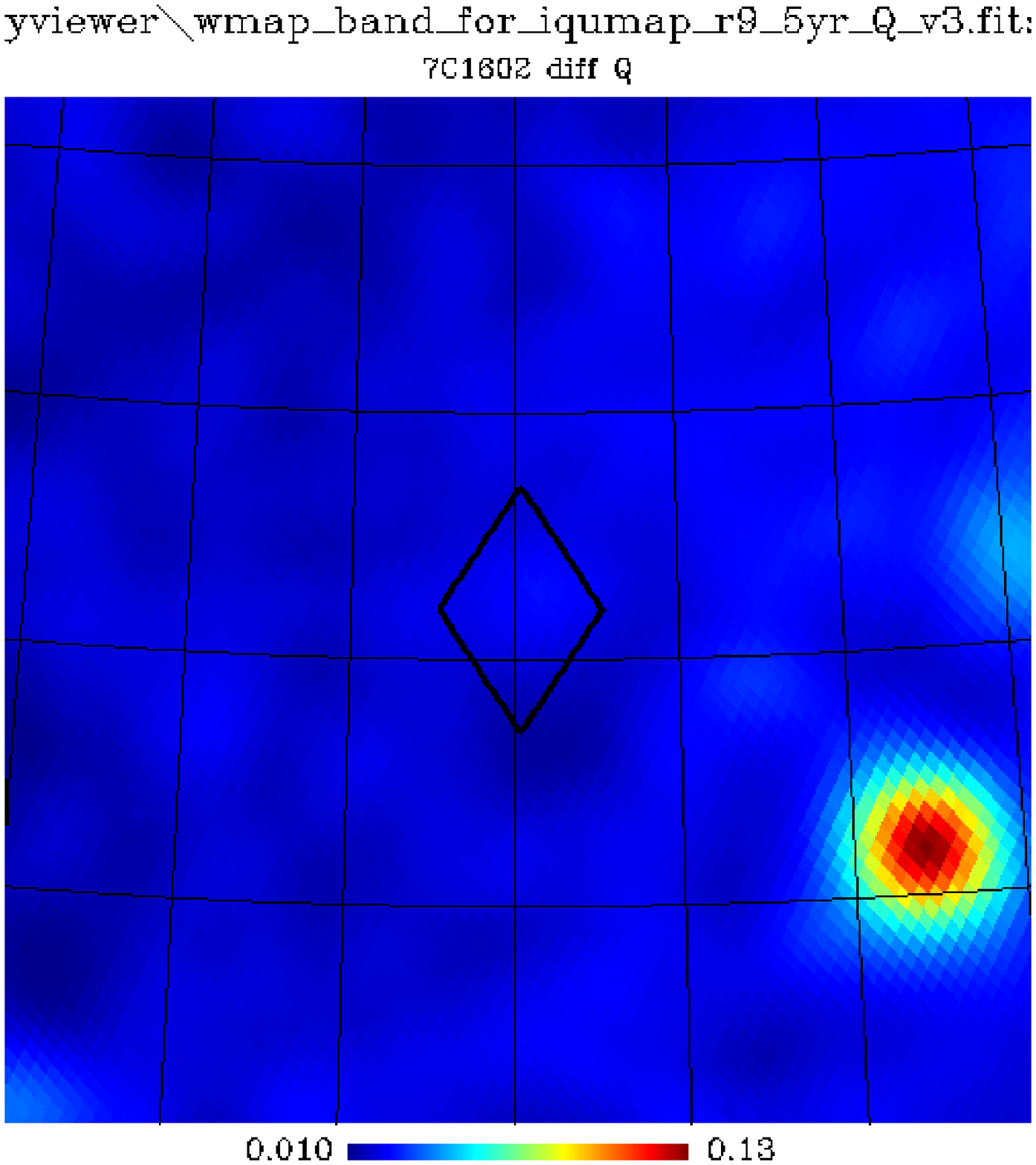,height=3.5cm,angle=0.0}
 \epsfig{file=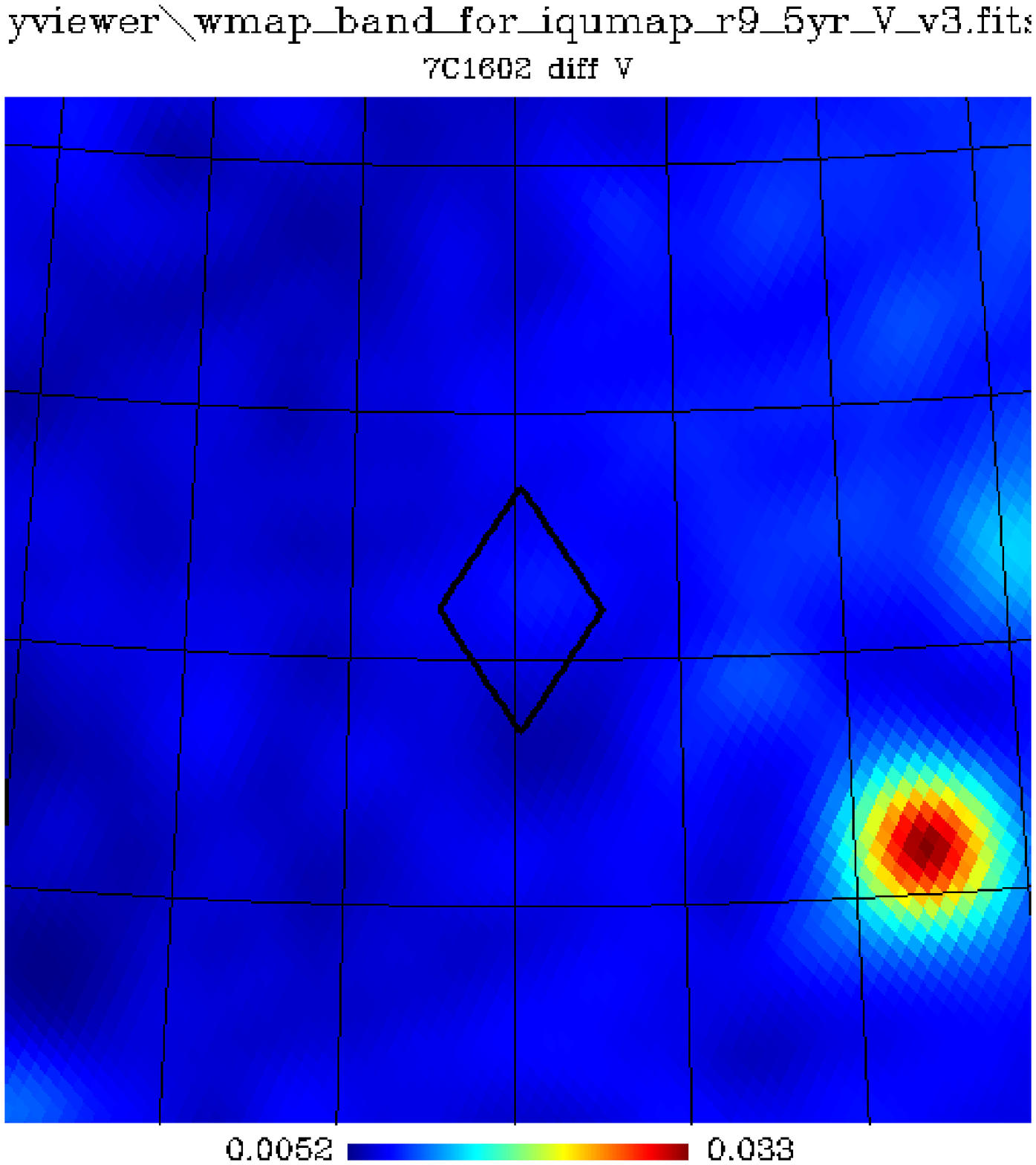,height=3.5cm,angle=0.0}
 \epsfig{file=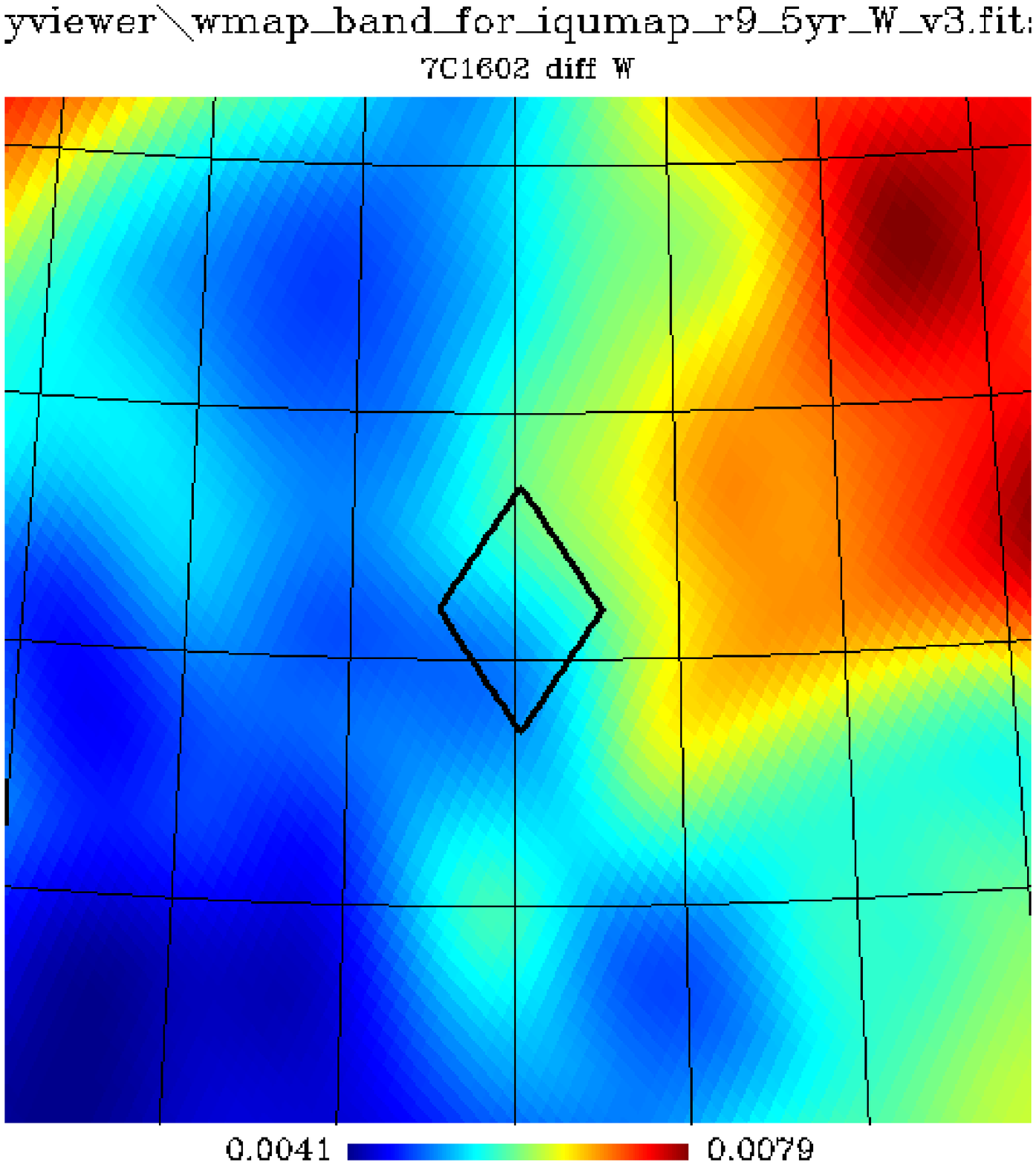,height=3.5cm,angle=0.0}
 }
}
\end{center}
 \caption{\footnotesize{CMB--subtracted WMAP Q (left), V (middle), and W (right) band
maps of the seven objects in Table \ref{tab.4}. From top to bottom:
3C 274.1, 3C 292, 3C 294, 7C 1602+3739.
 }}
 \label{fig.sz.wmap}
\end{figure}

\begin{figure}[ht]
\begin{center}
\vbox{
 \hbox{
 \epsfig{file=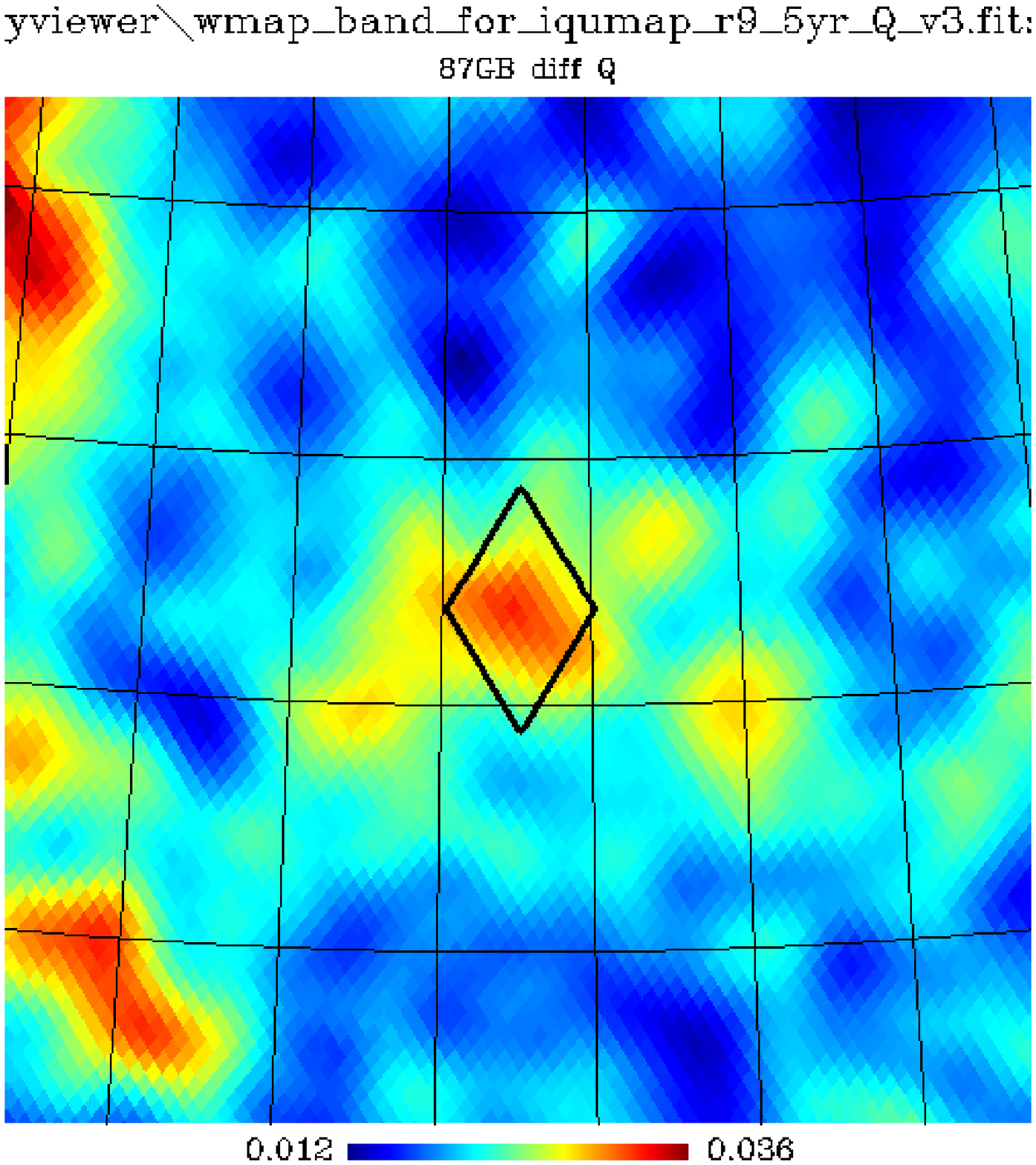,height=3.5cm,angle=0.0}
 \epsfig{file=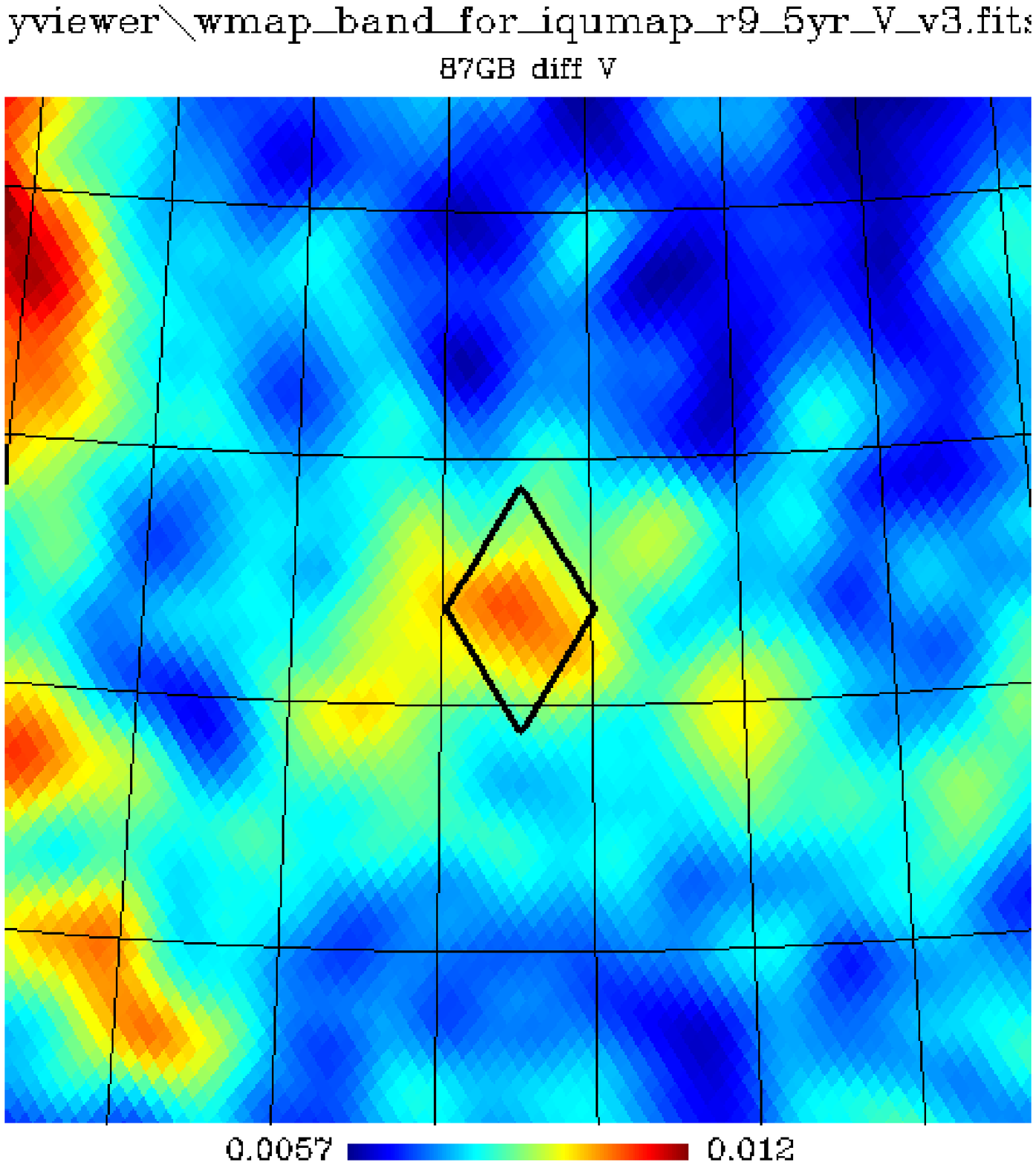,height=3.5cm,angle=0.0}
 \epsfig{file=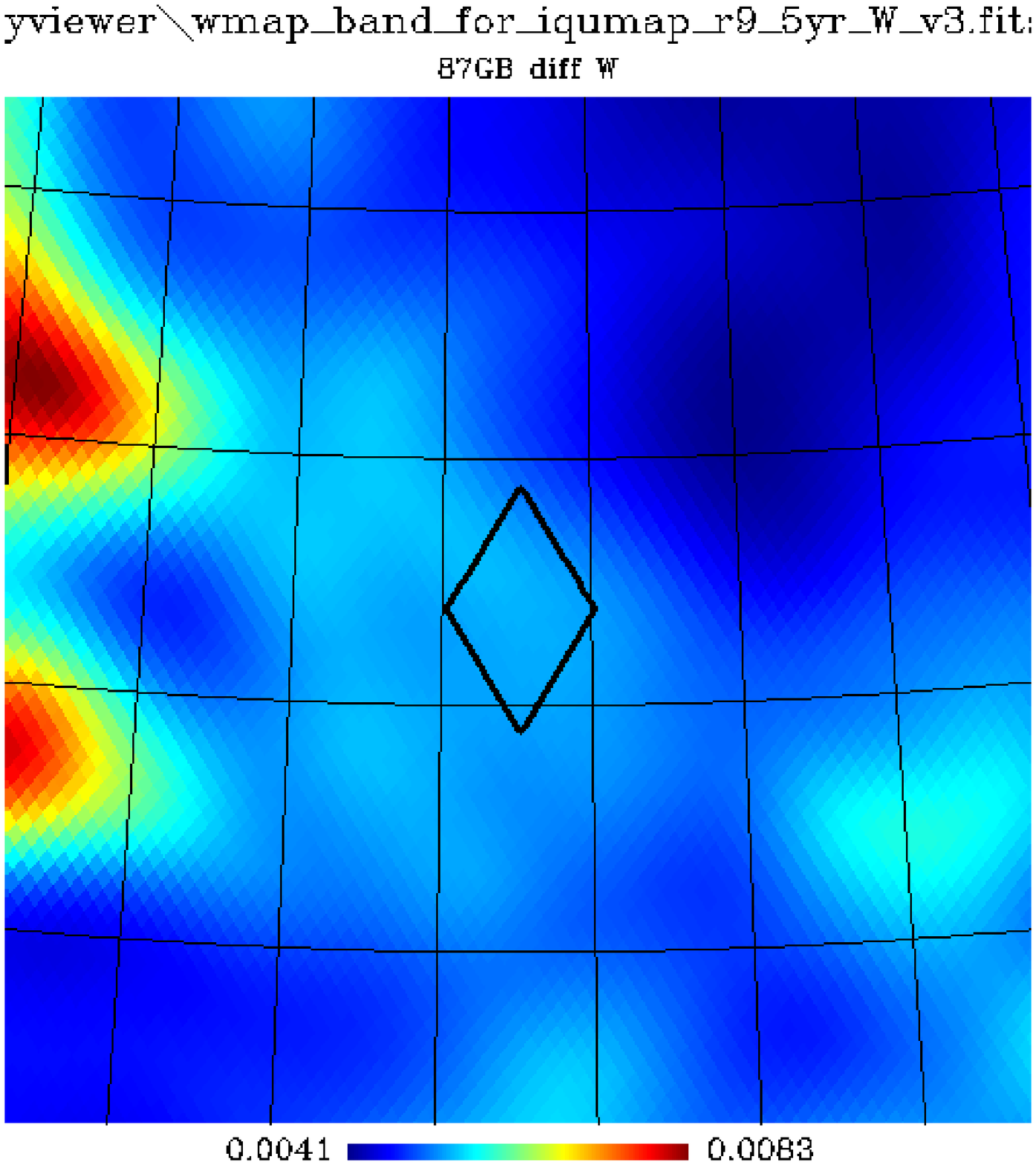,height=3.5cm,angle=0.0}
 }
 \hbox{
 \epsfig{file=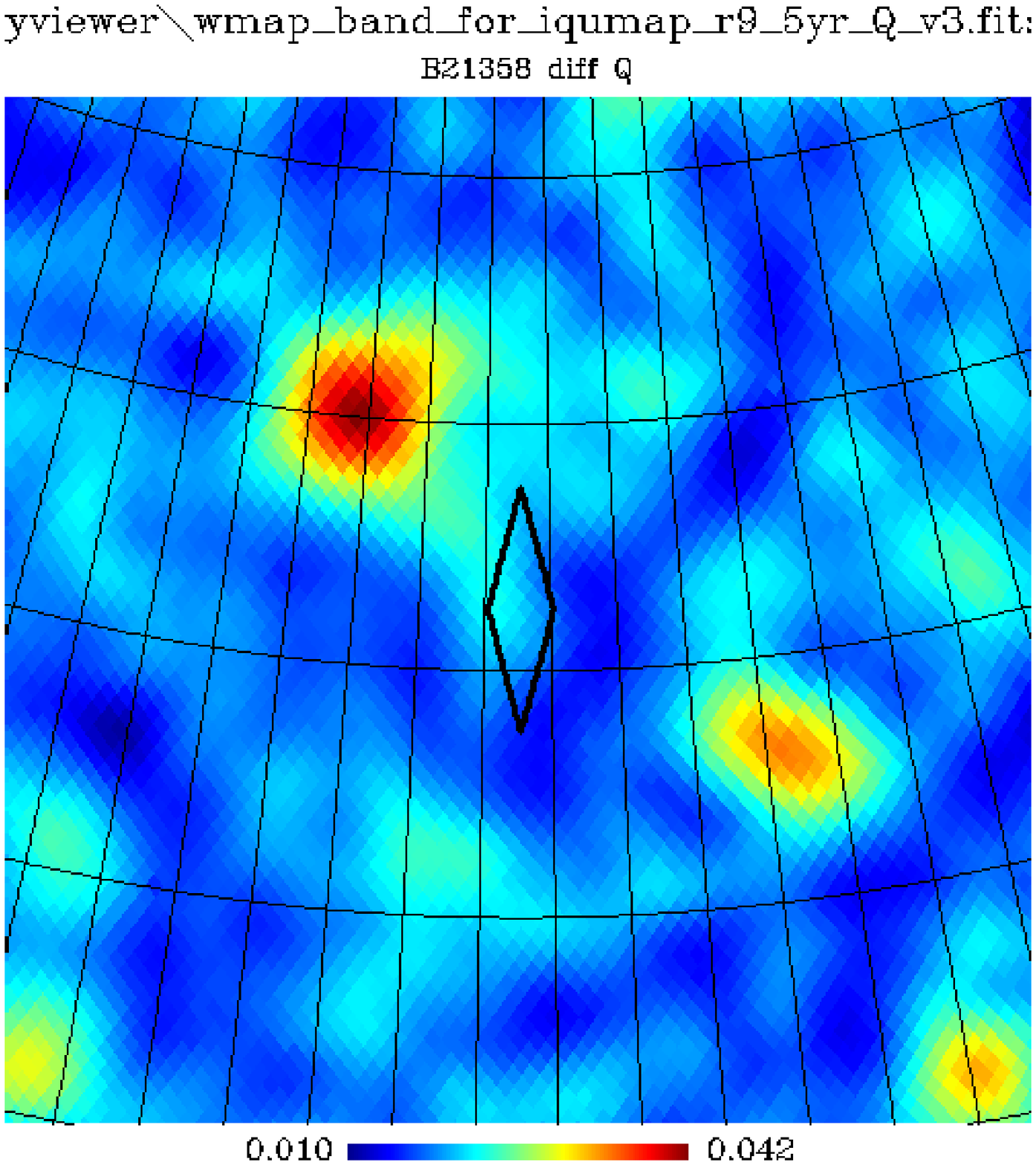,height=3.5cm,angle=0.0}
 \epsfig{file=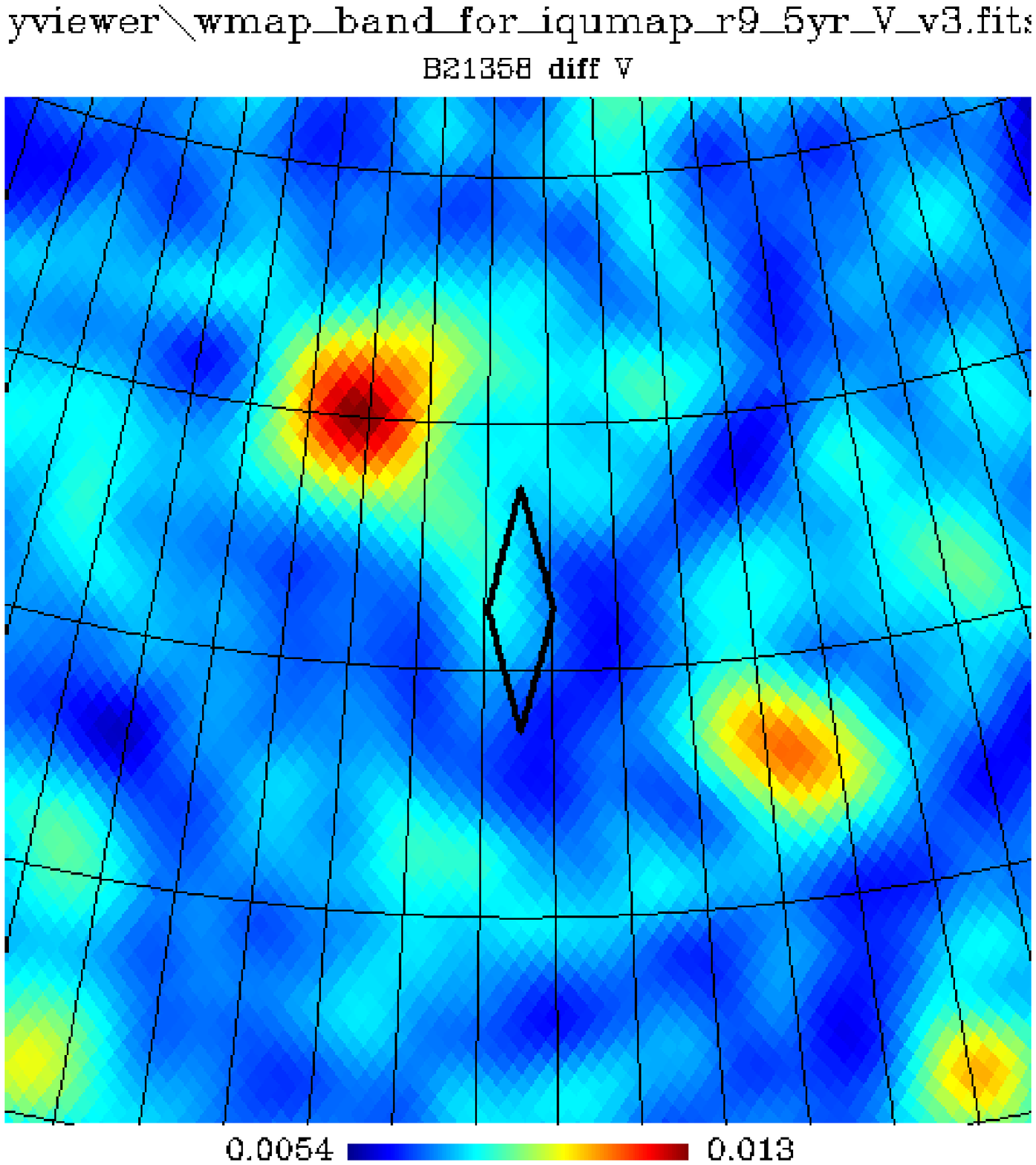,height=3.5cm,angle=0.0}
 \epsfig{file=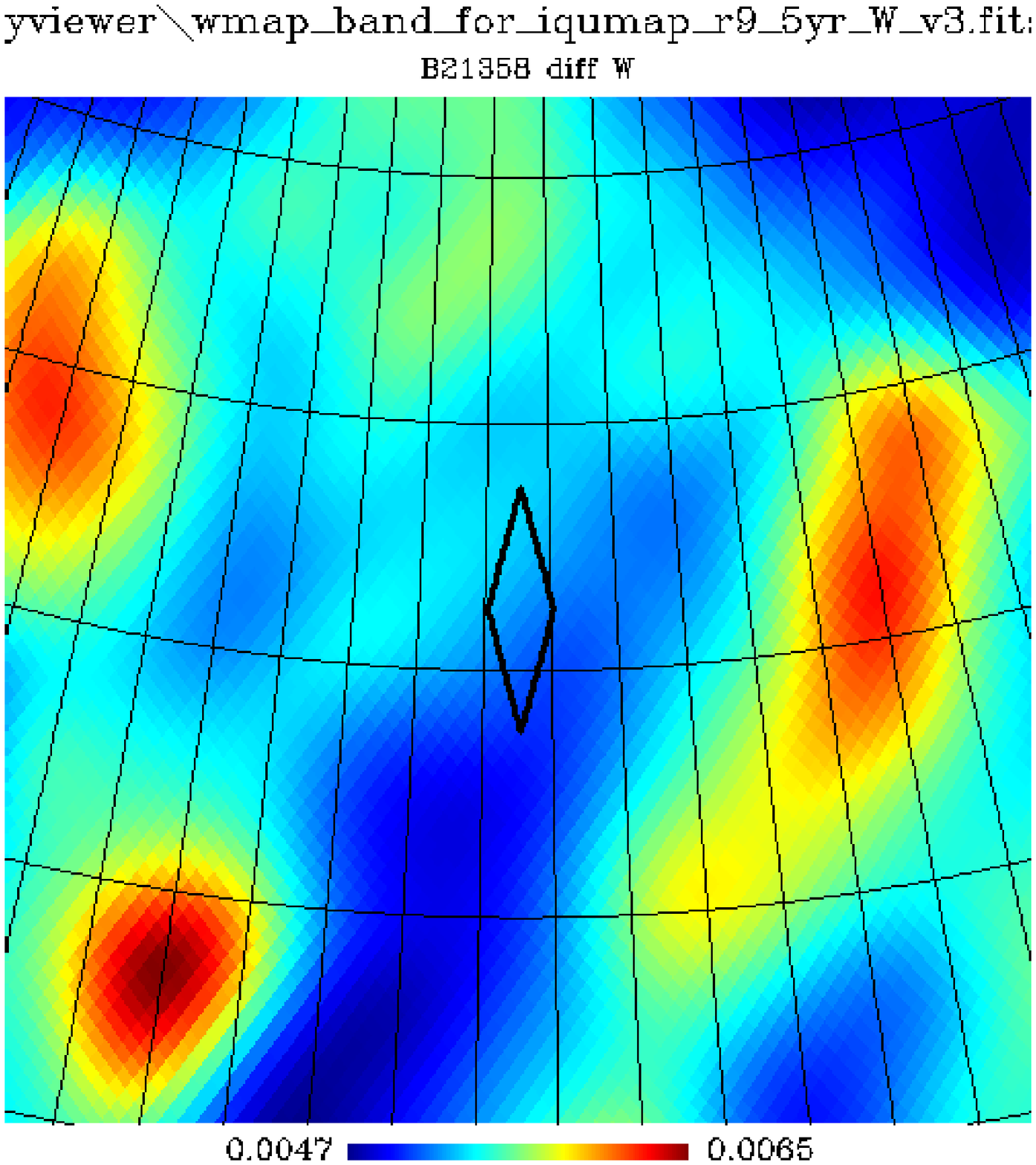,height=3.5cm,angle=0.0}
 }
 \hbox{
 \epsfig{file=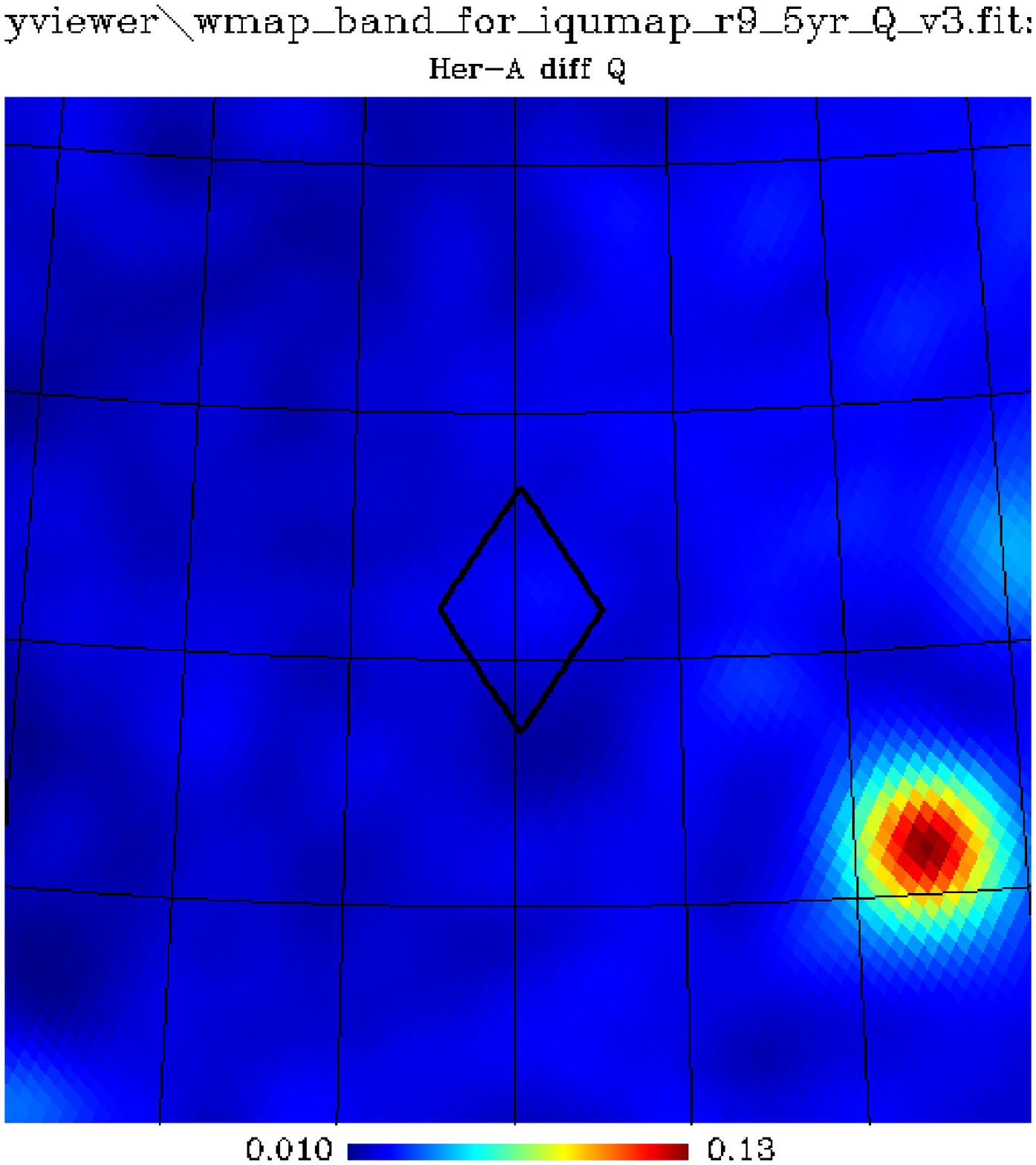,height=3.5cm,angle=0.0}
 \epsfig{file=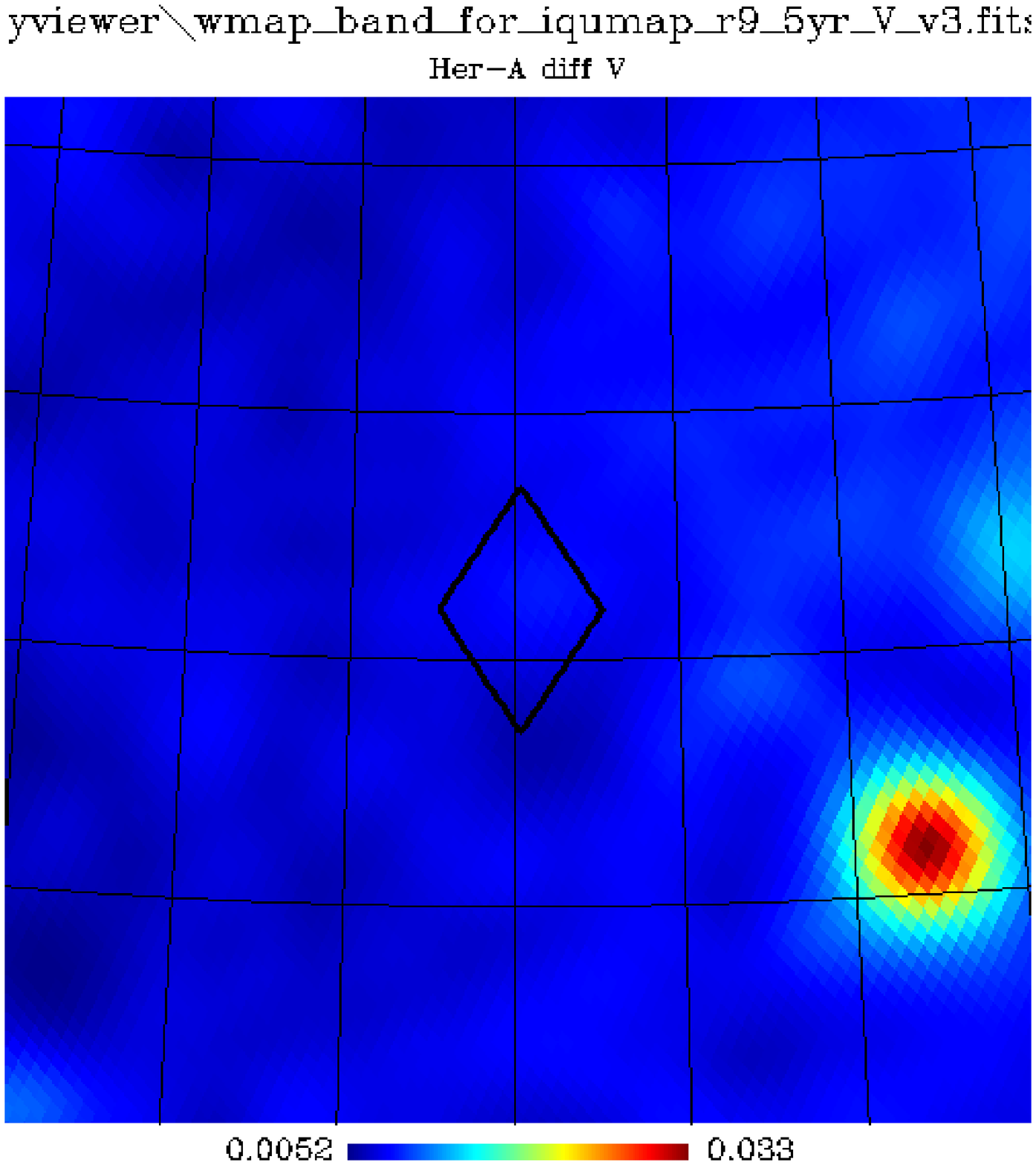,height=3.5cm,angle=0.0}
 \epsfig{file=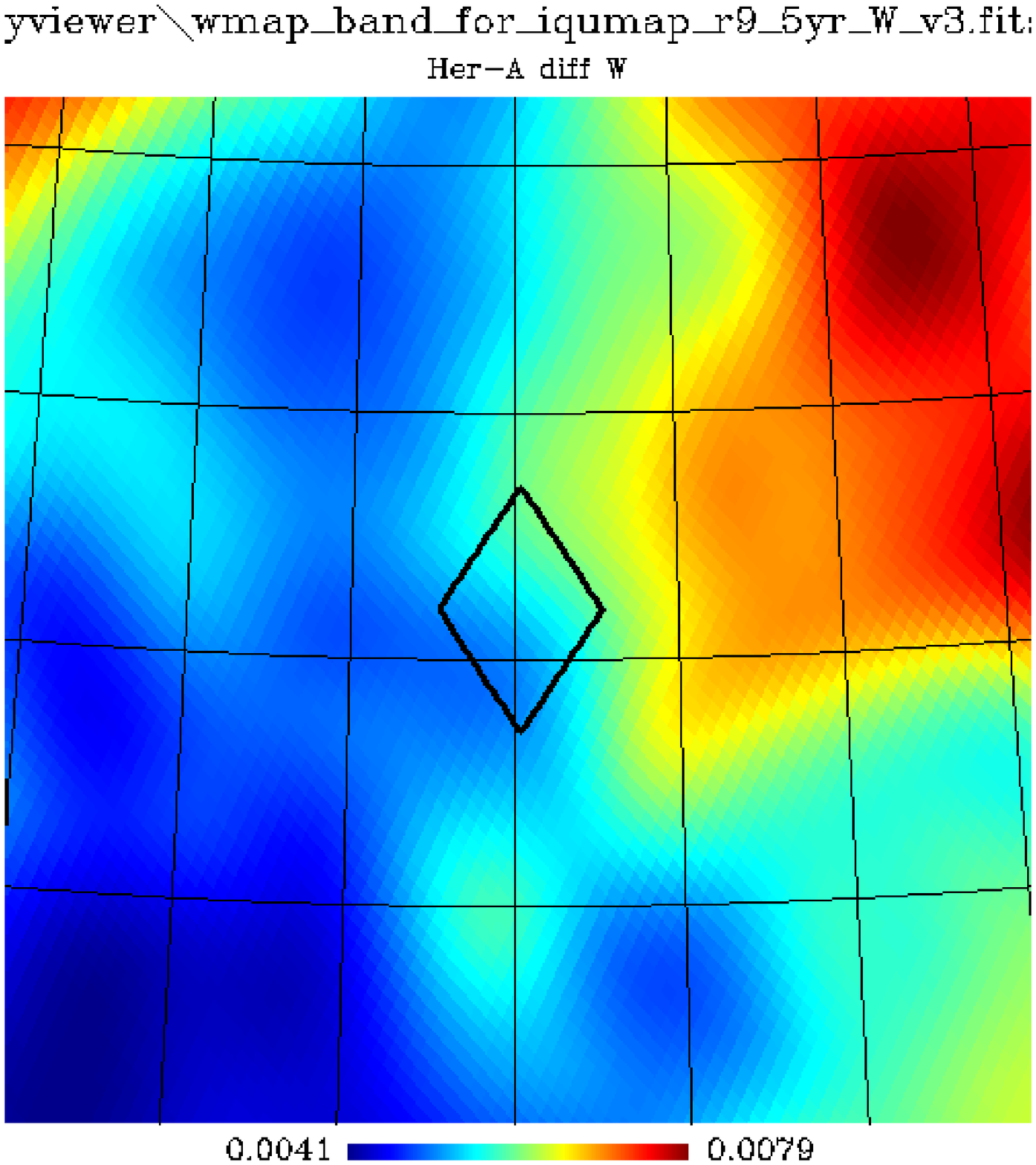,height=3.5cm,angle=0.0}
 }
}
\end{center}
 \caption{\footnotesize{CMB--subtracted WMAP Q (left), V (middle), and W (right) band
maps of the seven objects in Table \ref{tab.4}. From top to bottom:
87GB 121815.5+635745, B2 1358+30C, Hercules A.
 }}
 \label{fig.sz.wmap.2}
\end{figure}

The WMAP Q, V, and W band maps centered on the seven
objects reported in Table \ref{tab.4} are shown in
Figs.\ref{fig.sz.wmap}-\ref{fig.sz.wmap.2}. The total flux of the
source is evaluated within a 1 deg$^2$ box centered on the RG lobe
center.

\begin{figure}[ht]
\begin{center}
 \epsfig{file=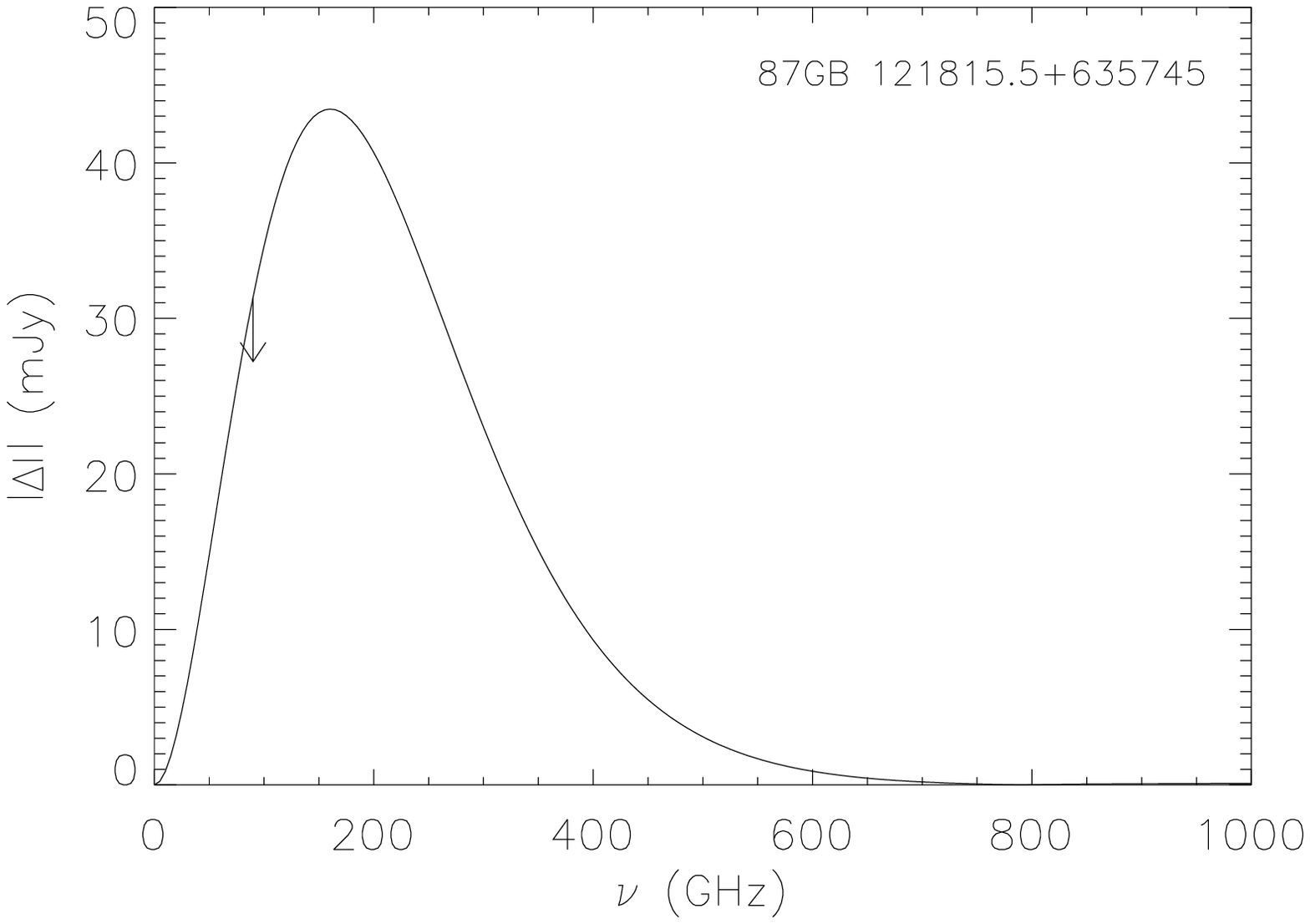,height=5.cm,width=9.cm,angle=0.0}
 \epsfig{file=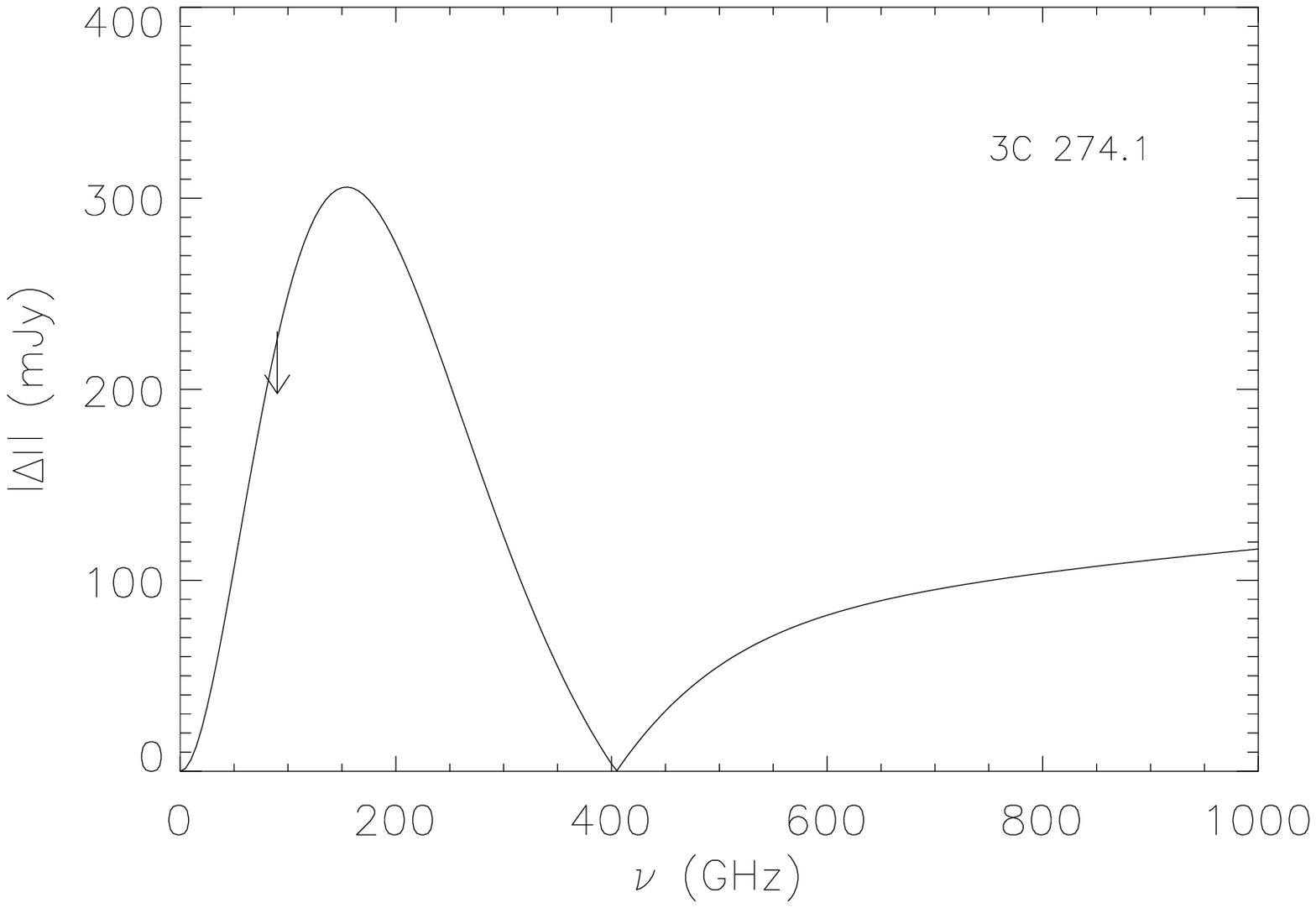,height=5.cm,width=9.cm,angle=0.0}
 \epsfig{file=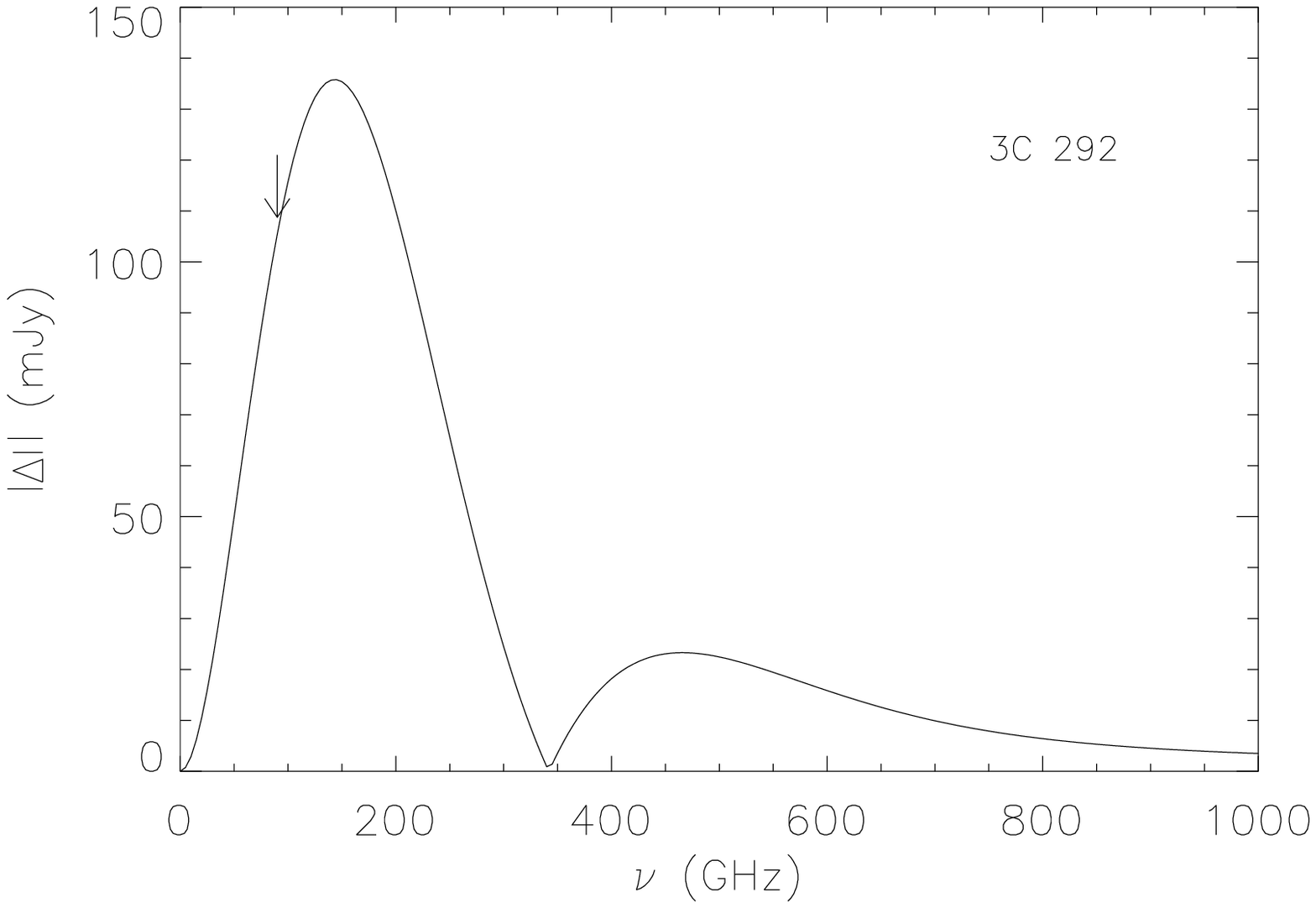,height=5.cm,width=9.cm,angle=0.0}
\end{center}
 \caption{\footnotesize{Absolute value of the SZE spectrum integrated over the whole
 source region for the RGs
 87GB 121815.1+635745 (top), 3C 274.1 (mid), and 3C 292 (bottom)
 calculated for a single power-law spectrum with the values
 of $p_1$ reported in Table \ref{tab.wmap}. We plot the upper limits
 derived from the WMAP W-band data.
 }}
 \label{fig.sz.limsup1}
\end{figure}

\begin{figure}[ht]
\begin{center}
 \epsfig{file=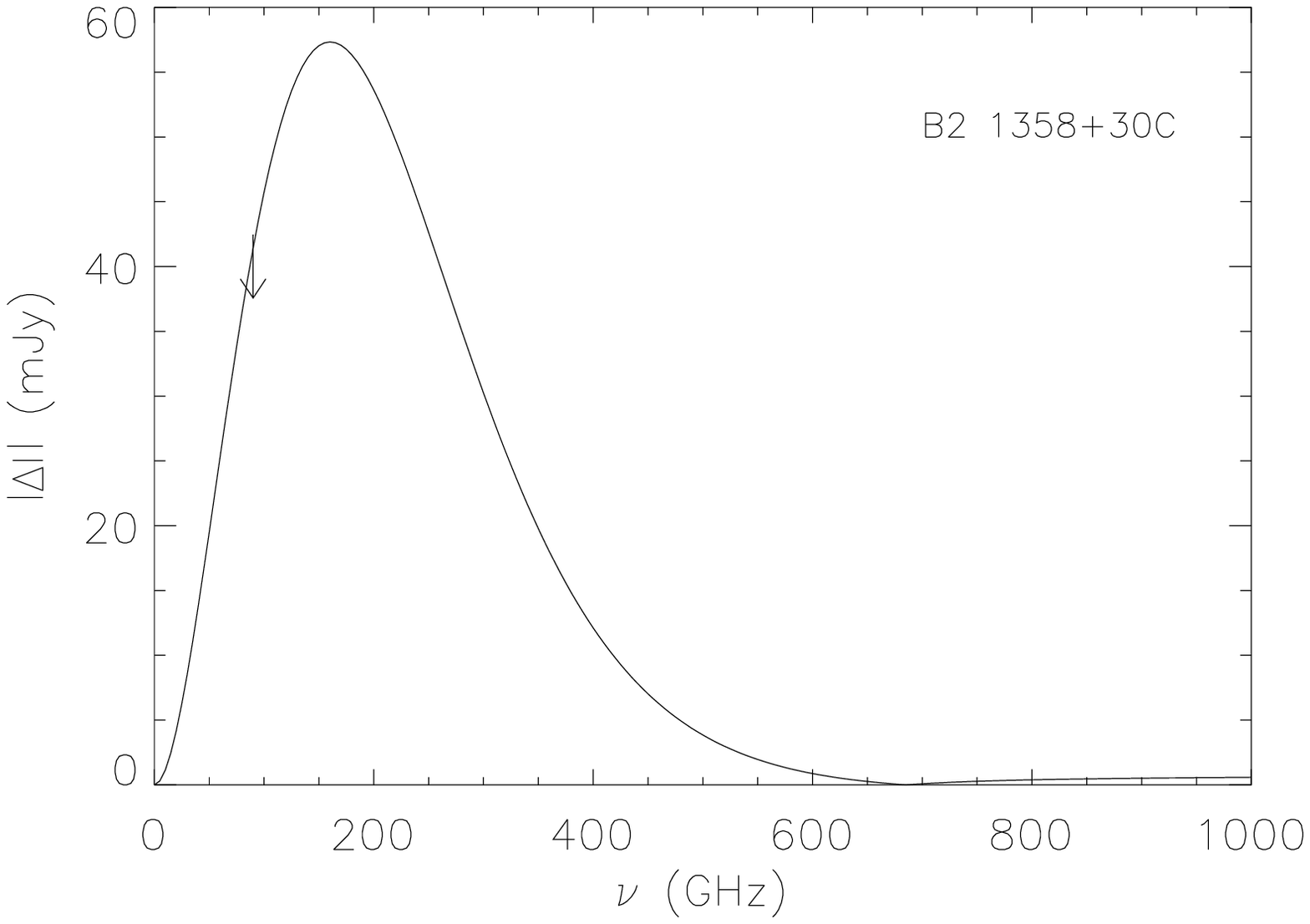,height=5.cm,width=9.cm,angle=0.0}
 \epsfig{file=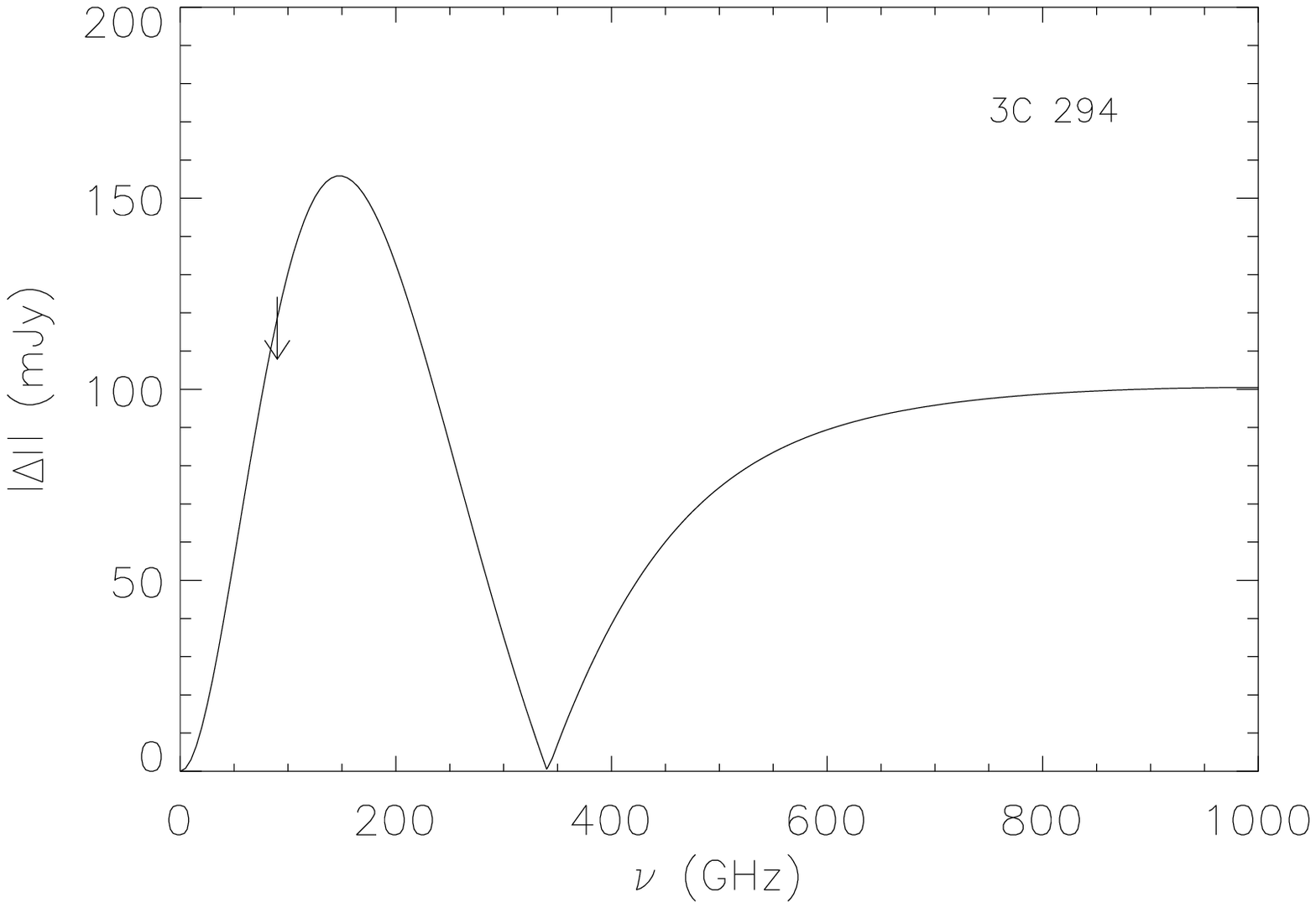,height=5.cm,width=9.cm,angle=0.0}
 \epsfig{file=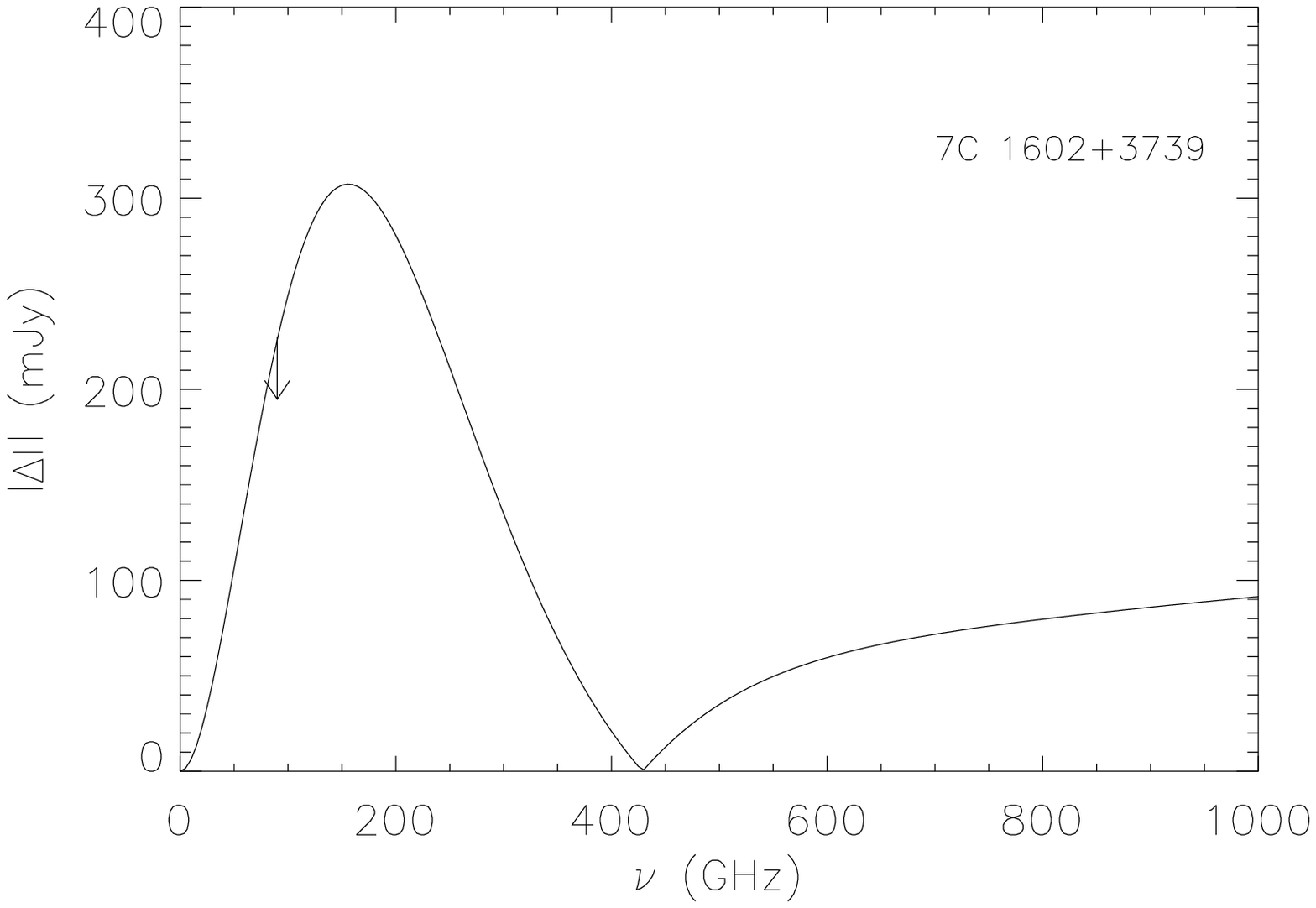,height=5.cm,width=9.cm,angle=0.0}
\end{center}
 \caption{\footnotesize{Same as Fig.\ref{fig.sz.limsup1}, but for the
 RGs B2 1358+300 (top), 3C 294 (mid), and 7C 1602+3739 (bottom).
 }}
 \label{fig.sz.limsup2}
\end{figure}

\begin{figure}[ht]
\begin{center}
 \epsfig{file=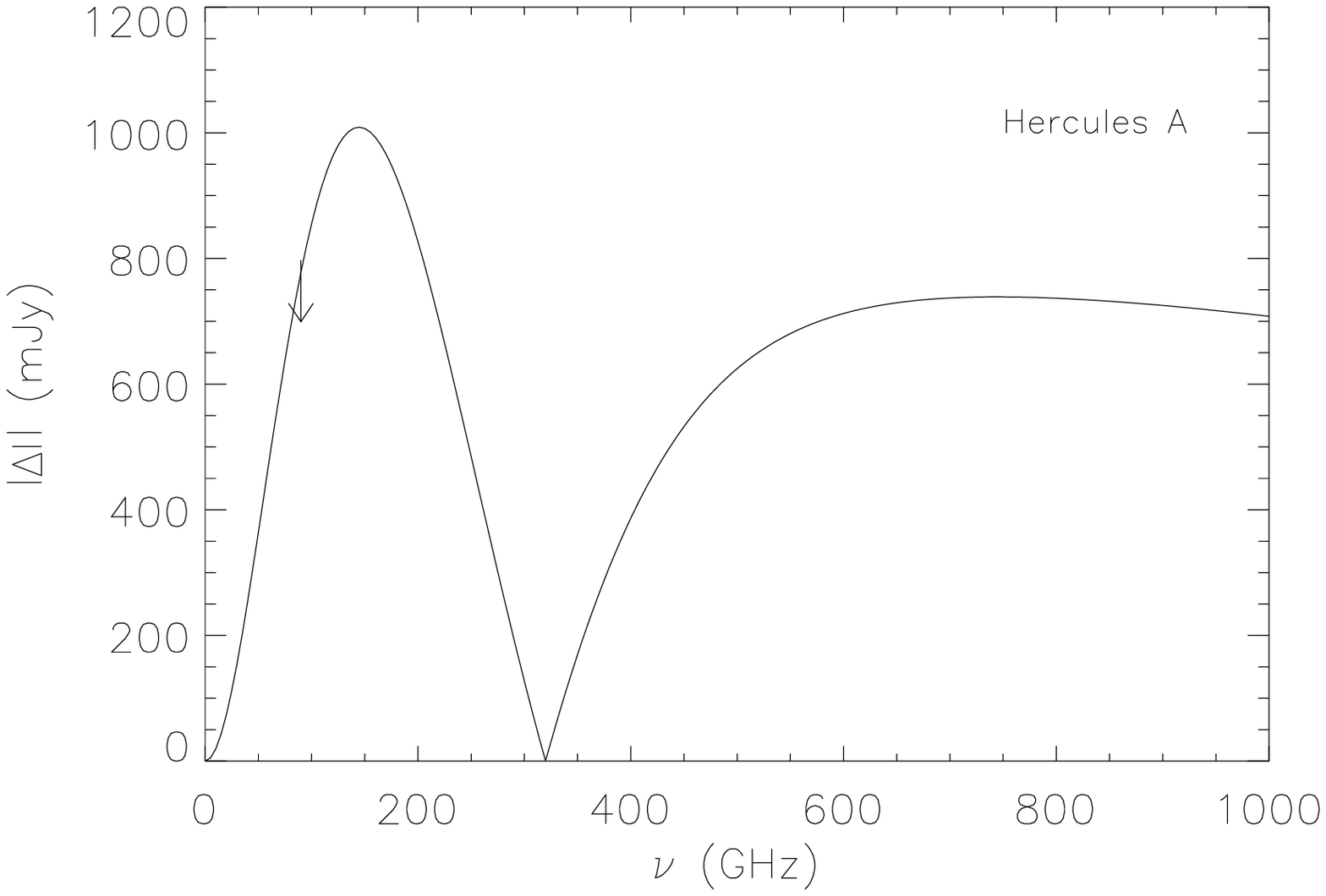,height=5.cm,width=9.cm,angle=0.0}
\end{center}
 \caption{\footnotesize{Same as Fig.\ref{fig.sz.limsup1}, but for the RG Hercules A.
 }}
 \label{fig.sz.limsup3}
\end{figure}

We derive an upper limit on the amplitude of the SZE only
in the W band at 90 GHz by subtracting the synchrotron emission
(extrapolated from lower-frequency radio observations) to the
limit derived from the WMAP 7yr maps and evaluated within a 1
deg$^2$ box centered on the RG lobe center. We considered the WMAP
W channel at 90 GHz because this is the frequency at which the SZE
signal is expected to be stronger and the synchrotron signal is
expected to be smaller.
The SZE flux are calculated using a value of the electron spectrum
normalization $k_0=2.6$ cm$^{-3}$ fixed for all the objects that
have no information in X-rays, and with the specific value of
$k_0$ derived in Sect.4 for the three objects with X-ray data,
i.e., 3C 292, 3C 294, and Hercules A. The SZE fluxes are then set
equal to the derived WMAP W-band limit by adjusting the value of
$p_1$.
This procedure allows the minimum value of $p_1$ (for a
single power-law spectrum) for which the total lobe flux does not
overcome the WMAP W-band limit to be derived.
We also derived, analogously, the minimum value of the break,
$p_b$, for a double power-law spectrum with $p_1=1$ and
$\alpha_1=0$ at low electron momenta). The resulting values of
$p_1$ and $p_b$ are reported in Table \ref{tab.wmap}.\\
The SZE spectra calculated for a single power-law electron
spectrum with the values of $p_1$ obtained from the upper limit of
WMAP at 90 GHz in Table \ref{tab.wmap}, are shown in Figs.
\ref{fig.sz.limsup1}--\ref{fig.sz.limsup3}.
Different values of $p_1$ produce different values of
the optical depth, pressure, and energy density of the electron
population, which is reflected in different spectral shapes of the SZE
in the RG lobes.
For objects with increasing values of $p_1$, we find that
the SZE spectrum is modified with respect to the one shown in
Figs.\ref{fig.87gb12}--\ref{fig.hera}, which was obtained with the same
electron spectrum energy range, both in amplitude and in spectral
shape. In particular, in agreement with the discussion in
Sect.3.2 and Fig.\ref{fig.3c294_p1}, the crossover
frequency $\nu_0$ is found at frequencies higher than the case
when $p_1=1$. In the cases in which its value is particularly
high (e.g., the cases of 87GB 121815.5+635745, B2 1358+30C), the
amplitude of the positive part of the SZE at $\nu \sim 800 - 1000$
GHz is dramatically reduced compared to the other cases.
This effect of $p_1$ on the spectral shape can be verified
by measurements of the SZE in the range 400-850 GHz obtainable
with Planck, OLIMPO, and Herschel-SPIRE. Therefore, the possible
detection (or even additional limits) of the positive part of the
SZE in these RG lobes will better constrain the electron spectrum
and can fully determine the physical parameters of the electron
plasma in the RG lobes (e.g., the value of $p_1$).

We also notice that the values of $p_1$ and $p_b$ found with the
WMAP limits correspond to quite low values of the electron
energies: the maximum value of the minimum momentum ($p_b$ in this
case) for 87GB 121815.5+635745 corresponds to a value of the
electron energy of $\sim 50$ MeV. For these energies, the
electrons in the radio lobes emit synchrotron radiation in the
radio frequencies and produce ICCMB emission in the X-ray
energies at frequencies (energies) lower than the frequency range
at which the lobe emission is actually observed. For the previous
case of $50$ MeV electrons and for a magnetic field of 0.21
$\mu$G (see Table \ref{tab.6}), the spectral break at radio
frequencies should be found at $\sim 9$ kHz, while the ICCMB
emission break should be found at $\sim 0.02$ keV. For all the
other objects we consider here, the spectral breaks should be
visible at even lower frequencies (energies) that cannot be
observed with radio and X-ray instruments. We can conclude,
therefore, that the SZE study of these RG lobes provides the best
method to determine the spectral break or the lower cutoff
$p_1$ of the electron population for these systems.
\begin{table*}[htb]{}
\vspace{2cm}
\begin{center}
\begin{tabular}{|*{6}{c|}}
\hline Object name  & $p_1$ & $p_b$ \\
\hline
 87GB 121815.5+635745 & 63  & 100 \\
 3C 274.1             & 2.8 & 4.3 \\
 3C 292               & 0.0048 & -- \\
 B2 1358+30C          & 28 & 45 \\
 3C 294               & 1.6 & 2.2 \\
 7C 602+3739          & 3.6 & 5.6 \\
 Hercules A           & 1.3 & 1.7 \\
 \hline
 \end{tabular}
 \end{center}
 \caption{\footnotesize{Values of $p_1$ and $p_b$ derived from
 the WMAP W-band limits.
 }}
 \label{tab.wmap}
 \end{table*}

\subsection{Visibility analysis}

We analyze in this section the detectability of the SZE signals
expected for the set of RG lobes selected from Tables 5 and 7 with
the experimental set-up provided by Planck, OLIMPO, and Herschel.
%
%
Table \ref{tab.predictions} shows the SZE signals expected from
these RG lobes in the frequency channels from 100 to 857 GHz
covered by the combination of Planck-HFI, OLIMPO, and
Herschel-SPIRE. For each experiment, we report the signals (in
units of mJy/beam) expected from each source in the frequency
channels of the various instruments. 
The values of the signal-to-noise (S/N) ratio for the various instruments is also given in parentheses at each frequency channel.
For each frequency channel, we
considered the appropriate beam of the instrument; for Planck, we
concentrated only on the HFI channels since these have the highest
spatial resolution. The expected signals were calculated
using the SZE spectra shown in Figs.
\ref{fig.sz.limsup1}--\ref{fig.sz.limsup3}, obtained with the
complete multi-$\nu$ limits at $\mu$waves, X-rays, and gamma-rays.

The flux error calculated for a 1-hour integration with OLIMPO is
1.61, 1.79, 0.54, and 0.29 mJy/beam at $\nu = 143, 217, 353$, and 450
GHz, respectively, and the confusion noise (assumed to be dust
dominated) is 0.2, 0.76, 3.32, and 7.19 mJy/beam at the same
frequency channels.
For the Planck-HFI instrument, we calculate a flux error for a 14-month
integration time of 3.97, 2.81, 3.88, 7.34, 14.29, and 16.67
mJy/beam at $\nu = 100, 143, 217, 353, 545,$ and 857 GHz,
respectively. The confusion noise (assumed to be dust dominated)
at the same Planck-HFI frequency channels is 0.09, 0.26, 0.95,
6.68, 37.94, and 231.99 mJy/beam.
For the Herschel-SPIRE instrument, the detector noise is
negligible with respect to the confusion noise levels of 6.80,
6.30, and 5.80 mJy/beam at $\nu = 600, 857, 1200$ GHz.
The S/N ratio for each instrument channel is
reported in parentheses after the value of the expected SZE flux.

We find that most of the RG lobes have an SZE well detectable up
to 353 GHz with both OLIMPO and Planck, while the S/N ratio
decreases to values of $\simlt$ a few or even below 1 at higher
frequencies, where both the intrinsic noise of the instruments and
the confusion noise largely increase. The Herschel-SPIRE
instrument is therefore crucial to detect the SZE signals from RG
lobes at higher frequencies $> 600$ GHz. For this instrument, the
expected S/N ratios are significantly larger than 1, except in the case
of RG 87GB 121815.5+635745. This RG lobe has a very steep electron
spectrum with a power-law index $\alpha=3.9$, the largest among
the selected RG lobes, which given the WMAP W band limit requires a
rather large value of $p_1$. This value depresses the SZE signal in
the high-frequency region.
As verification of this trend we notice that the RG with the
smaller electron spectrum cutoff (i.e., 3C 294 with $p_1=1.4$) shows
the largest expected SZE signal in all the frequency channels of
the Herschel-SPIRE instrument.

\subsection{Confusion and biases}

Possible sources of confusion of the SZE signal on the spatial
scales of the RG lobes we consider here include the synchrotron
emission from the same RG lobes, CMB fluctuations, point-like
sources and the diffuse Galaxy emission.
The synchrotron emission from the RG lobes does not contribute
substantially to the level of confusion of the SZE signals at 150
GHz, and even less at higher frequency $\simgt 500$ GHz (see
Tables \ref{tab.2} and \ref{tab.5}).
The CMB fluctuations on the angular scales ($\sim 1$ to $15$
arcmin) of RG lobes do not provide a sensitive level of confusion.
In addition they have a flat spectrum that
can be efficiently separated from the RG lobe SZE signal.\\
The Galaxy emission is the major source of contamination. However,
since most of the RG lobes are selected to be at high galactic
latitudes, we can reasonably think that this source of confusion
could be also separated from the SZE signals by knowing its
spectral behavior.\\
The level of unresolved point source contamination is more
difficult to establish and depends also on the model assumed for
the sub-mm and mm source evolution with redshift.
However, the recent HERSCHEL-SPIRE results provide an estimate of
the confusion at the levels of 5.8, 6.3, and 6.8 mJy/beam at 250,
350, and 500 $\mu$m, respectively. The beams of the instrument are
18.1, 24.9, and 36.6 arcsec (FWHM) at 250, 350, and 500 $\mu$m
(N'guyen et al. 2010).\\
In the estimate of the SZE signals expected for
the seven RG lobes with Planck, OLIMPO, and Herschel-SPIRE (see Table
\ref{tab.predictions}), we considered the confusion caused by unresolved
dusty galaxies; this confusion is more relevant for OLIMPO and Planck than
for Herschel-SPIRE.

We conclude that the level of confusion of the RG lobe SZE signal
is marginal at low frequencies (150 GHz), relevant at the level of
intermediate frequencies ($\sim 300-400$ GHz where the crossover
is found), and dominant at high frequencies ($\simgt 500$
GHz). At these high frequencies, most of the astrophysical information on the RG lobes
can be extracted by studying the relative SZE spectral shape.
We find that the RG lobes whose ICS spectra require low values of
the minimum momentum cutoff $p_1$, which are consistent with
observations and upper limits at multi-frequency, are the objects
more easily detectable in all the frequency channels of the three
experiments we considered here. Therefore, they offer the best
targets to recover their overall SZE spectrum from microwaves to
sub-mm ranges.
%
\begin{table*}[htb]{}
\vspace{2cm}
\begin{center}
\tiny{
\begin{tabular}{|*{9}{c|}}
 \hline
                      & 100 & 143 & 217 & 353 & 450 & 600 & 857 & 1200 \\
                      & GHz & GHz & GHz & GHz & GHz & GHz & GHz & GHz  \\
 \hline
 OLIMPO               &     &     &     &     &     &     &     &      \\
 \hline
 87GB 121815.5+635745 &       & -18.8 (11.6)& -8.88 (4.6) & -1.26 (0.4) & -0.31 (0)& & & \\
 3C 274.1             &       & -304 (186.9)& -248 (127.5)& -43.3 (12.9)& 36    (5.0)& & & \\
 3C 292               &       & -136 (83.6)& -92.2 (47.4) & 7.46 (2.2)& 23.2 (3.2)& & & \\
 B2 1358+30C          &       & -55.3 (34.0)& -39.9 (20.5)& -9.5  (2.8) & -2.45 (0.3)& & & \\
 3C 294               &       & -156  (95.9)& -114 (58.6) & 14.2 (4.2)  & 62.8 (8.7)& & & \\
 7C 1602+3739         &       & -306 (188.2)& -254 (130.5)& -58.6 (17.4)& 16.8 (2.3) & & & \\
 Hercules A           &       & -1009 (620.5)& -693 (356.2)& 205 (60.9)& 458 (63.6) & & & \\
 \hline
 \hline
 Planck               &      &    &     &     &      &     &    &       \\
 \hline
 87GB 121815.5+635745 & -31.3 (7.9)& -30.2 (10.7)& -15.5 (3.9)& -5.95 (0.6)& -0.73 (0)& & 0.022 (0) & \\
 3C 274.1             & -249 (62.7)& -304 (107.9)& -254 (63.6)& -51.4 (5.2)& 69.7 (1.7)& & 108 (0.5)& \\
 3C 292               & -116 (29.2)& -135 (47.9)& -95.8 (24.0)& 4.78 (0.5)& 19.8 (0.5)& & 5.19 (0)& \\
 B2 1358+30C          & -45.7 (11.5)& -56.5 (20.0)& -48.4 (12.1)& -18.5 (1.9)& -2.03 (0.1)& & 0.45 (0)& \\
 3C 294               & -130 (32.7)& -156 (55.4)& -118 (29.5) & 9.26 (0.9) & 82.7 (2.0) & & 99.7 (0.4) & \\
 7C 1602+3739         & -249 (62.7) & -305 (108.2) & -260 (65.1) & -66.2 (6.7)& 48.4 (1.2) & & 83.3 (0.4) & \\
 Hercules A           & -855 (215.3) & -1009 (358.0)& -720 (180.2) & 184 (18.5) & 676 (16.7)  & & 731 (3.1)  & \\
 \hline
 \hline
 HERSCHEL             &     &     &      &     &      &    &    &        \\
 \hline
 87GB 121815.5+635745 &     &       &       &       &       & -0.48 (7.3) & 0.016 (0.3) & 0.047 (1) \\
 3C 274.1             &     &       &       &       &       & 77.2 (70.4) & 106 (112.7) & 126 (158) \\
 3C 292               &     &       &       &       &       & 17.5 (13.8) & 5.65 (5.2)   & 2.48 (2.7) \\
 B2 1358+30C          &     &       &       &       &       & -1.3 (7.6)  & 0.44 (3.0) & 0.7 (5.6)  \\
 3C 294               &     &       &       &       &       & 87 (35.8)   & 99.3 (47.6)  & 99.5 (56.3)  \\
 7C 1602+3739         &     &       &       &       &       & 55.4 (59.1) & 81.8 (101.7) & 101 (148.2)  \\
 Hercules A           &     &       &       &       &       & 700 (1752.4)& 734 (2140.8)  & 664 (2284.9)  \\
 \hline
 \end{tabular}
 }
 \end{center}
 \caption{\footnotesize{SZE signals for the restricted set of RG lobes expected in the Planck, OLIMPO, and Herschel-SPIRE instruments.
 }}
 \label{tab.predictions}
 \end{table*}

\section{Discussion and conclusions}

We have presented in this paper the first extensive study of the
relativistic SZE in RG lobes that makes use of the available
observational constraints over a wide frequency range, from radio
and microwave to X-ray and gamma-ray frequency bands.
The constraints on the electron spectral slope and on the
intensity at microwaves (i.e., at 90 GHz from the WMAP W-band),
X-ray, and gamma-ray (i.e., Fermi-LAT) allow limits to be derived for the
value of the electron spectrum normalization and the minimum
electron momentum $p_1$, and in turn for the amplitude and
spectral shape of the SZE in RG lobes. The inclusion of X-ray
observations from some of the RG lobes in our sample (assuming
that the X-ray flux data are produced by ICCMB) further allow
realistic estimates on the overall value of the magnetic
field in the RG lobes to be established. They also enable
accurate determination of
the spectrum of the relativistic electrons over a wide range of
energies that fit the radio synchrotron emission, the ICCMB
X-ray, and the gamma-ray emission limits from RG lobes.
With these constraints, we determined the realistic spectrum and
amplitude of the non-thermal SZE in the RG lobes considered in our
analysis.

A detailed visibility analysis of the SZE for the selected RG
lobes showed that balloon-borne and space-borne
experiments like Planck-HFI, OLIMPO, and Herschel-SPIRE will be
able to determine accurately the SZE spectral characteristics
in many RG lobes and open the way to the vast
astrophysical and cosmological use of this specific SZE signal.\\
The largest expected SZE signals from RG lobes occur at microwaves
(in the region of its minimum at $\sim 150-170$ GHz) and in the
sub-mm range (in the region of the quite flat maximum at $\nu
\simgt 600$ GHz, which also could be extended at higher frequencies
depending on the electron spectral cutoff $p_1$).
The combination of PLANCK, Herschel-SPIRE, and OLIMPO provides an
excellent framework to study the SZE from giant RG lobes and
demonstrates the importance of future experiments in the mm and
sub-mm range with spatially resolved spectroscopic capabilities.

Future SZE observations of RG lobes will be crucial for the
determination of the energetics, pressure, and spatial distribution
of relativistic electrons in the RG lobes, and hence for the
description of the very high-E phenomena associated with these
systems.
Knowing the SZE spectra will allow the total energy and
spectrum of the relativistic electrons in RG lobes
to be measured, because the
SZE depends on the total pressure/energy density of the electronic
population along the los through the lobe. SZE measurements
provide much more accurate estimates of the electron
pressure/energy density than techniques like ICCMB X-ray
emission or synchrotron radio emission, since the former can only
provide an estimate of the electron energetics in the high-energy
part of the electron spectrum, and the latter is sensitive to the
degenerate combination of the electron spectrum and of the
magnetic field in the radio lobes (see Colafrancesco \&
Marchegiani 2011).

We have shown that the combination of the SZE and radio
observations of RG lobes also allow the
magnetic field structure in RG lobes
to be determined more reliably than the combination of ICCMB
X-ray (or gamma-ray) and radio emission. This is because the SZE
is a measure of the total energy density, i.e., an integral over
the whole electron spectrum. X-ray observations in a fixed
energy band, however, can provide an estimate of the electron spectrum in a
slice of the electron spectrum that does not fit with the electron
energy range probed by radio synchrotron observations $E_e \approx
7.9 \,GeV\, B_{\mu}^{-1/2} (\nu/GHz)^{1/2}$ ($B_\mu$ is the
magnetic field expressed in $\mu$G) unless very high B-fields are
present in the radio lobes (see Colafrancesco \& Marchegiani
2011). Specifically, an electron of energy $E_e$ can produce both
ICS and synchrotron emission for a value of the magnetic field
\begin{equation}
 B_{\mu} = \bigg({7.9 \over 0.35}\bigg)^2 \bigg({\nu \over GHz}\bigg)
 \bigg({E_X \over keV}\bigg)^{-1},
 \label{eq.energy_radio_ics}
\end{equation}
which requires magnetic field values of order of $\sim 255$ to
$2550$ $\mu$G to perform observations at an X-ray energy of
$E_X=0.2$ keV and for radio observations in the range $0.1 - 1$
GHz.
The assumption of a constant slope of the electron spectrum from
the energies probed by radio observations (i.e., $\sim$ several to
many GeV) down to the energies probed by ICS X-ray observations
(i.e., $\sim$ tenths to a GeV) is required to use the relation
$F_{ICS}/F_{synch} = {\cal E}_{CMB}/{\cal E}_B$, which provides the
estimate for the magnetic field energy density ${\cal E}_B$, given
the known value of the CMB radiation energy density ${\cal
E}_{CMB}$.
For three of the seven RGs that we consider in detail, this technique
allows the value of the (uniform) B-field to be derived at the level of
$\approx$ 2.7 (for 3C 292), $\simgt$ 3.0 (for Hercules A), and
$\simgt$ 0.41 (for 3C 294) $\mu$G.\\
The spatially resolved study of the SZE and the synchrotron
emission in the lobes of RGs can provide indications on
the radial behavior of both the leptonic pressure and the
magnetic field from the inner parts to the boundaries of the
lobes.
The study of the pressure evolution in radio lobes can give
crucial information on the transition from radio-lobe environments
to the atmospheres of giant cavities observed in galaxy cluster
atmospheres (see, e.g., McNamara \& Nulsen 2007 for a review),
which seem naturally related to the penetration of RG
jets/lobes into the intra-cluster medium.\\
The detection of the SZE from RG lobes also has a
relevant cosmological implication since it can be used, in
combination with ICS X-ray emission measurements, to measure the
evolution of the CMB temperature at each RG redshift (as proposed
by Colafrancesco 2008).
The possibility to detect RG lobes at high redshifts then allows
the evolution of $T_{CMB}(z)$ to be probed in the early ages of the
cosmic history. This possibility, coupled to the study of
$T_{CMB}(z)$ at low and moderate redshifts by using galaxy
clusters (see, e.g., Battistelli et al. 2003) allows a reconstruction of
a large part of the cosmic evolution of the CMB radiation field
and of the possible energy injections at early and late epochs.

Preliminary studies of the SZE in RG lobes are limited so far to a
few cases at radio frequency (see, e.g., Yamada et al. 2010). The
low frequency of these observation do not permit
non-thermal SZE to be detected because of the low expected brightness of these
signals in this part of the e.m. spectrum. However, the next
generation radio telescopes like SKA with sensitivity $\simlt
\mu$Jy and wide spectral coverage, $\sim 0.5 - 45$ GHz, will be
able to provide crucial information on the low-$\nu$ part of the
SZE spectrum and its polarization level (Colafrancesco et al. 2012
in preparation).
The SZE from RG lobes could be also observed with mm arrays like
ALMA, especially in small-size lobes with dimensions matching the
ALMA field of view. Even though the specific sources we selected in
this paper are located in the northern hemisphere and thus not
visible by SKA and/or ALMA, it will be relevant for the detailed
study of the RG lobes to have a simultaneous observation of RG
lobes from $\sim 0.4$ GHz to $\sim 40$ GHz, as well as in the ALMA
1-3 bands, with arcsec spatial resolution in order to disentangle
the diffuse synchrotron emission from the relative ICS emission.
This strategy will allow the spatial and energy
distribution of the relativistic electron population
to be separated from that of
the magnetic field in RG lobes. We will address this issue in more
detail for SKA and ALMA in a forthcoming paper (Colafrancesco et
al. 2012 in preparation).

We finally stress the strong complementarity between SZE
observations and high-E observations of the ICCMB emission in RG
lobes, which are possible mainly in the X-rays (Chandra), hard
X-rays (NuSTAR, Astro-H) and in the gamma-rays (Fermi, HESS, CTA).
The preliminary studies of the $\mu$wave--mm and X-ray --
gamma-ray emission from RG lobes will allow to determine the full
energy spectrum of the electronic population in RG lobes and their
interaction properties with external radiation and magnetic
fields. We will address this important issue in a future study.

\begin{acknowledgements}
We thank an anonymous referee for several useful comments and
suggestions that allowed us to improve the discussion and the
presentation of our results. S.C. acknowledges support by the
South African Research Chairs Initiative of the Department of
Science and Technology and National Research Foundation and by the
Square Kilometre Array. Our research was also
supported by contracts  OLIMPO I/004/07/1 and PLANCK-HFI
I/030/10/0 of the Italian Space Agency.
\end{acknowledgements}


\end{document}